\documentclass[aps,prx,showpacs,twocolumn]{revtex4-1}

\usepackage{xcolor}
\usepackage{amsmath}
\usepackage{amssymb}
\usepackage{graphicx}
\usepackage{hyperref}
\usepackage{comment}

\begin{document}
\title{Generic transport formula for a system driven by Markovian reservoirs}
\author{Tony Jin}
\author{Michele Filippone}
\author{Thierry Giamarchi}

\affiliation{Department of Quantum Matter Physics, Ecole de Physique University
of Geneva, Quai Ernest-Ansermet 24, CH-1211 Geneva 4, Switzerland}

\begin{abstract}
We present a generic, compact formula for the current flowing in interacting and non-interacting systems which are driven out-of-equilibrium by biased reservoirs described by Lindblad jump operators. We show that, in the limit of high temperature and chemical potential, our formula is equivalent to the well-known Meir-Wingreen formula, which describes the current flowing through a system connected to fermionic baths, therefore bridging the gap between the two formalisms.
Our formulation gives a systematic way to address the transport properties of  correlated systems strongly driven out of equilibrium. As an illustration, we provide explicit calculations of the current in three cases : {\it i)} a single-site impurity {\it ii)} a free fermionic chain {\it iii)} a fermionic chain with loss/gain terms along the chain. In this last case, we find that  the current across the system has the same behavior for loss or gain terms and depends on the loss/gain rate in a non-monotonic way.
\end{abstract}
\maketitle

\section{Introduction}

The formulation by Landauer and B\"uttiker~\cite{landauer_electrical_1970,buttiker_four-terminal_1986,lesovik_scattering_2011} of the current through mesoscopic regions underpins our understanding of  electron transport in quantum-coherent systems. It makes explicit the connection between the current and the local properties of the finite region (its transmission coefficients) and the distribution functions of connected reservoirs, and has been extremely successful to deal with transport in non-interacting systems, such as disordered systems or Fermi liquids. Moving from the transport in non-interacting systems to the understanding of strongly correlated systems  remains one of the most challenging and not yet fully achieved tasks in quantum Physics. Beyond  the well-established and practical interest in the context of transport measurements in bulk solid-state systems~\cite{ashcroft_solid_2003,ziman_electrons_1960,coleman_introduction_2015} and nanoscopic devices~\cite{bruus_many-body_2004,nazarov_quantum_2009,akkermans_mesoscopic_2007,moskalets_scattering_2011}, the recent realization of novel experimental platforms, probing stationary quantum transport in synthetic quantum matter systems  relying on circuit-QED~\cite{fitzpatrick_observation_2017,chiaro_direct_2020,ma_dissipatively_2019,dutta_out--equilibrium_2020}, quantum dot arrays~\cite{zajac_scalable_2016,hensgens_quantum_2017,mills_shuttling_2019} and  atomtronics~\cite{amico_roadmap_2020,amico_quantum_2005,seaman_atomtronics_2007,stadler_observing_2012,*brantut_conduction_2012,*brantut_thermoelectric_2013,*krinner_observation_2015,lebrat_band_2018,jepsen_spin_2020,ShunEsslingerGiamarchi_superfluidpointcontact,eckel_interferometric_2014,*eckel_contact_2016,cominotti_optimal_2014,*gutman_cold_2012,*Filippone2016b,*papoular_increasing_2012,*simpson_one-dimensional_2014,*Salerno2019,*Filippone2019,*Greschner2019,*nietner_transport_2014,*rancon_bosonic_2014}, paves the way to accessing novel and unexplored transport regimes also far away from equilibrium.

\begin{figure}[b]
\begin{centering}
 \includegraphics[width=0.9\columnwidth]{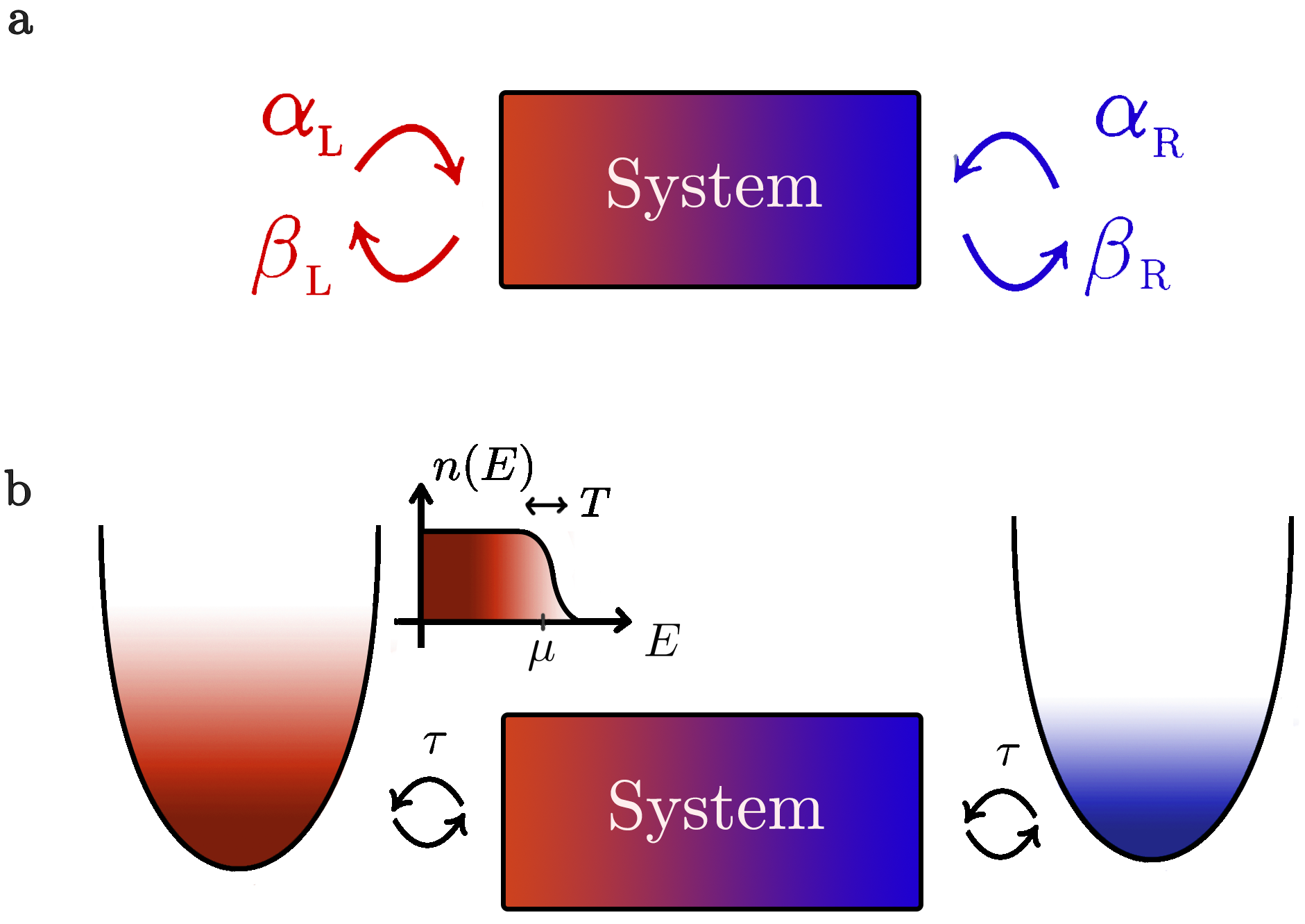}
\end{centering}
\caption{\label{fig:The-two-situations} The two type of systems considered in this study. \textbf{a.} System coupled to the exterior via Lindblad-type creation and annihilation operators.
There is no knowledge of the structure of the environment. All memory effects are discarded in this setting. \textbf{b.} Explicit coupling to a fermionic bath described at equilibrium by the Fermi-Dirac distribution.}
\end{figure}

One step towards the understanding of such transport properties in interacting systems, was provided by a remarkable generalization, by Meir and Wingreen~\cite{MeirWingreenformula} (MW), of the Landauer-B\"uttiker formula to the case of an interacting system.
This generalization, expressing the current in terms of local Green's functions of the system in presence of the reservoirs, provided a unified framework in which understanding the transport was akin to finding approximate (or exact) ways of computing such
Green's function of the system in presence of two fermionic reservoirs (see Fig.~\ref{fig:The-two-situations}). Indeed in such class of systems a stationary current is usually generated by letting the system exchange particles at different rates with two (or multiple) reservoirs.

An alternative way to view the coupling of a quantum system to the external world, used in particular routinely in the context of quantum optics \citep{BreuerPetruccione_book,GardinerZoller_quantumnoise}, is to describe the evolution of the system by
Lindblad-type generators~\citep{Lindblad_seminalpaper,Gorini1976}. Such description consists of Markovian processes by which the system has a non-unitary evolution due to some coupling to the external world. The Lindblad description where the operators would either
inject or absorb particles could thus replace the coupling to external fermionic reservoirs in order to generate a steady state current through a quantum system, as depicted in Fig.~\ref{fig:The-two-situations}. The Lindblad evolution, fully Markovian, is a priori
simpler, although of course not equivalent to the fermionic reservoirs, and as such has been widely used coupled with a Liouvillian formalism~\citep{Breuer_HeattransportCPmaps,Bertini2020,Znidaric_MPS_XX_open,Znidaric__XXdeph,Prosen_OpenXXZ,Kehrein_Lindblad_fullcounting,Schutz_MPSXXZLindblad,ProsenBuca_Countingstatistics,Guimaraes_NonequilibriummultisiteLindblad,Poletti_Quadraticopensystems,znidaric_nonequilibrium_2019,BernardJin_QSSEP,debnath_nonequilibrium_2017,Kurchan_duality,damanet_reservoir_2019,damanet_controlling_2019,Dubi_DisorderInteractionQtransport},
to tackle out of equilibrium issues. Beyond experimental interest, for which the Lindblad coupling is the proper microscopic description, on the theory side, this approach has
allowed to unveil  non-trivial properties of highly excited and correlated systems: integrable structures, traditionally
restrained to closed systems in the quantum realm \citep{Prosen_thirdquantization,ProsenEssler_Mapping,Essler_YBLindblad,BernardJin_SolutionQSSEPcontinue}; the existence of ballistics spin-transport~\cite{Zotos1997,Zotos1999,Prosen2011} and anomalous diffusion~\cite{Gopalakrishnan2019,Znidaric2011,Ljubotina2017} in the integrable XXZ model, thus allowing for the discovery of Kardar-Parisi-Zhang correlations~\cite{kardar_dynamic_1986,kriecherbauer_pedestriantextquotesingles_2010} in the quantum realm~\cite{ljubotina_kardar-parisi-zhang_2019,de_nardis_universality_2020,JinKrajenbrinkBernard_QKPZ,PierreDenis_QKPZ}. Additionally, it has allowed to characterize the anomalous transport properties of disordered ~\cite{znidaric_diffusive_2016,mendoza-arenas_asymmetry_2019} and quasi-periodic~\cite{znidaric_interaction_2018} interacting systems, the persistence of ballistic transport in the presence of level repulsion induced by single impurities~\cite{Brenes2018,Brenes2020,Brenes2020b} and ballistic-to-diffusive transition induced by integrability-breaking in finite-sized systems~\cite{znidaric_weak_2020,ferreira_ballistic--diffusive_2020}.

However, the study of transport with Lindblad boundary conditions is mostly done on a case by case basis, and it remains unclear which properties of the interacting region determine the current in systems driven by Lindblad  reservoirs.
In a similar way, although some connection between the fermionic reservoir description and the Lindblad one can be found in the literature \cite{dorda_optimized_2017,Arrigoni_Keldysh}, generic consequences for the transport properties have not been carried out.
In particular,  an equivalent of the Meir and Wingreen's formulation~\cite{MeirWingreenformula} for Lindblad boundary conditions was not worked out yet, and it is yet unclear whether a systematic evaluation of transport properties in arbitrary Markovian settings is possible.

We address this question in the present paper and develop such formalism. We show, by using a Keldysh description~\cite{kamenev_fieldtheorybook,Diehl_KeldyshLindblad} of an arbitrary system in presence of Lindblad boundary conditions injecting and extracting particles,
that one can derive a generic transport formula in the spirit of the one of Meir-Wingreen.
Quite remarkably, this formula relates the transport properties of an arbitrary system uniquely  to its  Keldysh Green's function and the injection/extraction rates of the Lindblad boundaries.
We also generalize the transport formula to the case when the system itself can have losses and gains of particles by coupling to other Lindblad reservoirs. We then illustrate the usefulness of our generic formula by deriving the current in various settings, summarized in Fig.~\ref{fig:setups}, and in particular a one dimensional tight-binding chain in presence of losses/gains in the bulk.

The paper is structured as follows. In Sec.~\ref{sec:model}, we introduce the general setup and its description in the Hamiltonian and Lindbladian formulation. In Sec.~\ref{sec:mapping}, we introduce the Keldysh formalism and compute
the effective contribution of reservoirs onto the system in both cases. We show how
they map onto each other in the limit of high chemical potential and temperature for the fermionic reservoirs.
In Sec.~\ref{sec:mw}, we apply our mapping to derive the generic transport formula for Lindblad type boundaries, and a generic system, potentially containing both interactions and dissipation.
In Sec.~\ref{sec:applications}, we show applications of our formula to three examples. The two first ones, a single level and a one-dimensional chain, were already studied in the literature by other methods
and serve to show how or formula allows to recover easily the previous results. The last example, a one-dimensional chain with dissipation, is to the best of our knowledge new and exhibits unusual properties of
the current in presence of the dissipative terms. Sec.~\ref{sec:conclusion} is the conclusion and perspectives. Finally, several technical points have been put in appendices.

%%%%%%%%%%%%%%%%%%%%%%%%%%%%%%%%%%%%%%%%%%%%%%%%%%%%%%%
% MODELS
%%%%%%%%%%%%%%%%%%%%%%%%%%%%%%%%%%%%%%%%%%%%%%%%%%%%%%%

\section{Models}\label{sec:model}

In this section, we detail the two sets of models that we examine in the present paper, as sketched in Fig.~\ref{fig:The-two-situations}a.
In the fist class of models, generically denoted by a subscript $\mathcal L$, Lindblad  jump operators inject and extract particles at the left ($L$) and right ($R$) edge of a generic  system at different rates.
In the second class, generically denoted by a subscript $\mathcal F$, the system exchanges particles with two fermionic baths, at the temperature $T$, with different chemical potential $\mu_{L,R}$.

\subsection{Lindblad boundary conditions}

For the Lindblad boundary condition, the contact with the environment is described
by the action of a Lindblad operator
${\cal L}$~\citep{Gorini1976,Lindblad_seminalpaper,BreuerPetruccione_book}. The non-unitary evolution of the  density matrix of the system $\rho_{\rm S}$ obeys the Lindblad master equation
\begin{equation}\label{eq:master}
 \frac{d \rho_{\rm S}}{dt}=-i[H_{{\rm S}},\rho_{{\rm S}}]+{\cal L}(\rho_{{\rm S}})\,,
\end{equation}
where $H_{{\rm S}}$ is the Hamiltonian of the system. In general,  $\mathcal L(\rho)=\sum_{n,m}\lambda_{m,n}(\Lambda_{n}\rho \Lambda_{m}^{\dagger}-\frac{1}{2}\{\Lambda_{m}^{\dagger}\Lambda_{n},\rho\})$
where $\{\}$ denotes anticommutation of generic jump  operators $\Lambda_{n}$, acting with rates $\lambda_{m,n}$ on the system.
We focus here for simplicity on spinless fermions on a lattice. The extension including additional degrees of freedom is straightforward. The experimentally relevant situation, sketched in Fig.~\ref{fig:The-two-situations}, involves an external  environment which injects particles at site $r=L/R$ at rates $\alpha_r$ and extracts them at rates $\beta_r$. Injecting and extracting particles at different rates at each end of the system allows to drive currents through it.

This situation is described by the Lindblad operator~\cite{Prosen_thirdquantization,ProsenEssler_Mapping,Essler_YBLindblad,
BernardJin_SolutionQSSEPcontinue,Prosen2011,Znidaric2011,Ljubotina2017,
ljubotina_kardar-parisi-zhang_2019,de_nardis_universality_2020,znidaric_diffusive_2016,mendoza-arenas_asymmetry_2019,znidaric_interaction_2018,Brenes2018,Brenes2020,Brenes2020b,znidaric_weak_2020,ferreira_ballistic--diffusive_2020}
\begin{equation}\label{eq:Lop}
\begin{split}
{\cal L}(\rho_{\rm S})  =\sum_{r=L,R}&\Big\{\alpha_r\Big[2c_{r}^{\dagger}\rho_{\rm S} c_{r}-\{c_{r}c_{r}^{\dagger},\rho_{\rm S}\}\Big]\\
 & \hspace{1em}+\beta_r\Big[2c_{r}\rho_{\rm S} c_{r}^{\dagger}-\{c_{r}^{\dagger}c_{r},\rho_{\rm S}\}\Big]\Big\}\,,
\end{split}
\end{equation}
in which ($c_r$,$c_r^{\dagger}$) are  fermionic annihilation and creation operators acting on the system on site $r$.

The description~\eqref{eq:Lop} of the reservoirs relies on the  Markovian approximation, which discards  all memory effects and correlations between the system
and the environment. This is visible in the fact that the action of ${\cal L}$ only depends on the current
state $\rho_{\rm S}$ of the system.

\subsection{Fermionic reservoirs}

The other canonical description, in particular heavily used in solid-state mesoscopic systems~\cite{bruus_many-body_2004,nazarov_quantum_2009,akkermans_mesoscopic_2007,moskalets_scattering_2011}, consists of coupling the system to two free fermionic  baths, see Fig.\ref{fig:The-two-situations}b.
These baths mimic large metallic contacts exchanging particles with the edges $r=L/R$ of the system. This setting is described by a Hamiltonian of the form
\begin{equation}\label{eq:ham}
 H=H_{{\rm S}}+H_{\rm F}+H_{\tau}\,,
\end{equation}
in which $H_{\rm S}$ is the Hamiltonian of the system and $H_{\rm F}$ the Hamiltonian describing the $L$- and $R$-reservoirs:
\begin{equation}\label{bathH}
H_{{\rm F}} =\sum_{k,r\in \{L,R \}}\Big(\epsilon_{k}-\mu_r\Big)a_{k,r}^{\dagger}a_{k,r}\,,
\end{equation}
where the $a_{k,r}$ denote the usual fermionic annihilation operators. We assume that these fermions have a continuous spectrum $\epsilon_k$ and chemical potential $\mu_r$. The exchange of fermions between system and reservoirs is described by the tunnelling Hamiltonian
\begin{equation}\label{eq:tunnel}
H_{\tau} = - \tau\sum_{k,r\in \{L,R \}}\Big(a_{k,r}^{\dagger}c_r+{\rm h.c.}\Big)\,.
\end{equation}

The systems described by (\ref{eq:master}--\ref{eq:Lop}) and~(\ref{eq:ham}--\ref{eq:tunnel}) correspond in general to different physical situations. In general they thus lead to different transport properties. For the case of the Lindblad boundaries the current is controlled by the asymmetry of the injection and extraction rates between the $L$ and $R$ sides. For the fermionic case the current is controlled by the chemical potential difference between the two reservoirs. It is thus interesting to connect these two situations, this would allow to gain  physical insight by transferring well-established results  obtained in each of the two formulations of the problem. 

%%%%%%%%%%%%%%%%%%%%%%%%%%%%%%%%%%%%%%%%%%%%%%%%%%%%%%%
% MAPPING
%%%%%%%%%%%%%%%%%%%%%%%%%%%%%%%%%%%%%%%%%%%%%%%%%%%%%%%

\section{Keldysh formulation}\label{sec:mapping}

In this section we give a Keldysh description~\cite{keldysh_1965} of the two situations of the previous section. Using the Keldysh formalism allows us to trace out in both cases the reservoirs and boundary conditions and provides a natural path for connecting the two
approaches. Given the fact that the reservoirs are coupled locally at each end of the system, it is enough to consider the case where reservoirs are coupled to a single site to see how its action gets modified.
In this case, the system is described by the simple Hamiltonian
\begin{equation}\label{eq:hsingle}
 H_{{\rm S}}=\epsilon_{0}c^{\dagger}c\,,
\end{equation}
in which $\epsilon_0$ describes the local chemical potential for the site occupation.

We operate in the Keldysh path-integral formalism~~\citep{kamenev_fieldtheorybook,altland_condensed_2006}, that we summarize here mainly to fix notations. The  object of interest is the Keldysh action $\mathcal S$, appearing in the partition function $\mathcal Z$ of the system:
\begin{equation}
\mathcal  Z=\int{\cal D}[\psi,\bar{\psi}]e^{i\mathcal S[\psi,\bar{\psi}]}\,.
\end{equation}
We assume implicitly the standard Keldysh matrix structure in which $\psi=(\psi^+,\psi^-)$ are vectors of fermionic Grassmann variables defined on the upper and lower Keldysh branches $\mathcal C_{\pm}$. We also follow the Larkin and Ovchinnikov convention~\citep{Larkin_vortices_supra} to perform the Keldysh rotation.

In this basis, the Keldysh action is expressed in terms of the retarded, advanced and Keldysh green functions
$G^{{\mathcal  R}}$, $G^{{\mathcal  A}}$ and $G^{{\mathcal  K}}$:
\begin{equation}
\mathcal  S=\int\frac{d\epsilon}{2\pi}(\bar{\psi}^{1},\bar{\psi}^{2})_{(\epsilon)}\begin{pmatrix}[G^{{\mathcal  R}}]^{-1} & [G^{-1}]^{{\mathcal K}}\\
 0 & [G^{{\mathcal  A}}]^{-1}
 \end{pmatrix}_{(\epsilon)}\begin{pmatrix}\psi^{1}\\
 \psi^{2}
 \end{pmatrix}_{(\epsilon)}\,.\label{eq:actionsystem}
\end{equation}
For the single-site Hamiltonian~\eqref{eq:hsingle}, initially at thermodynamic equilibrium of temperature $T_{\rm S}$ and chemical potential $\mu_{\rm S}$, the Green functions read
\begin{align}
[G_{{\rm S}}^{{\mathcal  R(\mathcal A)}}]^{-1} & =\epsilon-\epsilon_{0}\pm i\eta\,,\\
G_{{\rm S}}^{{\mathcal  K}}&=-2\pi i  \tanh\left(\frac{\epsilon-\mu_{\rm S}}{2T_{{\rm S}}}\right)\delta(\epsilon-\epsilon_0)\,,\label{eq:GKS}
\end{align}
in which $\eta$ is an infinitesimally small quantity. We remind that $[G^{-1}]^{\rm K}=2i\eta \tanh[(\varepsilon-\mu)/2T_{\rm S}]$ is also infinitesimal and formally keeps memory about the initial state of the system~\cite{kamenev_fieldtheorybook,altland_condensed_2006}. As we are going to illustrate below, by comparing the effect of adding Lindblad and fermionic reservoirs on the single level,  this infinitesimal term can be neglected as soon as the system is coupled to external baths.

%%%%%%%%%%%%%%%%%%%%%%%%%%%%%%%%%%%%%%%%%%%%%%%%%%%%
% LINDBLAS RESERVOIR
%%%%%%%%%%%%%%%%%%%%%%%%%%%%%%%%%%%%%%%%%%%%%%%%%%%%

\subsection{Lindblad reservoirs}

In the case of reservoirs described by Lindblad operators as in Eq.~\eqref{eq:Lop}, the description within the Keldysh formalism is given in Appendix~\ref{sec:derivationLindbladaction}, following the method outlined in \cite{Diehl_KeldyshLindblad}. A single reservoir, injecting and extracting particles with rates $\alpha$ and $\beta$, leads to an additional contribution to the action~\eqref{eq:actionsystem}, which reads
\begin{align}
S_{{\cal L}} & =i\int\frac{d\epsilon}{2\pi}(\bar{\psi}^{1},\bar{\psi}^{2})_{(\epsilon)}\begin{pmatrix}\alpha+\beta & -2(\alpha-\beta)\\
0 & -(\alpha+\beta)
\end{pmatrix}\begin{pmatrix}\psi^{1}\\
\psi^{2}
\end{pmatrix}_{(\epsilon)}.\label{actionLindblad}
\end{align}
Notice that the upper right element of the action is now finite and the contribution from the initial Keldysh component~\eqref{eq:GKS} can be neglected. As an important consequence, the energy of the single level $\epsilon_0$ fully disappears from the action, by making the shift $\epsilon\to \epsilon+ \epsilon_0$ in the integral. This has direct consequences for a certain number of physical properties of the system which will become fully controlled by the bath.
For instance, for the case of a single site, the level occupation in the stationary state  is given, in terms of  Green functions, by
\begin{equation}\label{eq:nK}
 \langle n\rangle_{\infty} = \frac{i}{2}\int\frac{d\epsilon}{2\pi}\Big[G^{{\mathcal  R}}(\epsilon)-G^{{\mathcal  A}}(\epsilon)-G^{{\mathcal  K}}(\epsilon)\Big]\,,
\end{equation}
where the Green functions are derived by inverting the full action obtained by adding~\eqref{eq:actionsystem} and~\eqref{actionLindblad}. One obtains, in the case of Lindblad boundaries~\cite{Arrigoni_Keldysh}:
\begin{equation}\label{eq:occupationL}
 \langle n\rangle_{\infty}^{{\cal L}}=\frac{\alpha}{\alpha+\beta},
\end{equation}
which exclusively depends on the injection and extraction rates $\alpha$ and $\beta$, irrespective of the local chemical potential $\epsilon_{0}$. The Lindblad case will thus ``erase'' certain characteristics of the system.

It is also interesting to note that in \eqref{actionLindblad} the retarded and advanced parts of the action depend only on the sum $\alpha+\beta$ and thus are insensitive on whether the boundary condition is injecting or extracting particles. The difference
between extraction and injection only appears in the Keldysh component. As we will see below, this has remarkable consequences on some transport properties of systems with losses.

%%%%%%%%%%%%%%%%%%%%%%%%%%%%%%%%%%%%%%%%%%%%%%%%%%%%
% FERMIONIC RESERVOIR
%%%%%%%%%%%%%%%%%%%%%%%%%%%%%%%%%%%%%%%%%%%%%%%%%%%%

\subsection{Fermionic reservoirs}
In the case of a fermionic bath, it is useful to obtain simple analytical expressions by making certain approximations on the properties of the reservoirs which correctly describe typical metallic contacts, without  loss of generality.
In particular, we assume in what follows that the reservoir has a constant density of states.
We also make the approximation that the tunnelling takes place on a single site of the system so that the momentum is not conserved during the tunnelling.

These two assumptions allow us to analytically
integrate over the reservoirs in \eqref{eq:ham} and obtain the effective boundary terms for the system.
The details of this derivation are given in Appendix~\ref{sec:Derivation-of-the}. The single level action thus becomes (we adopt the convention $e=\hbar=k_B=1$):
\begin{equation}
 S_{\mathcal F}=i\int\frac{d\epsilon}{2\pi}(\bar{\psi}^{1},\bar{\psi}^{2})_{(\epsilon)}\Delta\begin{pmatrix}1 & 2\tanh(\frac{\epsilon-\mu}{2T})\\
 0 & -1
 \end{pmatrix}\begin{pmatrix}\psi^{1}\\
 \psi^{2}
 \end{pmatrix}_{(\epsilon)}\,,\label{eq:actionbain}
\end{equation}
where we introduced the hybridization constant $\Delta=\tau^{2}/v_{F}$, in which $v_{F}$ is the Fermi velocity of the reservoirs and thus a direct measure of the density of states.

\subsection{Mapping between the two boundary conditions}

The  comparison between \eqref{actionLindblad} and \eqref{eq:actionbain} shows that the main difference lies in the $\epsilon$--dependence of the Keldysh component of the
action for $S_{\mathcal F}$. This energy dependence has for consequence that the fermionic boundary term is non-local in time and thus encodes the memory
effects of the fermionic bath. Contrarily, the absence of such energy dependence in the action for the Lindblad boundaries $S_{{\mathcal  L}}$ is directly encoding the Markovian aspect.

It is possible to get rid of the $\epsilon$ dependence of the fermionic reservoirs
by taking the limit $\mu\to\infty$, $T\to\infty$, while keeping the ratio $\mu/T$
fixed. One thus obtains the mapping $S_{\mathcal F}\to S_{{\cal L}}$  by making the identification
\begin{align} \label{eq:correspondence}
\begin{split}
 \alpha & =\frac{1}{2}\Delta\left[1+\tanh\left(\frac{\mu}{2T}\right)\right]\,,\\
 \beta & =\frac{1}{2}\Delta\left[1-\tanh\left(\frac{\mu}{2T}\right)\right]\,.
\end{split}
\end{align}
Such limiting mapping between the fermionic reservoirs and the Lindblad boundaries was already noted in general terms in the literature \cite{BreuerPetruccione_book,dorda_optimized_2017}. This precise mapping between the two formalisms allows us to derive the transport properties, as will be done in the next section.

In connection with this limit, it is instructive to consider the single level occupation for the fermionic reservoirs. It is readily derived relying on \eqref{eq:nK}:
\begin{align}\label{eq:densiteKeldysh}
 \langle n\rangle_{\infty}^{{\mathcal F}} & =\int\frac{d\epsilon}{2\pi}\left[1-\tanh\left(\frac{\epsilon-\mu}{2T}\right)\right]\frac{\Delta}{(\epsilon-\epsilon_{0})^{2}+\Delta^{2}}\,.
\end{align}

First we note that in the limit $\Delta \to 0$ the coupling to the reservoirs becomes extremely small. The density then becomes:
\begin{equation}
 \lim_{\Delta\to0}\langle n\rangle_{\infty}^{{\mathcal F}}=\frac{1}{1+e^{\frac{\epsilon_{0}-\mu}{T}}}\,.
\end{equation}
In that case we recover the Fermi-Dirac distribution for a single site with energy $\epsilon_0$
and at a temperature of the \emph{reservoirs}, showing that in this limit the only effect of the reservoirs is to thermalize the single site.
Note that in our formalism we always implicitly take \emph{first} the limit of infinitely long time and assume that we have reached a stationary state before taking other limits.

On the other hand, if we take the infinite chemical potential and temperature limit \eqref{eq:correspondence}, we get back the Lindblad result \eqref{eq:occupationL}, as can be expected.
It is however important to stress that the spectrum of the system has to be bounded to allow to take such limit.
In the specific case of the single site case considered in \eqref{eq:densiteKeldysh}, the condition  $\mu,T \gg \Delta,\epsilon_{0}$ is sufficient to enforce the correspondence with \eqref{eq:occupationL}.

%%%%%%%%%%%%%%%%%%%%%%%%%%%%%%%%%%%%%%%%%%%%%%%%
% MR FORMULA
%%%%%%%%%%%%%%%%%%%%%%%%%%%%%%%%%%%%%%%%%%%%%%%%

\section{Generic transport formula for Lindblad boundaries}\label{sec:mw}

We are now in a position to tackle the main question of the paper, namely a generic transport formula for Lindblad ($\mathcal{L}$) type boundaries.

\subsection{Generic formula}

Let us thus consider a generic quantum system driven out of equilibrium by a a $L$- and $R$- fermionic reservoir (see Fig.~\ref{fig:The-two-situations}).
Since we want to be able to address the more general case in which the system itself can potentially lose or gain particles, we define two currents
\begin{equation}
 \frac{d n_L}{dt} = -J_L, \quad \frac{d n_R}{dt} = J_R\,,
\end{equation}
in which $n_{L/R}$ is the occupation of the sites $L/R$  attached to the reservoirs $L/R$. As a consequence, $J_L$ is the current leaving the left reservoir, while $J_R$ is the one entering the right reservoir. As a result two generic currents can be defined: $J = (J_L + J_R)/2$ is the current going through the system, while $J_D = (J_L-J_R)$ is the current representing the loss (or gain) of particles in the bulk. In the absence of such extraction or injection of charges in the system $J_D=0$ and $J$ is the usual conserved current. Let $\boldsymbol{G^{a}}$ with $a\in\{\mathcal{R},\mathcal{A},\mathcal{K}\}$, be the matrices whose elements are the different Green's function in a given basis. A generic derivation of the current for Lindblad boundary conditions ($\mathcal{L}$) is presented in Appendix~\ref{sec:Alternative-derivation-of} and leads to
\begin{widetext}
\begin{align}\label{eq:lindbladtransport}
 J_{{\cal L}} &= \frac{1}{2}\left\{\big(\alpha_{L}-\beta_{L}-\alpha_{R}+\beta_{R}\big) +\frac{i}{2}\int\frac{d\epsilon}{2\pi}{\rm Tr}\Big[\Big((\alpha_{L}+\beta_{L})\boldsymbol{\gamma_{L}}-(\alpha_{R}+\beta_{R})\boldsymbol{\gamma_{R}}\Big)\boldsymbol{G^{{\mathcal  K}}}\Big]\right\}\,,\\
J_{D,{\cal L}}&=\frac{i}{2}\int\frac{d\epsilon}{2\pi}{\rm Tr}\Big[\Big((\alpha_{L}+\beta_{L})\boldsymbol{\gamma_{L}}+(\alpha_{R}+\beta_{R})\boldsymbol{\gamma_{R}}\Big)\boldsymbol{G^{\mathcal{K}}}\Big].\label{eq:Lindbladtransport2}
\end{align}
\end{widetext}
with $\gamma_{r,ij}=2\delta_{i,r}\delta_{i,j}$ in the position basis.
This is the main result of the paper. 

While Appendix~\ref{sec:Alternative-derivation-of} presents the full derivation of (\ref{eq:lindbladtransport}--\ref{eq:Lindbladtransport2}), we show in the main text how one can also derive \eqref{eq:lindbladtransport} by using the mapping~\eqref{eq:correspondence} and the MW formula for fermionic reservoirs ($\mathcal{F}$).
We consider the Hamiltonian of the system in the form $H_{\rm S}= H_0+V$, in which $H_0=\sum_{i,j}c^\dagger_ih_{i,j}c_j$ is the kinetic part of the Hamiltonian written in position basis and $h_{i,j}$ describes hopping between sites $i$ and $j$. The operator $V$ describes  many-body interactions. For such a system the MW formula for the current reads~\citep{MeirWingreenformula}
\begin{widetext}
\begin{equation}\label{eq:MW}
\begin{split}
 J_{{\mathcal F}} & =\frac{i}{2}\int\frac{d\epsilon}{2\pi}{\rm Tr}\left\{\left[\left(f_{L}(\epsilon)-\frac{1}{2}\right)\boldsymbol{\Gamma_{L}}-\left(f_{R}(\epsilon)-\frac{1}{2}\right)\boldsymbol{\Gamma_{R}}\right](\boldsymbol{G^{{\mathcal  R}}}-\boldsymbol{G}^{\boldsymbol{{\mathcal  A}}})+\frac{1}{2}(\boldsymbol{\Gamma}_{L}-\boldsymbol{\Gamma}_{R})\boldsymbol{G^{{\mathcal  K}}}\right\}
\end{split}
\end{equation}
\end{widetext}
where the matrices $\Gamma_{r,ij}=\Delta_{r}\gamma_{r,ij}$ in position basis describe the coupling between the system and the baths at the edge site $r$, and $f_{r}(\epsilon)$ is the Fermi-Dirac distribution associated to the $r$-bath.

Using the correspondence \eqref{eq:correspondence} in \eqref{eq:MW} one notices that, after performing the transformation, the term between square brackets becomes constant as a function of energy.
This allows to use the additional relation
\begin{equation}
 i\int\frac{d\epsilon}{2\pi}(\boldsymbol{G^{{\mathcal  R}}}-\boldsymbol{G^{{\mathcal  A}}})=\boldsymbol{\mathbb{I}}\,,
\end{equation}
which follows from the fermionic anticommutation relations. Using that $\mbox{Tr}[\mathbf \Gamma^{r}]=2\Delta_{r}$, we obtain the formula \eqref{eq:lindbladtransport} giving the current for a generic system driven by Lindblad boundary conditions.
Note that this result is also valid in presence of dissipation in the system (see Appendix~\ref{sec:Alternative-derivation-of}).

One of the remarkable properties of the Lindblad boundary condition is the fact that the current is \emph{fully} determined by the Keldysh component of the local Green's function $\boldsymbol{G^{{\mathcal  K}}}$.
Note that the first term in \eqref{eq:lindbladtransport} depends on the difference between injection and extraction rates $\alpha-\beta$ for each of the reservoirs. We could naively expect that this difference plays a similar role to the voltage or chemical potential difference for fermionic reservoirs and thus control the current flow. Nevertheless, this term is not sensitive to the properties of the system, which are encoded in the Keldysh Green's function appearing in the second term. The second term is also sensitive to the sum of the injection and extraction $\alpha+\beta$ rates of each of the reservoirs. %The difference between injection and extraction appears only via the Keldysh Green's function.
 These considerations  apply also for the current~\eqref{eq:Lindbladtransport2}, which quantifies dissipative gains and losses in the system. The Keldysh Green function $\boldsymbol{G^{{\mathcal  K}}}$ is  thus the central object to understand  transport in dissipative systems driven by Lindblad boundaries,  as we will examine in the examples of the next section.

\subsection{Non-interacting systems}

As for the case of fermionic reservoirs, a non-interacting and non-dissipative system allows for further simplifications of the transport formula. In that case, we can rely on two additional relations~\citep{CaroliNozieres_tunneling_current}:
\begin{align}
\begin{split}
 \frac{1}{2}(&\boldsymbol{G^{{\mathcal  A}}-G^{{\mathcal  R}}+G^{{\mathcal K}}})  =\\&if_{L}(\epsilon)\boldsymbol{G^{{\mathcal  R}}}\boldsymbol{\Gamma_{L}} \boldsymbol{G^{{\mathcal  A}}}+if_{R}(\epsilon)\boldsymbol{G ^{{\mathcal  R}}}\boldsymbol{\Gamma_{R}}\boldsymbol{G^{{\mathcal  A}}}\,,\label{eq:Noziere1}
 \end{split}
\\
 \boldsymbol{G^{{\mathcal  R}}}-&\boldsymbol{G^{{\mathcal  A}}} =i\boldsymbol{G^{{\mathcal  R}}}(\boldsymbol{\Gamma_{L}}+\boldsymbol{\Gamma_{R}})\boldsymbol{G^{{\mathcal  A}}}\,,\label{eq:Noziere2}
\end{align}
leading to the following expression for the current in the fermionic and Lindblad setting
\begin{align}
 J_{{\mathcal  F}} & =\int\frac{d\epsilon}{2\pi}\Big[f_L(\epsilon)-f_R(\epsilon)\Big]{\rm Tr}\Big[\boldsymbol{\Gamma}_{\boldsymbol R}\boldsymbol{G^{{\mathcal  R}}}\mathbf \Gamma_{\boldsymbol L}\boldsymbol{G^{{\mathcal  A}}}\Big],\label{eq:freecurrentbath}\\
 J_{{\cal L}} & =\int\frac{d\epsilon}{2\pi}\Big[\alpha_L\beta_R-\beta_L\alpha_R\Big]{\rm Tr}\Big[\boldsymbol{\gamma_RG^{{\mathcal  R}}\gamma_LG^{{\mathcal  A}}}\Big].\label{eq:freecurrentlindblad}
\end{align}
In position basis these expressions become:
\begin{align}
J_{{\mathcal  F}} & =\int\frac{d\epsilon}{2\pi}\Big[f_L(\epsilon)-f_R(\epsilon)\Big]4\Delta_R\Delta_L|G_{L,R}^{{\mathcal  R}}|^{2}\,,\label{eq:freecurrentbath-1}\\
J_{{\cal L}} & =\int\frac{d\epsilon}{2\pi}\Big[\alpha_L\beta_R-\beta_L\alpha_R\Big]4|G_{L,R}^{{\mathcal  R}}|^{2}\,,\label{eq:freecurrentlindblad-1}
\end{align}
where we used the fact that $\boldsymbol{G^{{\mathcal  R}*}=G^{{\mathcal  A}}}$ and $\boldsymbol{G^{{\mathcal  R}}}$ is
symmetric.

For the fermionic reservoirs, Eq.~\eqref{eq:freecurrentbath-1} reproduces Landauer-B\"uttiker formula, where the current is directly related to the probability of transmission through the system at a given energy $\epsilon$. The transmission probability is given by the Green's function $G^{\mathcal R}_{L,R}$ connecting the two reservoirs. The expressions of the currents for the Lindblad and the fermionic bath boundaries both depend on this transmission probability. These probabilities coincide for these two cases, as can be seen from Eqs. \eqref{eq:actionsystem} and \eqref{eq:actionbain}, by making the identification $\Delta=\alpha+\beta$. It is thus possible to draw a connection between the two driving protocols in terms of transport coefficients. 

Indeed, the corresponding formula for Lindblad boundaries allows further simplifications. For non-interacting and non-dissipative systems, the retarded Green function  $\boldsymbol{G^{\mathcal R}}$ only depends on the hybridization coefficients $\Delta_r =\alpha_r + \beta_r$.
This has the remarkable consequence in \eqref{eq:freecurrentlindblad-1} that the current $J_{\cal L}$ is \emph{always linear} in the bias $\alpha_L\beta_R-\beta_L\alpha_R$. This allows us to exactly connect the linear response for fermionic systems to the
Lindblad driving. For the fermionic case,  we consider the linear response limit of  (\ref{eq:freecurrentbath}--\ref{eq:freecurrentbath-1}), in which $\delta \mu=\mu_L-\mu_R\rightarrow0$ and $T_R=T_L=T$. In this limit, $f_L(\epsilon)-f_R(\epsilon)\sim - \partial_\varepsilon f(\varepsilon)\delta\mu/T$, which, in the limit of large temperatures, scales as $ \delta\mu/4T$. We stress again that such limit makes sense if the transmission amplitude ${\rm Tr}\Big[\boldsymbol{\Gamma}_{\boldsymbol R}\boldsymbol{G^{{\mathcal  R}}}\mathbf \Gamma_{\boldsymbol L}\boldsymbol{G^{{\mathcal  A}}}\Big]$ is non-zero only on a finite energy window. In such a high temperature limit the conductance becomes
\begin{equation}\label{eq:g}
\begin{split}
g(T\rightarrow\infty)&=\lim_{\delta\mu\rightarrow 0}\frac {J_{\mathcal F}}{\delta\mu}=\frac{1}{4T}\int\frac{d\epsilon}{2\pi}{\rm Tr}\Big[\boldsymbol{\Gamma}_{\boldsymbol R}\boldsymbol{G^{{\mathcal  R}}}\mathbf \Gamma_{\boldsymbol L}\boldsymbol{G^{{\mathcal  A}}}\Big]\,,
\end{split} 
\end{equation}
which vanishes as the inverse temperature $T$, as expected. Comparing (\ref{eq:g}) with (\ref{eq:freecurrentlindblad}), one gets
\begin{equation}\label{eq:lb}
J_{\cal L}=c\,Tg(T\rightarrow\infty)\,,
\end{equation}
where $c=(\alpha_L\beta_R-\beta_L\alpha_R)/4\Delta_R\Delta_L$ is a constant which depends on the choice of the Lindblad driving. There is thus a perfect connection between the large temperature conductance of a fermionic system and the transport measured with
Lindblad driving. Note that the condition of large chemical potential  necessary  to derive the mapping~\eqref{eq:correspondence} is not required here, the limit of large temperature is sufficient. Whether such an exact connection applies in the presence of interactions  remains an open question, which is left for further investigations.

%%%%%%%%%%%%%%%%%%%%%%%%%%%%%%%%%%%%%%%%%%%%%%%%%%%%%%%%%%%%%%
% APPLICATIONS
%%%%%%%%%%%%%%%%%%%%%%%%%%%%%%%%%%%%%%%%%%%%%%%%%%%%%%%%%%%%%%

\section{Applications}\label{sec:applications}

\begin{figure}
\begin{centering}
 \includegraphics[scale=0.6]{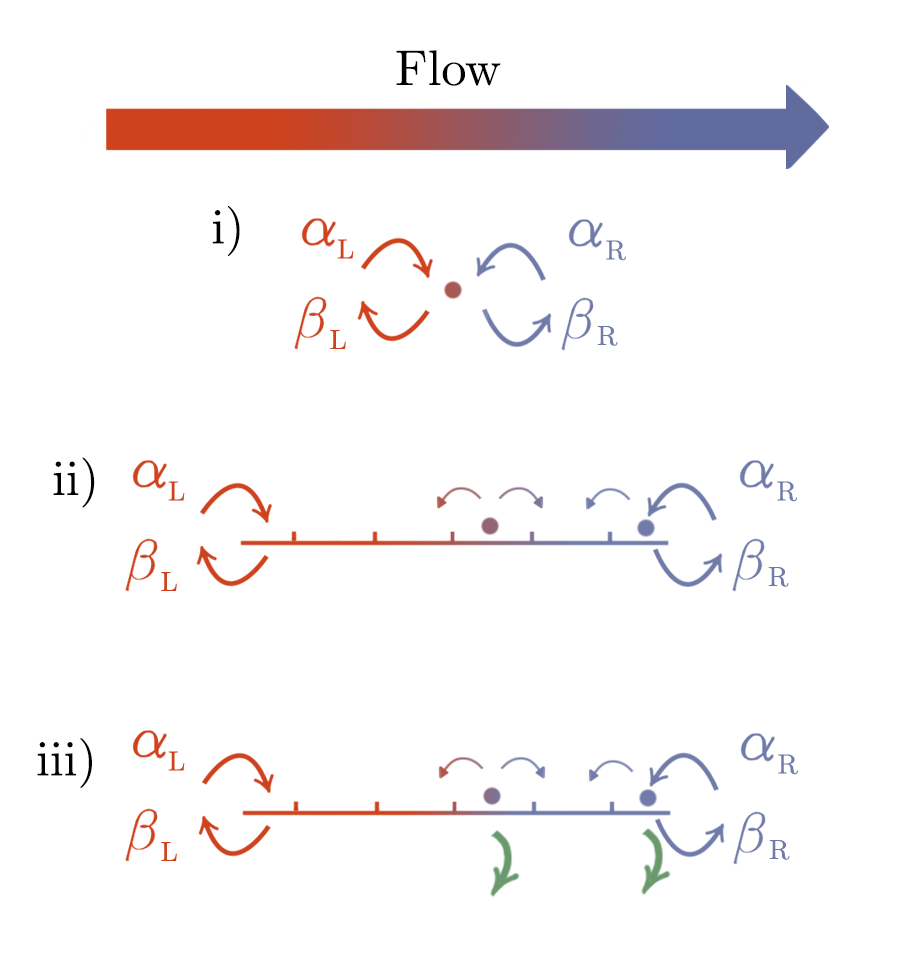}
\end{centering}
\caption{\label{fig:setups} The different situations for which the current is computed with the transport formulae (\ref{eq:lindbladtransport}--\ref{eq:Lindbladtransport2}): {\it i)} Single level coupled to reservoirs; {\it ii)} Free fermionic chain; {\it  iii)} Fermionic chain with loss terms along the chain.}
\end{figure}

We provide some applications of the general formula~\eqref{eq:lindbladtransport} for the systems sketched in Fig.~\ref{fig:setups}. We will compute the current for : i) a single site of energy $\epsilon_0$ as described by the Hamiltonian~\eqref{eq:hsingle}; ii) a free fermionic chain of $N$ sites;  iii) the same free-fermionic chain but with loss or gain terms modelled by injecting or extracting Lindblad terms acting throughout the chain.
\renewcommand\thesubsection{\roman{subsection}}

\subsection{Single site connected to biased reservoirs}\label{singlesite} 

A single site connected to biased reservoirs leads to standard Breit-Wigner resonances~\cite{breit_capture_1936}. In this  case, $\Gamma_{L/R}=2\Delta_{L/R}$ and the Green function reads, in the case of fermionic bath boundaries, $G^{{\mathcal  R(\mathcal A)}}=(\epsilon-\epsilon_{0}\pm i(\Delta_{R}+\Delta_{L}))^{-1}$. Equation~\eqref{eq:freecurrentbath} leads to the well known result for the current~\cite{nazarov_quantum_2009}:
\begin{equation}\label{eq:JBsinglesite}
 J_{{\mathcal  F}}=\int\frac{d\epsilon}{2\pi}\frac{4\Delta_{R}\Delta_{L}[f_L(\epsilon)-f_R(\epsilon)]}{(\epsilon-\epsilon_{0})^{2}+(\Delta_{L}+\Delta_{R})^{2}}\,.
\end{equation}

For Lindblad type boundaries we have instead
\begin{equation}
 G^{{\mathcal{ R(A)}}}=(\epsilon-\epsilon_{0}\pm i(\alpha_{R}+\beta_{R}+\alpha_{L}+\beta_{L}))^{-1}\,.
\end{equation}
Using the relation \ref{eq:freecurrentlindblad-1} for the current leads to
\begin{equation}
J_{{\cal L}}=2e\frac{(\alpha_{L}\beta_{R}-\beta_{L}\alpha_{R})}{\alpha_{L}+\beta_{L}+\alpha_{R}+\beta_{R}}\label{eq:JLsinglesite}\,.
\end{equation}
Comparing~\eqref{eq:JBsinglesite} and~\eqref{eq:JLsinglesite} one
sees that, as can be expected, all dependence  of the current on $\epsilon_{0}$ is lost for the Lindblad
case. Thus, the current is fixed entirely by the boundary conditions, in analogy to the occupation~\eqref{eq:occupationL} of the impurity in the presence of  a single Lindblad reservoir.  On the other hand, it is clear that for the fermionic baths case, the conductivity
depends on the relative values of the chemical potentials of the bath
and the Fermi energy of the system.
\begin{figure}
\includegraphics[width=\columnwidth]{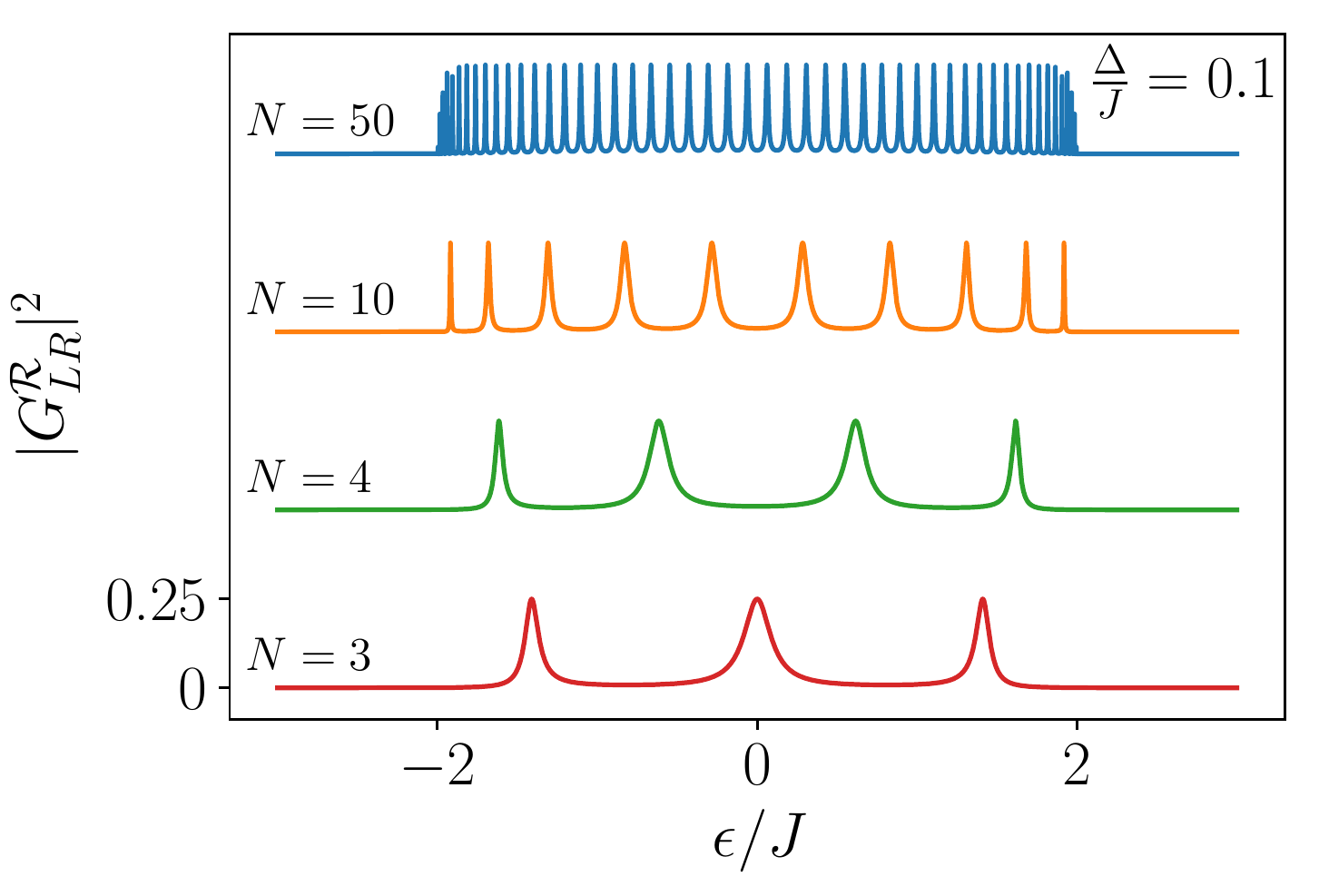}
\caption{\label{fig:resonances} Resonance spectrum~\eqref{eq:glr} of the tight-binding chain as a function of energy and different system sizes $N$. We consider $\Delta/J=0.1$. The plots are shifted vertically for different $N$, for readability.}
\end{figure}

\subsection{1D free fermionic chain}\label{freefermionicchain}
The current flowing through a non-interacting fermionic chain attached to Lindblad reservoirs has been recently derived relying on variational~\citep{Znidaric_MPS_XX_open} and third-quantization methods~\cite{Prosen_thirdquantization,Poletti_Quadraticopensystems}. Our formulation allows to derive the above results in a systematic way relying on standard techniques.
Let $N\geq2$ be the size of the system.  The
Hamiltonian is $H_{\rm S, chain}=J\sum_{j=1}^{N-1}[c_{j}^{\dagger}c_{j+1}+\mbox{h.c.}]$.
According to \eqref{eq:JBsinglesite} and \eqref{eq:JLsinglesite},
we need to compute for fermionic baths boundaries
\begin{equation}
{\rm Tr}(\boldsymbol{\Gamma_R G^{{\mathcal  R}}\Gamma_LG^{{\mathcal  A}}})=4\Delta_R\Delta_L|G_{L,R}^{{\mathcal  R}}|^{2}\label{eq:tracemain}
\end{equation}
where the indices $L,R$ refer to the sites where the left and right reservoir are connected, i.e $1$ and $N$. For Lindblad
boundaries we have similarly
\begin{equation}
{\rm Tr}(\boldsymbol{\gamma_{R}G^{{\mathcal  R}}\gamma_LG^{{\mathcal  A}}})=4|G_{L,R}^{{\mathcal  R}}|^{2}\,.
\end{equation}
In the position basis, for fermionic bath boundaries these functions
are given by tridiagonal matrices of the form
\begin{equation}
[\boldsymbol{G}^{\boldsymbol{{\mathcal  R(\mathcal A)}}}]^{-1}=\begin{pmatrix}\epsilon+i\Delta_L & J & \cdots & 0 & 0\\
J & \epsilon & J & \cdots & 0\\
\vdots & J & \ddots & J & \vdots\\
0 & \cdots & J & \epsilon & J\\
0 & 0 & \cdots & J & \epsilon+i\Delta_R
\end{pmatrix}\,,
\end{equation}
in which the presence of boundaries affects  only the first and last
diagonal term. The Green functions for Lindblad type boundaries
are simply obtained by making the substitution $\Delta_{L}\to\alpha_{L}+\beta_{L}$,
$\Delta_{R}\to\alpha_{R}+\beta_{R}$. To simplify notations, we will suppose
in what follows that $\Delta_{L}=\Delta_{R}=\Delta$ and $|J|=1$.
The inverse of such tridiagonal matrices has been derived in Ref.~\citep{Tan_InverseTridiagonal}.
For the element $(i,j)$ with $i<j$, the Green function reads:
\begin{equation}
G_{i,j}^{{\mathcal{R(A)}}}=(-J)^{i+j}\frac{B_{i-1}^\mathcal{R(A)}B_{N-j}^\mathcal{R(A)}}{(\epsilon\pm i\Delta)B_{N-1}^\mathcal{R(A)}-B_{N-2}^\mathcal{R(A)}}\,, \label{eq:inverseG}
\end{equation}
where $B_{i}^\mathcal{R(A)}=\frac{(r_{+}\pm i\Delta)r_{+}^{i}-(r_{-}\pm i\Delta)r_{-}^{i}}{(r_{+}\pm i\Delta)-(r_{-}\pm i\Delta)}$ (one has to take the $+$ sign for $\mathcal{R}$ and the $-$ sign for $\mathcal{A}$),
and $r^{\pm}_\epsilon=(\epsilon\pm\sqrt{\epsilon^{2}-4J^{2}})/2$.
Equations~(\ref{eq:tracemain}-\ref{eq:inverseG}) lead to
\begin{widetext}
\begin{equation}\label{eq:glr}
|G^\mathcal R_{L,R}(\varepsilon,\Delta)|^2=\frac{4|(\epsilon^{2}-4)|}{\bigg|(\epsilon-\Delta^{2}\epsilon+4i\Delta)\big[(r^{+}_\epsilon)^{N}-(r^{-}_\epsilon)^{N}\big]+(\Delta^{2}+1)\sqrt{\epsilon^{2}-4}\big[(r^{+}_\epsilon)^{N}+(r^{-}_\epsilon)^{N}\big]\bigg|^{2}}\,,
\end{equation}
\end{widetext}
which,  inserted in~(\ref{eq:freecurrentbath}-\ref{eq:freecurrentlindblad}), gives the current flowing through the system. Equation~\eqref{eq:glr} is plotted in Fig.~\ref{fig:resonances} as a function of energy $\epsilon$. It features $N$ peaks whose width is controlled by the hybridization constant $\Delta$. These peaks  correspond to the $N$ single-particle resonances of a chain of $N$ sites, appearing within an energy band of width $4J$.

\begin{comment}
\begin{widetext}
\begin{equation}
J_{{\rm B}}=\int\frac{d\epsilon}{2\pi}(f_{L}(\epsilon)-f_{R}(\epsilon))16\Delta^{2}\frac{|(\epsilon^{2}-4)|}{\bigg|\left((\epsilon-\Delta^{2}\epsilon+4i\Delta)((r^{+})^{N}-(r^{-})^{N})+(\Delta^{2}+1)\sqrt{\epsilon^{2}-4}((r^{+})^{N}+(r^{-})^{N})\right)\bigg|^{2}}\,,\label{eq:JBmultisite}
\end{equation}
\begin{equation}
J_{{\cal L}}=\int\frac{d\epsilon}{2\pi}16(\alpha_{L}\beta_{R}-\beta_{L}\alpha_{R})\frac{|(\epsilon^{2}-4)|}{\bigg|\left((\epsilon-\Delta^{2}\epsilon+4i\Delta)((r^{+})^{N}-(r^{-})^{N})+(\Delta^{2}+1)\sqrt{\epsilon^{2}-4}((r^{+})^{N}+(r^{-})^{N})\right)\bigg|^{2}}\,.\label{eq:JLmultisite}
\end{equation}
\end{widetext}
\end{comment}

For the fermionic baths, there is no simple way of carrying the integral
in general but, for large values of $N$, there is a simple way of bounding
$J_{\mathcal F}$ up to corrections of order $e^{-N}$:
\begin{equation}
 J_3<J_{\mathcal F}<J_1\,,
\end{equation}
with
\begin{align}
 J_\alpha & =\int_{-\pi/2}^{\pi/2}\frac{d\theta}{\pi}\Big[f_{L}(2\sin\theta)-f_{R}(2\sin\theta))\cos^\alpha(\theta)\Big]
\end{align}
where we took $\Delta=1$ to simplify and made the trigonometric change of variable $\epsilon \to 2 \sin \theta $.

\begin{figure}
\includegraphics[width=\columnwidth]{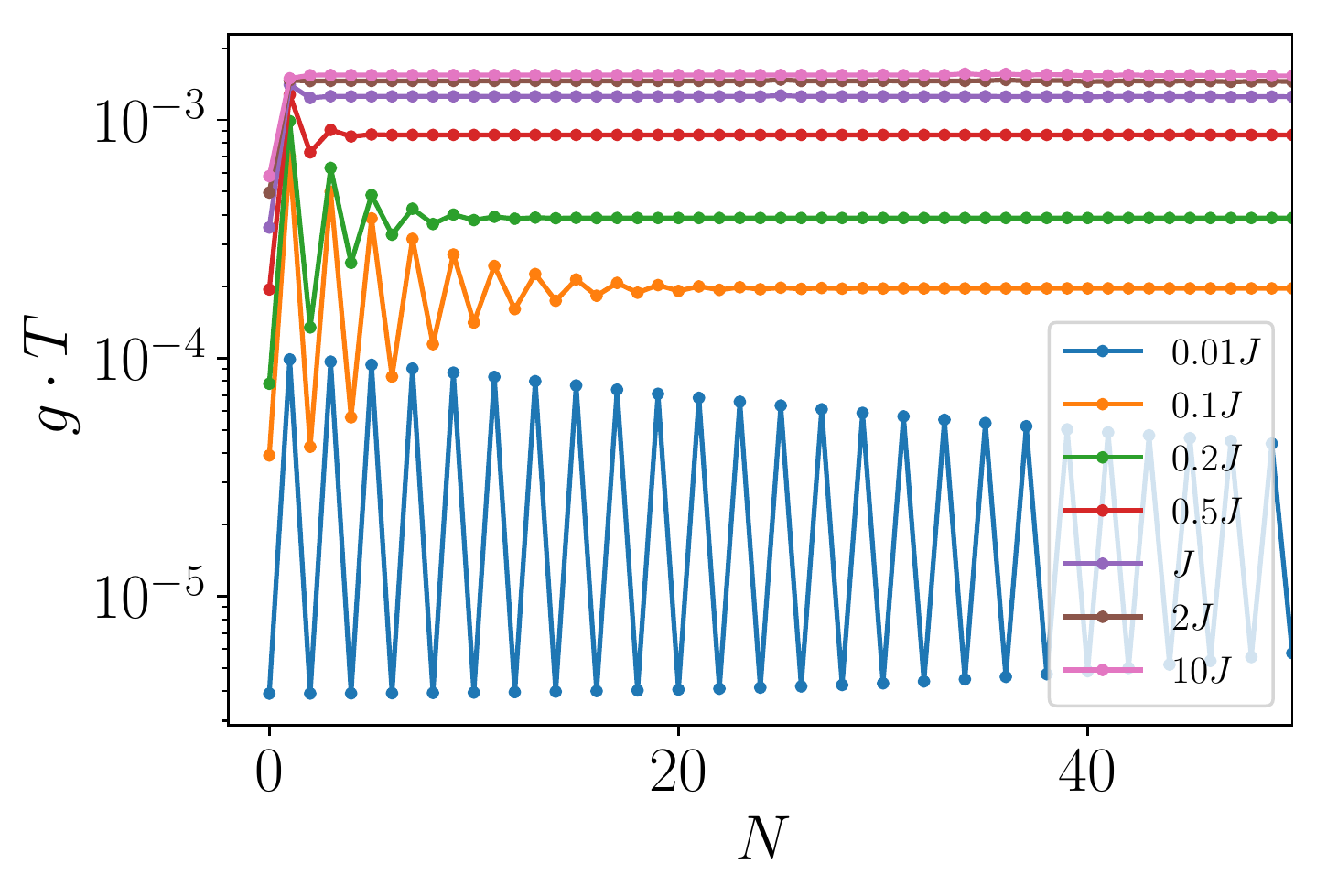}
\caption{\label{fig:conductance} Linear conductance $g$~\eqref{eq:g}, multiplied by the temperature $T$, as function of the chain size $N$, for $\Delta/J=0.1$ and $\mu=0$ for fermionic baths boundaries. Different colors correspond to different  temperatures $T$, listed in the legend.}
\end{figure}

For Lindblad boundaries, a numerical evaluation of the expression~\eqref{eq:freecurrentlindblad-1} shows that the current is independent of the system size $N$, namely
\begin{equation}
 J_{{\cal L}}=\frac{(\alpha_{L}\beta_{R}-\beta_{L}\alpha_{R})}{\Delta(1+\Delta^{2})}\,,
\end{equation}
in agreement with the result of Ref.~\citep{Znidaric_MPS_XX_open}. As it may be expected for ballistic systems, this result does not depend on the size of the system $N$. Nevertheless, this should be contrasted with the case in which the system is coherently driven by fermionic reservoirs. In this case, the current depends in general on the system size $N$, as it is shown in Fig.~\ref{fig:conductance}. This is particularly clear when considering the  conductance as defined in Eq.~\eqref{eq:g}, but at zero temperature ($T\to0$). In this limit,  the conductance corresponds to the transmission probability $|G^{\mathcal R}_{L,R}(\mu,\Delta)|^2$, taken at the  chemical potential $\mu$. When plotted as a function of the system size $N$, it displays pronounced even/odd oscillations corresponding to the appearance/disappearance of  Fabry-Perot resonances at the energy $\mu=0$ (see also Fig.~\ref{fig:resonances}). As expected, and accordingly to the Lindblad limit exemplified by Eq.~\eqref{eq:lb}, the size dependence disappears in the $T\rightarrow \infty$ limit, in which the temperature acts as if effectively broadening the  single-particle peaks, and making them indistinguishable.

\subsection{1D fermionic chain with loss/gain terms}\label{freefermionicchainloss}
We restrict in this case to Linblad type boundary conditions and add on top of the free fermionic Hamiltonian $H_{S,{\rm chain}}$ the following Lindblad terms in the dynamics $\nu\sum_{j}2c_{j}\rho_{t}c_{j}^{\dagger}-\{c_{j}^{\dagger}c_{j},\rho_{t}\}$,
which describes the loss of particles at rate $\nu$ on each site. The gain of particles is instead described by a Lindblad term of the form $\nu\sum_{j}2c^\dagger_{j}\rho_{t}c_{j}-\{c_{j}c_{j}^{\dagger},\rho_{t}\}$. Remarkably, the advanced and retarded Green's
functions in the position basis do not discriminate between losses and gains and read, in both cases \cite{DordaArrigoni_Impurity2014}:
\begin{equation}
\begin{split}
 & [\boldsymbol{G}^{\boldsymbol{\mathcal{R}(\mathcal{A})}}]^{-1}=\\
 & \begin{pmatrix}\epsilon\pm i\Delta_{L}\pm i\nu & J & \cdots & 0 & 0\\
J & \epsilon\pm i\nu & J & \cdots & 0\\
\vdots & J & \ddots & J & \vdots\\
0 & \cdots & J & \epsilon\pm i\nu & J\\
0 & 0 & \cdots & J & \epsilon\pm i\Delta_{R}\pm i\nu
\end{pmatrix}
\end{split}
\end{equation}
The equivalence of gains and losses for the advanced and retarded Green functions is also apparent in Eq.~\eqref{actionLindblad}, where the injection/extraction rates $\alpha/\beta$ appears with the same sign on the diagonal.

Importantly, this is not the case for the Keldysh Green function, which, in the case of losses, reads
\begin{equation}
\begin{split}
 & [\boldsymbol{G}^{\boldsymbol{\mathcal{K}}}]^{-1}=\\
 & \begin{pmatrix}-2i((\alpha_{L}-\beta_{L})-\nu) & 0 & \cdots & 0\\
0 & 2i\nu &  & \vdots\\
\vdots & 0 & \ddots & 0\\
0 & \cdots & 0 & -2i((\alpha_{R}-\beta_{R})-\nu)
\end{pmatrix}
\end{split}
\end{equation}
while the case with gains is obtained by making the substition $\nu\to -\nu$. Differently from the retarded and advanced components, the Keldysh Green's function discriminates between losses and gains in the bulk, with important consequences on transport. For instance,  the presence of loss/gain terms invalidates the relations (\ref{eq:Noziere1},\ref{eq:Noziere2}). Thus, standard identities for non-interacting fermion do not apply  anymore, but the complexity remains manageable because the loss/gain terms are quadratic in fermionic operators. We thus compute
the current from (\ref{eq:lindbladtransport}). More specifically, we have to compute $G_{1,1}^{\mathcal{K}}$
and $G_{N,N}^{\mathcal{K}}$ (recall that the left site index $L$ is $1$ and the right-site index is $N$ here). As before, we will consider the symmetric
case. In addition, to lighten the final formulae we choose $\alpha_{R/L}+\beta_{R/L}=\Delta$ and take $|J|=1$. We also suppose
that $\alpha_{L}-\beta_{L}=-(\alpha_{R}-\beta_{R})=\delta \tilde \mu$, to be distinguished from the bias in chemical potential $\delta\mu$ in Eq.~\eqref{eq:g}. We also recall the
relation $\boldsymbol{G^{\mathcal{K}}}=-\boldsymbol{G^{\mathcal{R}}[G^{\mathcal{K}}]^{-1}G^{\mathcal{A}}}$~\citep{kamenev_fieldtheorybook}.
Since the loss term simply adds a term on the diagonal, we get the
inverse of $[\boldsymbol{G}^{\boldsymbol{\mathcal{R}(\mathcal{A})}}]^{-1}$
by making the substitution $\epsilon\to\epsilon\pm i\nu$ in the
expression of the inverse for the free case in Section~\ref{freefermionicchain}.

\begin{figure}
\begin{centering}
\includegraphics[width=\columnwidth]{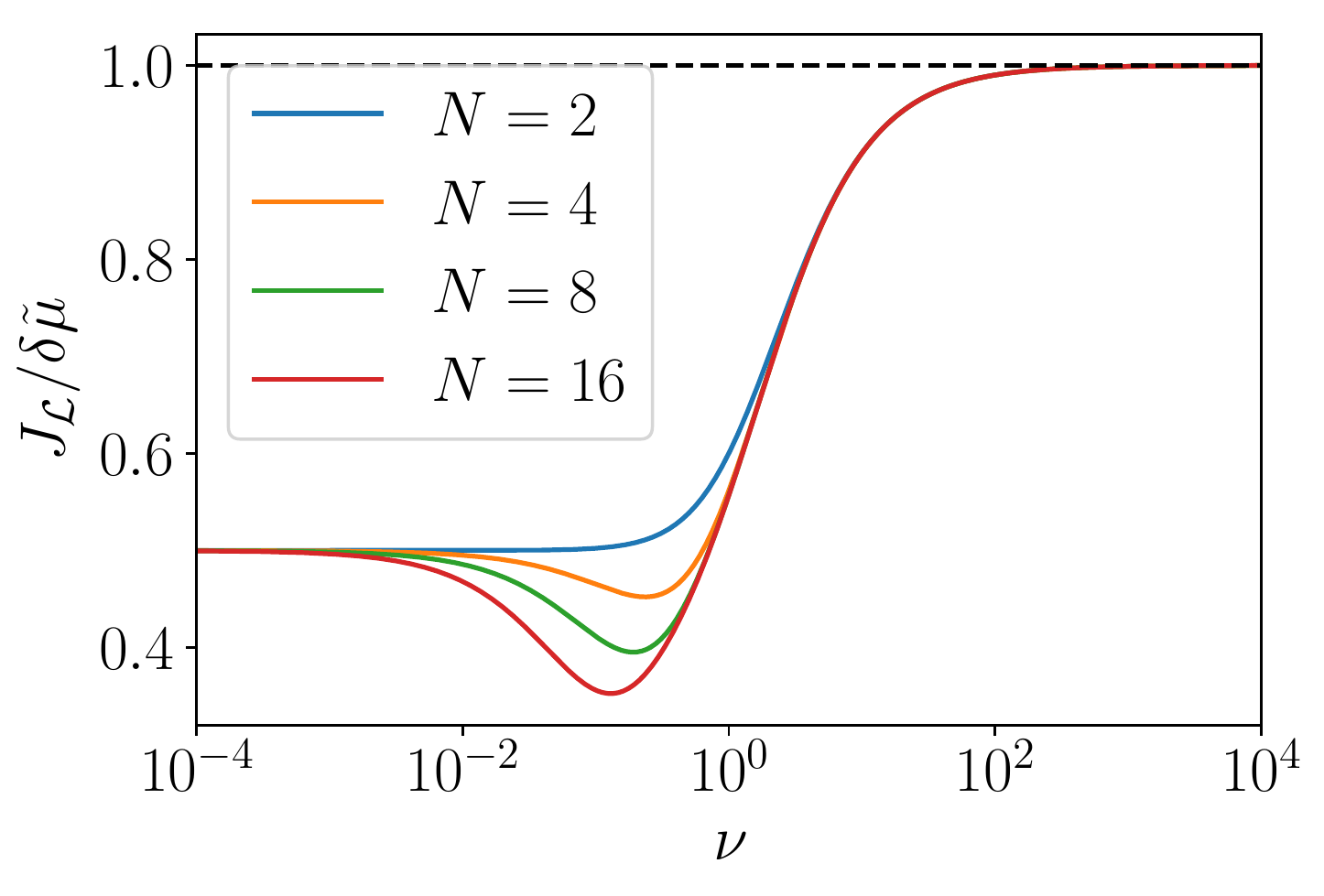}
\end{centering}
\caption{\label{fig:currentloss}Plot of the behavior of the current as a function of the loss of particles $\nu$ in the system for Lindblad boundary conditions. The different
colors correspond to different system sizes $N$. Here $\Delta/J=1$.}
\end{figure}

As it is shown in App.~\ref{app:detailsiii}, it is also in this case possible to reexpress the current as a function of $\boldsymbol{G^\mathcal{R}}$, but differently from the Landauer-B\" uttiker like expressions~\eqref{eq:freecurrentbath-1} and~\eqref{eq:freecurrentlindblad-1}, namely:
\begin{equation}\label{eq:jloss}
J_{{\cal L}}=\delta\tilde\mu\left[1+2\Delta\int\frac{d\epsilon}{2\pi}\bigg(-|G_{1,1}^{{\rm \mathcal{R}}}|^{2}+|G_{1,N}^{\mathcal{R}}|^{2}\bigg)\right]\,.
\end{equation}
$G_{i,j}^{\mathcal{R}}$ is given by the expression (\ref{eq:inverseG})
where one made the substitution $\epsilon\to\epsilon+ i\nu$. The behavior of the current with respect to $\nu$ is shown in Fig.~\ref{fig:currentloss} for different system sizes. In this case the current \emph{does} depend on the system size and,
for $N>2$, we observe first a decrease of the current and then an
\emph{increase }with respect to the loss of particles $\nu$. For large values of $\nu$
all the curves collapse towards the same value $J_{\mathcal L}=\alpha_R-\alpha_L=\delta\tilde \mu$. The reason for this counter-intuitive non-monotonous behavior is due to the fact that, in the large $\nu$ limit, the bulk dynamic doesn't matter anymore: a particle injected from the left never reaches the right reservoir. As a consequence, the current  $J_{L}$ injected from the left is equal to the rate of injection of the left reservoir $2\alpha_L$ and, inversely, $J_R$ to minus the rate of injection at the right reservoir $2\alpha_R$.

Another counterintuitive feature of this model, which is not apparent in Fig.~\ref{fig:currentloss}, is that the behavior of the current $J_{\mathcal L}$ \emph{does not} discriminate between loss and gain terms. Indeed, $J_{\cal{L}}$ is entirely expressed in terms of elements of $\boldsymbol{G^\mathcal{R}}$ (see App.~\ref{app:detailsiii}) which, as previously discussed, has the same expression whether we inject particles at rate $\nu$ along the chain or we extract them. This interesting dependence of the current show that the dissipative model deserves further scrutiny, which will be done in future studies.

\begin{figure}
\begin{centering}
\includegraphics[width=\columnwidth]{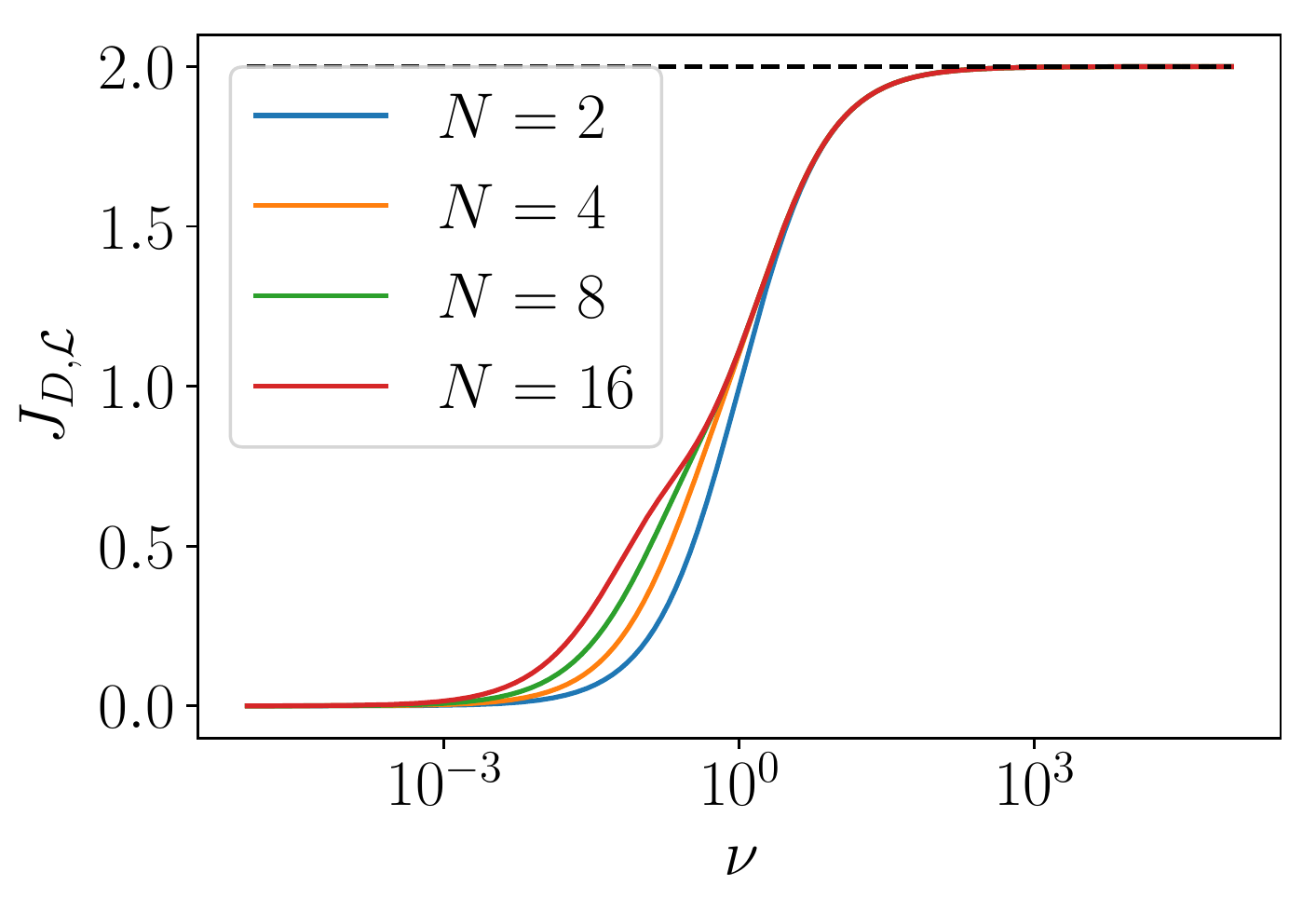}
\end{centering}
\caption{\label{fig:JD} Plot of $J_{D,{\cal L}}$  for Lindblad boundary conditions for different system sizes. We see that as $\nu$
increases, the current leaking from the system to the environment
also increases as expected.}
\end{figure}

Additionally, we can use the formula~\eqref{eq:Lindbladtransport2} to compute the current of
particles going from the system to the environment. Again, we provide the proof in  App.~\ref{app:detailsiii}. It leads to
\begin{equation}
J_{D,{\cal L}}=\pm4\int\frac{d\epsilon}{2\pi}\sum_{j=1}^N\nu\,|G_{1,j}^{\mathcal{R}}|^{2}\,,
\end{equation}
where the $+/-$ sign corresponds to the situation in which we consider a loss/gain term weighted by $\nu$ in the Lindbladian. Differently from the direct current $J_{\mathcal L}$, the current $J_{D,{\cal L}}$ discriminates between injection
or extraction of particles, as expected. For loss terms, we show the
behavior of $J_{D,{\cal L}}$ for different system sizes on Fig.~\ref{fig:JD}. For
a gain term, we would just have gotten the symmetric of this curve
with respect to the horizontal axis. Notice that $J_{D,\mathcal L}$ does not depend on the bias $\delta\tilde \mu$ and that it tends  towards $2$ in the $\nu\rightarrow \infty$ limit.

\section{Conclusion and perspectives}\label{sec:conclusion}

In this work, we derived a generic expression for the stationary current flowing in a general interacting system, potentially with loss or gain of particles, driven by Lindblad jump operators. This expression is the equivalent for Markovian reservoirs of the Meir-Wingreen formula for the fermionic reservoirs. These two situations are related by a limit of high temperature and chemical potential for the fermionic reservoirs. In the non-interacting regime, an additional number of simplifications are possible in this generic transport formula.
Using our general transport formula, we showed that, for Lindblad boundaries, the current is always in a linear regime no matter the values of the injecting and extracting rates. We illustrated how our approach can be systematically  applied on three different examples concerning non-interacting fermions. For each application, we  witnessed drastic differences in the behavior of the current between fermionic bath boundaries and Lindblad ones. Of particular interest is the example of a fermionic tight-binding chain with loss of gain of particles. This system shows a non monotonic 
dependence of the current in the strength of the dissipation. In addition it has the very counterintuitive feature that the behavior of the current is independent on whether one injects or extracts particles in the bulk.

Our work thus provides a path to tackle transport properties of open systems driven by Lindblad boundaries, with methodology similar to the one used in for mesoscopic systems in contact 
with fermionic reservoirs and should help bridge the gap between the phenomena in these two situations. 

There are of course several directions in which it would be interesting to extend our study.  

Concerning the reservoirs, the study of the present paper has been done for Lindbladians that described injection and extraction of particles. It would be interesting to understand whether similar mappings exist for other types of Lindblad action and whether it can be done in a systematic way. For instance, it is possible to generate diffusive behaviors in lattice models of spins or fermions by putting local, independent dephasing noise at each site whose mean actions are described by quartic Lindblad terms \cite{Znidaric_dephasing,Znidaric__dephasingXXZ,BauerBernardJin_Stoqdissipative,dolgirev_non-gaussian_2020}. In this case, the noise is introduced by hand and is supposed to model interaction with an external environment. It would be interesting to make this description more physically grounded by seeing it as emerging from the interaction with an actual environment taken within a certain limit.

Concerning the systems themselves, the models for which we provided explicit evaluations of the transport formulas~\eqref{eq:lindbladtransport} and~\eqref{eq:Lindbladtransport2} here were only non-interacting models. It would be natural
to apply our formula to interacting models and give a general interpretation in terms of Keldysh Green functions of the emergence of ballistic and anomalous diffusion in integrable systems~\cite{Zotos1997,Zotos1999,Prosen2011,Gopalakrishnan2019,Znidaric2011,Ljubotina2017}. In this spirit, future and interesting research directions could be concerned with single impurity problems~\cite{hewson_kondo_1993,bulla_numerical_2008,Gull_qimpurity}. Given the simplifications that we observed in the studies of this paper for the case of Markovian reservoirs, one can have 
the hope that tackling such systems will be simpler than for their fermionic counterparts. Finally, extending our  
approach to describe transport of either matter or energy in interacting systems driven by Lindbladians~\cite{MendozaArenas2019} beyond the strictly one-dimensional case could be tackled by addressing the transport properties in quantum ladder systems.

\begin{acknowledgments}
The authors thank Enrico Arrigoni, Jo\~ ao Ferreira, Pierre Le Doussal and Marko {\v{Z}}nidari{\v{c}} for useful discussions. This work has been
supported by the Swiss National Science Foundation under Division II. M. F. acknowledges support from the FNS/SNF AmbizioneGrant PZ00P2\_174038.
\end{acknowledgments}

\appendix

\section{Derivation of the action associated to Lindblad type boundaries}\label{sec:derivationLindbladaction}

Here, for completeness, we briefly show how to derive (\ref{actionLindblad})
from the main text. This derivation was carried out in \citep{Diehl_KeldyshLindblad}
and we simply transcribe it here.

Let $\mathfrak{L}$ be the Liouvillian generating the total evolution,
i.e $\mathfrak{L}(\rho)=-i[H,\rho]+{\cal L}(\rho)$. where we choose
for ${\cal L}$ the injecting and extracting terms introduced in the
main text : ${\cal L}(\bullet)=\alpha(2c^{\dagger}\bullet c-\{cc^{\dagger},\bullet\}_{+})+\beta(2c\bullet c^{\dagger}-\{c^{\dagger}c,\bullet\}_{+}$.

By definition
\begin{equation}
\rho_{t}=\lim_{N\to\infty}(\mathbb{I}+\delta t{\cal \mathfrak{L}})^{N}\rho_{0}
\end{equation}
with $N\delta t=t$. To get the Keldysh action, we have to insert
$2N$ identity resolutions in the above equation, $N$ in the forward
direction of time and $N$ in the backward direction of time. The
contribution of ${\cal L}$ to the action for an elementary time step
is :
\begin{align}
 & \left\langle \psi_{n+1}^{+}\right|{\cal L}(\left|\psi_{n}^{+}\right\rangle \left\langle -\psi_{n}^{-}\right|)\left|-\psi_{n+1}^{-}\right\rangle \\
 & =-\left\langle \psi_{n+1}^{+}|\psi_{n}^{+}\right\rangle \left\langle \psi_{n}^{-}|\psi_{n+1}^{-}\right\rangle \nonumber \\
 & \hspace{1em}\bigg(\alpha(2\bar{\psi}_{n+1}^{+}\psi_{n+1}^{-}+(2-\bar{\psi}_{n+1}^{+}\psi_{n}^{+}-\bar{\psi}_{n}^{-}\psi_{n+1}^{-}))\nonumber \\
 & \hspace{1em}+\beta(2\psi_{n}^{+}\bar{\psi}_{n}^{-}+(\bar{\psi}_{n+1}^{+}\psi_{n}^{+}+\bar{\psi}_{n}^{-}\psi_{n+1}^{-}))\bigg)\nonumber
\end{align}
where $\psi_{n}^{\pm}\bar{\psi}_{n}^{\pm}$ are the usual Grassman
fields. Summing over all elementary steps and performing the Keldysh
rotation one gets the contribution to the action :
\begin{align}
S_{{\cal{L}}} & =\int\frac{d\epsilon}{2\pi}(\bar{\psi}_{1},\bar{\psi}_{2})_{(\epsilon)}\begin{pmatrix}\alpha+\beta & 2(\beta-\alpha)\\
0 & -(\alpha+\beta)
\end{pmatrix}\begin{pmatrix}\psi_{1}\\
\psi_{2}
\end{pmatrix}_{(\epsilon)}
\end{align}

\section{Derivation of the effective action for fermionic bath boundary condition}\label{sec:Derivation-of-the}

The starting point is the Hamiltonian (\ref{bathH}) from the main
text :
\begin{align}
H & =H_{{\rm S}}+H_{{\rm F}}-\mu N_{{\rm F}}+H_{{\rm \tau}},\nonumber \\
H_{{\rm F}} & =\sum_{k}\epsilon_{k}\tilde{c}_{{\rm F},k}^{\dagger}\tilde{c}_{{\rm F},k},\\
H_{\tau} & =-\tau(c_{{\rm F},0}^{\dagger}c_{\rm F,j}+{\rm h.c}).\nonumber
\end{align}
As explained in the main text, we take the continuous limit in $k$
and the dispersion relation is linearized around the Fermi points
\begin{equation}
\epsilon_{k}=\begin{cases}
v_{F}(k-k_{F}) & -\Lambda\leq k\leq\Lambda\\
-v_{F}(k+k_{F}) & -\Lambda\leq k\leq\Lambda
\end{cases}
\end{equation}
The total Keldysh action of the full system is composed of three
parts,
\begin{equation}
S_{{\rm tot}}=S_{{\rm S}}+S_{{\rm F}}+S_{{\rm \tau}}
\end{equation}
The goal is to get an effective action for the system alone by integrating
over the degrees of freedom of the bath contained in $S_{{\rm F}}$
and $S_{{\rm \tau}}$.

The Keldysh action for the reservoir is \citep{kamenev_fieldtheorybook}
\begin{align}
 & S_{\rm F}=\nonumber \\
 & \int_{-\Lambda}^{\Lambda}\frac{dk}{2\pi}\int\frac{d\epsilon}{2\pi}(\bar{\tilde{\psi}}_{{\rm F}}^{1},\bar{\tilde{\psi}}_{{\rm F}}^{2})_{(\epsilon,k)}\begin{pmatrix}[G_{{\rm F}}^{{\mathcal R}}]^{-1} & [G_{{\rm F}}^{-1}]^{{\mathcal K}}\\
0 & [G_{{\rm F}}^{{\mathcal A}}]^{-1}
\end{pmatrix}_{(\epsilon,k)}\begin{pmatrix}\tilde{\psi}_{{\rm F}}^{1}\\
\tilde{\psi}_{{\rm F}}^{2}
\end{pmatrix}_{(\epsilon,k)}
\end{align}
with
\begin{align}
G_{{\rm F}}^{{\mathcal {R(A)}}}(\epsilon,k) & =\frac{1}{\epsilon-(\epsilon_{k}-\mu)\pm i\delta},\\
G_{{\rm F}}^{{\mathcal K}}(\epsilon,k) & =-2\pi i\tanh(\frac{\epsilon-\mu}{2T})\delta(\epsilon-(\epsilon_{k}-\mu)).
\end{align}
where $\delta=0^{+}$.

The Keldysh action associated to $H_{\tau}$ is
\begin{align}
 & S_{\tau}=\nonumber \\
 & \tau\int_{-\Lambda}^{\Lambda}\frac{dk}{2\pi}\int\frac{d\epsilon}{2\pi}\bigg((\bar{\tilde{\psi}}_{{\rm F}}^{1},\bar{\tilde{\psi}}_{{\rm F}}^{2})_{(\epsilon,k)}\begin{pmatrix}\psi_{j}^{1}\\
\psi_{j}^{2}
\end{pmatrix}_{(\epsilon)}\\
 & +(\bar{\psi}_{j}^{1},\bar{\psi}_{j}^{2})_{(\epsilon,k)}\begin{pmatrix}\tilde{\psi}_{{\rm F}}^{1}\\
\tilde{\psi}_{{\rm F}}^{2}
\end{pmatrix}_{(\epsilon)}\bigg)
\end{align}
Since the theory is quadratic, we can use usual Gaussian integrals
formula to integrate over $\tilde{\psi}_{{\rm F}},\bar{\tilde{\psi}}_{{\rm F}}$
\begin{align}
 & \int\prod d\bar{\psi}_{j}d\psi_{j}e^{-\sum_{ij}\bar{\psi}_{i}M_{ij}\psi_{j}+\sum_{j}\bar{\psi}_{j}\chi_{j}+\bar{\chi}_{j}\psi_{j}}\\
 & =\det(M)e^{\sum\bar{\chi}_{i}M_{ij}^{-1}\chi_{j}}
\end{align}
identifying $\psi\to i\begin{pmatrix}\psi^{1}\\
\psi^{2}
\end{pmatrix}$, $M\to i\begin{pmatrix}[G_{{\rm F}}^{{\mathcal R}}]^{-1} & [G_{{\rm F}}^{-1}]^{{\mathcal K}}\\
0 & [G_{{\rm F}}^{{\mathcal A}}]^{-1}
\end{pmatrix}_{(\epsilon,k)}$, $\chi\to -i\tau\begin{pmatrix}\psi^{1}\\
\psi^{2}
\end{pmatrix}_{(\epsilon)}$in the above formula and using $\det(M)=1$ yields the effective action
of the bath on the system :
\begin{equation}
S_{\mathcal F}=-\int_{-\Lambda}^{\Lambda}\frac{dk}{2\pi}\int\frac{d\epsilon}{2\pi}\tau^{2}(\bar{\psi}_{j}^{1},\bar{\psi}_{j}^{2})_{(\epsilon)}\begin{pmatrix}G_{{\rm F}}^{{\mathcal R}} & G_{{\rm F}}^{{\mathcal K}}\\
0 & G_{\rm F}^{{\mathcal A}}
\end{pmatrix}_{(\epsilon,k)}\begin{pmatrix}\psi_{j}^{1}\\
\psi_{j}^{2}
\end{pmatrix}_{(\epsilon)}
\end{equation}
So we need to compute $\int_{-\Lambda}^{\Lambda}\frac{dk}{2\pi}G_{{\rm F}}^{{\mathcal R}({\mathcal A})}(\epsilon,k)$
and $\int_{-\Lambda}^{\Lambda}\frac{dk}{2\pi}G_{{\rm F}}^{{\mathcal K}}(\epsilon,k)$:
\begin{align}
 & \int_{-\Lambda}^{\Lambda}\frac{dk}{2\pi}G_{{\rm F}}^{{\mathcal R}({\mathcal A})}(\epsilon,k)\nonumber \\
 & =\int_{-\Lambda}^{\Lambda}\frac{dk}{2\pi}\frac{1}{\epsilon-(\epsilon_{k}-\mu)\pm i\delta}\nonumber \\
 & =\int_{-\Lambda v_{F}}^{\Lambda v_{F}}\frac{d\epsilon_{k}}{\pi v_{F}}\frac{1}{\epsilon-(\epsilon_{k}-\mu)\pm i\delta}\nonumber \\
 & =\int_{-\Lambda v_{F}}^{\Lambda v_{F}}\frac{d\epsilon_{k}}{\pi v_{F}}\big({\rm vp}(\frac{1}{\epsilon-(\epsilon_{k}-\mu)})\mp i\pi\delta(\epsilon-(\epsilon_{k}-\mu))\big)\nonumber \\
 & =\frac{1}{\pi v_{F}}(\ln\frac{\epsilon-(-\Lambda v_{F}-\mu)}{\epsilon-(\Lambda v_{F}-\mu)}\mp i\pi)+O(\frac{1}{\Lambda})\nonumber \\
 & =_{\lim\Lambda\to\infty}\mp\frac{i}{v_{F,L}}.
\end{align}
\begin{align}
 & \int_{-\Lambda}^{\Lambda}\frac{dk}{2\pi}G_{{\rm F}}^{{\mathcal K}}(\epsilon,k)\nonumber \\
 & =-\int_{-\Lambda}^{\Lambda}\frac{dk}{2\pi}2\pi i\tanh(\frac{\epsilon-\mu}{2T})\delta(\epsilon-(\epsilon_{k}-\mu))\nonumber \\
 & =-\int_{-\Lambda v_{F}}^{\Lambda v_{F}}\frac{d\epsilon_{k}}{\pi v_{F}}2\pi i\tanh(\frac{\epsilon-\mu}{2T})\delta(\epsilon-(\epsilon_{k}-\mu))\nonumber \\
 & =_{\lim\Lambda\to\infty}-\frac{2i}{v_{F}}\tanh(\frac{\epsilon-\mu}{2T}),
\end{align}
which proves (\ref{eq:actionbain})
\begin{equation}
S_{{\mathcal F}}=i\int\frac{d\epsilon}{2\pi}(\bar{\psi}_{j}^{1},\bar{\psi}_{j}^{2})_{(\epsilon)}\Delta\begin{pmatrix}1 & 2\tanh(\frac{\epsilon-\mu}{2T})\\
0 & -1
\end{pmatrix}\begin{pmatrix}\psi_{j}^{1}\\
\psi_{j}^{2}
\end{pmatrix}_{(\epsilon)}
\end{equation}

\section{Derivation of the transport formula for Lindblad boudaries}\label{sec:Alternative-derivation-of}

In this appendix we give the full derivation of the equation for the current (\ref{eq:lindbladtransport}) for a generic system, potentially with interactions and loss/gain of particles.
The dynamical evolution of the density $n_{L}={\rm tr}(\rho_{t}c_{L}^{\dagger}c_{L})$
of the site connected to the left reservoir is given by
\begin{equation}
\frac{dn_{L}}{dt}=2\alpha_{L}(1-n_{L})-\beta_{L}n_{L}+\mathfrak{L}^*(n_L)
\end{equation}
$\mathfrak{L}^*$ is the dual of the Liouvillian, generating the time evolution of operators in the Heisenberg picture.
The first term corresponds to the current from the reservoir to the site, the second term represents interaction with the rest of the system and/or other external degrees of freedom. In the stationary state, these two are equal. We have a similar equation for the right site:
\begin{equation}
\frac{dn_{R}}{dt}=2\alpha_{R}(1-n_{R})-\beta_{R}n_{R}+\mathfrak{L}^*(n_R)
\end{equation}
Hence the stationary current is given by
\begin{equation}
J_{{\cal L}}=\frac{1}{2}(J_{L}+J_{R})\label{eq:JLLindblad}
\end{equation}
with $J_{L}=2\alpha_{L}(1-n_{L})-\beta_{L}n_{L}$, $J_{R}=-(2\alpha_{R}(1-n_{R})-\beta_{R}n_{R})$.

The expression of the density at site $j$ in terms of Green's functions is given by $n_{j}=\frac{i}{2}\int\frac{d\epsilon}{2\pi}(G_{j,j}^{{\mathcal R}}-G_{j,j}^{{\mathcal A}}+G_{j,j}^{{\mathcal K}})$. We then have
\begin{align}
n_{L} & =\frac{i}{4}\frac{d\epsilon}{2\pi}{\rm tr}(\boldsymbol{\gamma^{L}}(\boldsymbol{G^{{\mathcal R}}}-\boldsymbol{G^{{\mathcal A}}}+\boldsymbol{G^{{\mathcal K}}}))\label{eqAppC:nL}\\
n_{R} & =\frac{i}{4}\frac{d\epsilon}{2\pi}{\rm tr}(\boldsymbol{\gamma^{R}}(\boldsymbol{G^{{\mathcal R}}}-\boldsymbol{G^{{\mathcal A}}}+\boldsymbol{G^{{\mathcal K}}}))\label{eqAppC:nR}
\end{align}
Inserting these two equations in (\ref{eq:JLLindblad}) and using
that $i\int\frac{d\epsilon}{2\pi}(\boldsymbol{G^{{\mathcal R}}}-\boldsymbol{G^{{\mathcal A}}})=\boldsymbol{\mathbb{I}}$,
we arrive at (\ref{eq:lindbladtransport}).

Similarly, one can get the current from the chain to the external environment $J_{D,\cal L}=(J_{L}-J_{R})$. When there is no loss in the system, this current is just $0$ by
conservation of the number of particle. If $J_{D,\cal L}>0$ it means, that
there are \emph{loss }of particles in the system\emph{ }while if $J_{D,\cal L}<0$,
it means that there is a\emph{ gain }of particles in the system. Using again (\ref{eqAppC:nL},\ref{eqAppC:nR}) and $i\int\frac{d\epsilon}{2\pi}(\boldsymbol{G^{{\mathcal R}}}-\boldsymbol{G^{{\mathcal A}}})=\boldsymbol{\mathbb{I}}$, we get after some elementary computations that
for Lindblad type boundary conditions:
\begin{equation}
J_{D,{\cal L}}=\frac{i}{2}\int\frac{d\epsilon}{2\pi}{\rm tr}(((\alpha_{L}+\beta_{L})\boldsymbol{\gamma_{L}}+(\alpha_{R}+\beta_{R})\boldsymbol{\gamma_{R}})\boldsymbol{G^{\mathcal{K}}}).
\end{equation}
which is Eq.(\ref{eq:Lindbladtransport2}) of the main text.
\section{Transport through dissipative fermionic chains}\label{app:detailsiii}

In this appendix, we provide more details on the calculations presented
in Sec. \ref{freefermionicchainloss}.

As explained in the main text, we compute the current from (\ref{eq:lindbladtransport}). For
that, we need to calculate the elements $G_{1,1}^{\mathcal{K}}$ and
$G_{N,N}^{\mathcal{K}}$. The starting point is the relation $\boldsymbol{G^{\mathcal{K}}}=-\boldsymbol{G^{{\cal R}}[G^{\mathcal{K}}]^{-1}G^{\mathcal{A}}}$:
\begin{align}
 & G_{k,l}^{{\rm \mathcal{K}}}=-\sum_{ij}G_{k,i}^{\mathcal{R}}[G^{\mathcal{K}}]_{i,j}^{-1}G_{j,l}^{\mathcal{A}}\nonumber \\
 & =\sum_{ij}G_{k,i}^{{\rm \mathcal{R}}}2i\delta_{i,j}\big((\alpha_{L}-\beta_{L})\delta_{i,1}\\
 & +(\alpha_{R}-\beta_{R})\delta_{i,L}-\nu\big)G_{j,l}^{\mathcal{A}}\nonumber
\end{align}
Thus,
\begin{align}
G_{1,1}^{\mathcal{K}}= & 2i\big(|G_{1,1}^{\mathcal{R}}|^{2}(\alpha_{L}-\beta_{L})+|G_{1,N}^{\mathcal{R}}|^{2}(\alpha_{R}-\beta_{R})\\
 & -\sum_{j}\nu|G_{1,j}^{\mathcal{R}}|^{2}\big)\nonumber \\
G_{N,N}^{\mathcal{K}}= & 2i\big(|G_{1,N}^{\mathcal{R}}|^{2}(\alpha_{L}-\beta_{L})+|G_{N,N}^{\mathcal{R}}|^{2}(\alpha_{R}-\beta_{R})\\
 & -\sum_{j}\nu|G_{j,N}^{\mathcal{R}}|^{2}\big)\nonumber
\end{align}
Now let us make the simplifying approximations $\alpha_{L}+\beta_{L}=\alpha_{R}+\beta_{R}=\Delta$,
$\alpha_{L}-\beta_{L}=-(\alpha_{R}-\beta_{R})=\delta\tilde{\mu}$. The
expression for the current then simplifies into :
\begin{align}
J_{{\cal L}} & =\delta\tilde{\mu}+\frac{i}{2}\Delta\int\frac{d\epsilon}{2\pi}\big(G_{1,1}^{\mathcal{K}}-G_{N,N}^{\mathcal{K}}\big)\,,\\
J_{D,{\cal L}} & =i\Delta\int\frac{d\epsilon}{2\pi}\big(G_{1,1}^{\mathcal{K}}+G_{N,N}^{\mathcal{K}}\big)
\end{align}
Substituting the found values for $G_{1,1}^{\mathcal{K}}$ and $G_{N,N}^{\mathcal{K}}$
in these last expressions, we arrive at
\begin{align}
J_{{\cal L}} & =\delta\tilde{\mu}-\Delta\int\frac{d\epsilon}{2\pi}\bigg[2\delta\tilde{\mu}(|G_{1,1}^{\mathcal{R}}|^{2}-|G_{1,N}^{\mathcal{R}}|^{2})\\
 & \hspace{1em}-\sum_{j}\nu(|G_{1,j}^{\mathcal{R}}|^{2}-|G_{j,N}^{\mathcal{R}}|^{2})\bigg]\,,\nonumber \\
J_{D,{\cal L}} & =2\Delta\int\frac{d\epsilon}{2\pi}\sum_{j}\nu\big(|G_{1,j}^{\mathcal{R}}|^{2}+|G_{j,N}^{\mathcal{R}}|^{2}\big)
\end{align}
Now from (\ref{eq:inverseG}) one can remark that $G_{1,N-j+1}^{\mathcal{R}}=G_{j,N}^{\mathcal{R}}$.
By re indexing the terms in the sum we have that $\sum_{j}|G_{1,j}^{\mathcal{R}}|^{2}=\sum_{j}|G_{j,N}^{\mathcal{R}}|^{2}$
so that :
\begin{align}
J_{{\cal L}} & =\delta\tilde{\mu}\big(1-2\Delta\int\frac{d\epsilon}{2\pi}(|G_{1,1}^{\mathcal{R}}|^{2}-|G_{1,N}^{\mathcal{R}}|^{2})\big)\,,\\
J_{D,{\cal L}} & =4\Delta\int\frac{d\epsilon}{2\pi}\sum_{j}\nu|G_{1,j}^{\mathcal{R}}|^{2}
\end{align}
For gain terms instead of loss terms, the proof follows exactly the
same lines. We end up with the same expression for $J_{{\cal L}}$
while
\begin{equation}
J_{D,{\cal L}}=-4\Delta\int\frac{d\epsilon}{2\pi}\sum_{j}\nu|G_{1,j}^{\mathcal{R}}|^{2}
\end{equation}

\bibliographystyle{apsrev4-2}
\bibliography{bibliotonybathlindblad,bib_michele,main_IB}

\begin{thebibliography}{104}%
\makeatletter
\providecommand \@ifxundefined [1]{%
 \@ifx{#1\undefined}
}%
\providecommand \@ifnum [1]{%
 \ifnum #1\expandafter \@firstoftwo
 \else \expandafter \@secondoftwo
 \fi
}%
\providecommand \@ifx [1]{%
 \ifx #1\expandafter \@firstoftwo
 \else \expandafter \@secondoftwo
 \fi
}%
\providecommand \natexlab [1]{#1}%
\providecommand \enquote  [1]{``#1''}%
\providecommand \bibnamefont  [1]{#1}%
\providecommand \bibfnamefont [1]{#1}%
\providecommand \citenamefont [1]{#1}%
\providecommand \href@noop [0]{\@secondoftwo}%
\providecommand \href [0]{\begingroup \@sanitize@url \@href}%
\providecommand \@href[1]{\@@startlink{#1}\@@href}%
\providecommand \@@href[1]{\endgroup#1\@@endlink}%
\providecommand \@sanitize@url [0]{\catcode `\\12\catcode `\$12\catcode
  `\&12\catcode `\#12\catcode `\^12\catcode `\_12\catcode `\%12\relax}%
\providecommand \@@startlink[1]{}%
\providecommand \@@endlink[0]{}%
\providecommand \url  [0]{\begingroup\@sanitize@url \@url }%
\providecommand \@url [1]{\endgroup\@href {#1}{\urlprefix }}%
\providecommand \urlprefix  [0]{URL }%
\providecommand \Eprint [0]{\href }%
\providecommand \doibase [0]{https://doi.org/}%
\providecommand \selectlanguage [0]{\@gobble}%
\providecommand \bibinfo  [0]{\@secondoftwo}%
\providecommand \bibfield  [0]{\@secondoftwo}%
\providecommand \translation [1]{[#1]}%
\providecommand \BibitemOpen [0]{}%
\providecommand \bibitemStop [0]{}%
\providecommand \bibitemNoStop [0]{.\EOS\space}%
\providecommand \EOS [0]{\spacefactor3000\relax}%
\providecommand \BibitemShut  [1]{\csname bibitem#1\endcsname}%
\let\auto@bib@innerbib\@empty
%</preamble>
\bibitem [{\citenamefont {Landauer}(1970)}]{landauer_electrical_1970}%
  \BibitemOpen
  \bibfield  {author} {\bibinfo {author} {\bibfnamefont {R.}~\bibnamefont
  {Landauer}},\ }\href@noop {} {\bibfield  {journal} {\bibinfo  {journal}
  {Philosophical magazine}\ }\textbf {\bibinfo {volume} {21}},\ \bibinfo
  {pages} {863} (\bibinfo {year} {1970})}\BibitemShut {NoStop}%
\bibitem [{\citenamefont {B{\"u}ttiker}(1986)}]{buttiker_four-terminal_1986}%
  \BibitemOpen
  \bibfield  {author} {\bibinfo {author} {\bibfnamefont {M.}~\bibnamefont
  {B{\"u}ttiker}},\ }\href {https://doi.org/10.1103/PhysRevLett.57.1761}
  {\bibfield  {journal} {\bibinfo  {journal} {Physical Review Letters}\
  }\textbf {\bibinfo {volume} {57}},\ \bibinfo {pages} {1761} (\bibinfo {year}
  {1986})},\ \bibinfo {note} {publisher: American Physical Society}\BibitemShut
  {NoStop}%
\bibitem [{\citenamefont {Lesovik}\ and\ \citenamefont
  {Sadovskyy}(2011)}]{lesovik_scattering_2011}%
  \BibitemOpen
  \bibfield  {author} {\bibinfo {author} {\bibfnamefont {G.~B.}\ \bibnamefont
  {Lesovik}}\ and\ \bibinfo {author} {\bibfnamefont {I.~A.}\ \bibnamefont
  {Sadovskyy}},\ }\href@noop {} {\bibfield  {journal} {\bibinfo  {journal}
  {Physics-Uspekhi}\ }\textbf {\bibinfo {volume} {54}},\ \bibinfo {pages}
  {1007} (\bibinfo {year} {2011})}\BibitemShut {NoStop}%
\bibitem [{\citenamefont {Ashcroft}(2003)}]{ashcroft_solid_2003}%
  \BibitemOpen
  \bibfield  {author} {\bibinfo {author} {\bibfnamefont {N.~W.}\ \bibnamefont
  {Ashcroft}},\ }\href@noop {} {\emph {\bibinfo {title} {Solid {State}
  {Physics}}}}\ (\bibinfo  {publisher} {Thomson Press},\ \bibinfo {address}
  {New Delhi},\ \bibinfo {year} {2003})\BibitemShut {NoStop}%
\bibitem [{\citenamefont {Ziman}(1960)}]{ziman_electrons_1960}%
  \BibitemOpen
  \bibfield  {author} {\bibinfo {author} {\bibfnamefont {J.~M.}\ \bibnamefont
  {Ziman}},\ }\href@noop {} {\emph {\bibinfo {title} {Electrons and phonons:
  the theory of transport phenomena in solids}}}\ (\bibinfo  {publisher}
  {Oxford university press},\ \bibinfo {year} {1960})\BibitemShut {NoStop}%
\bibitem [{\citenamefont {Coleman}(2015)}]{coleman_introduction_2015}%
  \BibitemOpen
  \bibfield  {author} {\bibinfo {author} {\bibfnamefont {P.}~\bibnamefont
  {Coleman}},\ }\href@noop {} {\emph {\bibinfo {title} {Introduction to
  many-body physics}}}\ (\bibinfo  {publisher} {Cambridge University Press},\
  \bibinfo {year} {2015})\BibitemShut {NoStop}%
\bibitem [{\citenamefont {Bruus}\ and\ \citenamefont
  {Flensberg}(2004)}]{bruus_many-body_2004}%
  \BibitemOpen
  \bibfield  {author} {\bibinfo {author} {\bibfnamefont {H.}~\bibnamefont
  {Bruus}}\ and\ \bibinfo {author} {\bibfnamefont {K.}~\bibnamefont
  {Flensberg}},\ }\href@noop {} {\emph {\bibinfo {title} {Many-{Body} {Quantum}
  {Theory} in {Condensed} {Matter} {Physics}. {An} {Introduction}}}}\ (\bibinfo
   {publisher} {Oxford University Press, USA},\ \bibinfo {year}
  {2004})\BibitemShut {NoStop}%
\bibitem [{\citenamefont {Nazarov}\ and\ \citenamefont
  {Blanter}(2009)}]{nazarov_quantum_2009}%
  \BibitemOpen
  \bibfield  {author} {\bibinfo {author} {\bibfnamefont {Y.~V.}\ \bibnamefont
  {Nazarov}}\ and\ \bibinfo {author} {\bibfnamefont {Y.~M.}\ \bibnamefont
  {Blanter}},\ }\href {https://doi.org/10.1017/CBO9780511626906} {\emph
  {\bibinfo {title} {Quantum {Transport}: {Introduction} to {Nanoscience}}}}\
  (\bibinfo  {publisher} {Cambridge University Press},\ \bibinfo {address}
  {Cambridge},\ \bibinfo {year} {2009})\BibitemShut {NoStop}%
\bibitem [{\citenamefont {Akkermans}\ and\ \citenamefont
  {Montambaux}(2007)}]{akkermans_mesoscopic_2007}%
  \BibitemOpen
  \bibfield  {author} {\bibinfo {author} {\bibfnamefont {E.}~\bibnamefont
  {Akkermans}}\ and\ \bibinfo {author} {\bibfnamefont {G.}~\bibnamefont
  {Montambaux}},\ }\href@noop {} {\emph {\bibinfo {title} {Mesoscopic {Physics}
  of {Electrons} and {Photons}}}}\ (\bibinfo  {publisher} {Cambridge University
  Press},\ \bibinfo {year} {2007})\BibitemShut {NoStop}%
\bibitem [{\citenamefont {Moskalets}(2011)}]{moskalets_scattering_2011}%
  \BibitemOpen
  \bibfield  {author} {\bibinfo {author} {\bibfnamefont {M.~V.}\ \bibnamefont
  {Moskalets}},\ }\href@noop {} {\emph {\bibinfo {title} {Scattering matrix
  approach to non-stationary quantum transport}}}\ (\bibinfo  {publisher}
  {World Scientific},\ \bibinfo {year} {2011})\BibitemShut {NoStop}%
\bibitem [{\citenamefont {Fitzpatrick}\ \emph {et~al.}(2017)\citenamefont
  {Fitzpatrick}, \citenamefont {Sundaresan}, \citenamefont {Li}, \citenamefont
  {Koch},\ and\ \citenamefont {Houck}}]{fitzpatrick_observation_2017}%
  \BibitemOpen
  \bibfield  {author} {\bibinfo {author} {\bibfnamefont {M.}~\bibnamefont
  {Fitzpatrick}}, \bibinfo {author} {\bibfnamefont {N.~M.}\ \bibnamefont
  {Sundaresan}}, \bibinfo {author} {\bibfnamefont {A.~C.~Y.}\ \bibnamefont
  {Li}}, \bibinfo {author} {\bibfnamefont {J.}~\bibnamefont {Koch}},\ and\
  \bibinfo {author} {\bibfnamefont {A.~A.}\ \bibnamefont {Houck}},\ }\href
  {https://doi.org/10.1103/PhysRevX.7.011016} {\bibfield  {journal} {\bibinfo
  {journal} {Phys. Rev. X}\ }\textbf {\bibinfo {volume} {7}},\ \bibinfo {pages}
  {011016} (\bibinfo {year} {2017})}\BibitemShut {NoStop}%
\bibitem [{\citenamefont {Chiaro}\ \emph {et~al.}(2020)\citenamefont {Chiaro},
  \citenamefont {Neill}, \citenamefont {Bohrdt}, \citenamefont {Filippone},
  \citenamefont {Arute}, \citenamefont {Arya}, \citenamefont {Babbush},
  \citenamefont {Bacon}, \citenamefont {Bardin}, \citenamefont {Barends},
  \citenamefont {Boixo}, \citenamefont {Buell}, \citenamefont {Burkett},
  \citenamefont {Chen}, \citenamefont {Chen}, \citenamefont {Collins},
  \citenamefont {Dunsworth}, \citenamefont {Farhi}, \citenamefont {Fowler},
  \citenamefont {Foxen}, \citenamefont {Gidney}, \citenamefont {Giustina},
  \citenamefont {Harrigan}, \citenamefont {Huang}, \citenamefont {Isakov},
  \citenamefont {Jeffrey}, \citenamefont {Jiang}, \citenamefont {Kafri},
  \citenamefont {Kechedzhi}, \citenamefont {Kelly}, \citenamefont {Klimov},
  \citenamefont {Korotkov}, \citenamefont {Kostritsa}, \citenamefont
  {Landhuis}, \citenamefont {Lucero}, \citenamefont {McClean}, \citenamefont
  {Mi}, \citenamefont {Megrant}, \citenamefont {Mohseni}, \citenamefont
  {Mutus}, \citenamefont {McEwen}, \citenamefont {Naaman}, \citenamefont
  {Neeley}, \citenamefont {Niu}, \citenamefont {Petukhov}, \citenamefont
  {Quintana}, \citenamefont {Rubin}, \citenamefont {Sank}, \citenamefont
  {Satzinger}, \citenamefont {Vainsencher}, \citenamefont {White},
  \citenamefont {Yao}, \citenamefont {Yeh}, \citenamefont {Zalcman},
  \citenamefont {Smelyanskiy}, \citenamefont {Neven}, \citenamefont
  {Gopalakrishnan}, \citenamefont {Abanin}, \citenamefont {Knap}, \citenamefont
  {Martinis},\ and\ \citenamefont {Roushan}}]{chiaro_direct_2020}%
  \BibitemOpen
  \bibfield  {author} {\bibinfo {author} {\bibfnamefont {B.}~\bibnamefont
  {Chiaro}}, \bibinfo {author} {\bibfnamefont {C.}~\bibnamefont {Neill}},
  \bibinfo {author} {\bibfnamefont {A.}~\bibnamefont {Bohrdt}}, \bibinfo
  {author} {\bibfnamefont {M.}~\bibnamefont {Filippone}}, \bibinfo {author}
  {\bibfnamefont {F.}~\bibnamefont {Arute}}, \bibinfo {author} {\bibfnamefont
  {K.}~\bibnamefont {Arya}}, \bibinfo {author} {\bibfnamefont {R.}~\bibnamefont
  {Babbush}}, \bibinfo {author} {\bibfnamefont {D.}~\bibnamefont {Bacon}},
  \bibinfo {author} {\bibfnamefont {J.}~\bibnamefont {Bardin}}, \bibinfo
  {author} {\bibfnamefont {R.}~\bibnamefont {Barends}}, \bibinfo {author}
  {\bibfnamefont {S.}~\bibnamefont {Boixo}}, \bibinfo {author} {\bibfnamefont
  {D.}~\bibnamefont {Buell}}, \bibinfo {author} {\bibfnamefont
  {B.}~\bibnamefont {Burkett}}, \bibinfo {author} {\bibfnamefont
  {Y.}~\bibnamefont {Chen}}, \bibinfo {author} {\bibfnamefont {Z.}~\bibnamefont
  {Chen}}, \bibinfo {author} {\bibfnamefont {R.}~\bibnamefont {Collins}},
  \bibinfo {author} {\bibfnamefont {A.}~\bibnamefont {Dunsworth}}, \bibinfo
  {author} {\bibfnamefont {E.}~\bibnamefont {Farhi}}, \bibinfo {author}
  {\bibfnamefont {A.}~\bibnamefont {Fowler}}, \bibinfo {author} {\bibfnamefont
  {B.}~\bibnamefont {Foxen}}, \bibinfo {author} {\bibfnamefont
  {C.}~\bibnamefont {Gidney}}, \bibinfo {author} {\bibfnamefont
  {M.}~\bibnamefont {Giustina}}, \bibinfo {author} {\bibfnamefont
  {M.}~\bibnamefont {Harrigan}}, \bibinfo {author} {\bibfnamefont
  {T.}~\bibnamefont {Huang}}, \bibinfo {author} {\bibfnamefont
  {S.}~\bibnamefont {Isakov}}, \bibinfo {author} {\bibfnamefont
  {E.}~\bibnamefont {Jeffrey}}, \bibinfo {author} {\bibfnamefont
  {Z.}~\bibnamefont {Jiang}}, \bibinfo {author} {\bibfnamefont
  {D.}~\bibnamefont {Kafri}}, \bibinfo {author} {\bibfnamefont
  {K.}~\bibnamefont {Kechedzhi}}, \bibinfo {author} {\bibfnamefont
  {J.}~\bibnamefont {Kelly}}, \bibinfo {author} {\bibfnamefont
  {P.}~\bibnamefont {Klimov}}, \bibinfo {author} {\bibfnamefont
  {A.}~\bibnamefont {Korotkov}}, \bibinfo {author} {\bibfnamefont
  {F.}~\bibnamefont {Kostritsa}}, \bibinfo {author} {\bibfnamefont
  {D.}~\bibnamefont {Landhuis}}, \bibinfo {author} {\bibfnamefont
  {E.}~\bibnamefont {Lucero}}, \bibinfo {author} {\bibfnamefont
  {J.}~\bibnamefont {McClean}}, \bibinfo {author} {\bibfnamefont
  {X.}~\bibnamefont {Mi}}, \bibinfo {author} {\bibfnamefont {A.}~\bibnamefont
  {Megrant}}, \bibinfo {author} {\bibfnamefont {M.}~\bibnamefont {Mohseni}},
  \bibinfo {author} {\bibfnamefont {J.}~\bibnamefont {Mutus}}, \bibinfo
  {author} {\bibfnamefont {M.}~\bibnamefont {McEwen}}, \bibinfo {author}
  {\bibfnamefont {O.}~\bibnamefont {Naaman}}, \bibinfo {author} {\bibfnamefont
  {M.}~\bibnamefont {Neeley}}, \bibinfo {author} {\bibfnamefont
  {M.}~\bibnamefont {Niu}}, \bibinfo {author} {\bibfnamefont {A.}~\bibnamefont
  {Petukhov}}, \bibinfo {author} {\bibfnamefont {C.}~\bibnamefont {Quintana}},
  \bibinfo {author} {\bibfnamefont {N.}~\bibnamefont {Rubin}}, \bibinfo
  {author} {\bibfnamefont {D.}~\bibnamefont {Sank}}, \bibinfo {author}
  {\bibfnamefont {K.}~\bibnamefont {Satzinger}}, \bibinfo {author}
  {\bibfnamefont {A.}~\bibnamefont {Vainsencher}}, \bibinfo {author}
  {\bibfnamefont {T.}~\bibnamefont {White}}, \bibinfo {author} {\bibfnamefont
  {Z.}~\bibnamefont {Yao}}, \bibinfo {author} {\bibfnamefont {P.}~\bibnamefont
  {Yeh}}, \bibinfo {author} {\bibfnamefont {A.}~\bibnamefont {Zalcman}},
  \bibinfo {author} {\bibfnamefont {V.}~\bibnamefont {Smelyanskiy}}, \bibinfo
  {author} {\bibfnamefont {H.}~\bibnamefont {Neven}}, \bibinfo {author}
  {\bibfnamefont {S.}~\bibnamefont {Gopalakrishnan}}, \bibinfo {author}
  {\bibfnamefont {D.}~\bibnamefont {Abanin}}, \bibinfo {author} {\bibfnamefont
  {M.}~\bibnamefont {Knap}}, \bibinfo {author} {\bibfnamefont {J.}~\bibnamefont
  {Martinis}},\ and\ \bibinfo {author} {\bibfnamefont {P.}~\bibnamefont
  {Roushan}},\ }\href {http://arxiv.org/abs/1910.06024} {\bibfield  {journal}
  {\bibinfo  {journal} {arXiv:1910.06024 [cond-mat, physics:quant-ph]}\ }
  (\bibinfo {year} {2020})},\ \bibinfo {note} {arXiv: 1910.06024}\BibitemShut
  {NoStop}%
\bibitem [{\citenamefont {Ma}\ \emph {et~al.}(2019)\citenamefont {Ma},
  \citenamefont {Saxberg}, \citenamefont {Owens}, \citenamefont {Leung},
  \citenamefont {Lu}, \citenamefont {Simon},\ and\ \citenamefont
  {Schuster}}]{ma_dissipatively_2019}%
  \BibitemOpen
  \bibfield  {author} {\bibinfo {author} {\bibfnamefont {R.}~\bibnamefont
  {Ma}}, \bibinfo {author} {\bibfnamefont {B.}~\bibnamefont {Saxberg}},
  \bibinfo {author} {\bibfnamefont {C.}~\bibnamefont {Owens}}, \bibinfo
  {author} {\bibfnamefont {N.}~\bibnamefont {Leung}}, \bibinfo {author}
  {\bibfnamefont {Y.}~\bibnamefont {Lu}}, \bibinfo {author} {\bibfnamefont
  {J.}~\bibnamefont {Simon}},\ and\ \bibinfo {author} {\bibfnamefont {D.~I.}\
  \bibnamefont {Schuster}},\ }\href {https://doi.org/10.1038/s41586-019-0897-9}
  {\bibfield  {journal} {\bibinfo  {journal} {Nature}\ }\textbf {\bibinfo
  {volume} {566}},\ \bibinfo {pages} {51} (\bibinfo {year} {2019})},\ \bibinfo
  {note} {number: 7742 Publisher: Nature Publishing Group}\BibitemShut
  {NoStop}%
\bibitem [{\citenamefont {Dutta}\ and\ \citenamefont
  {Cooper}(2020)}]{dutta_out--equilibrium_2020}%
  \BibitemOpen
  \bibfield  {author} {\bibinfo {author} {\bibfnamefont {S.}~\bibnamefont
  {Dutta}}\ and\ \bibinfo {author} {\bibfnamefont {N.~R.}\ \bibnamefont
  {Cooper}},\ }\href {http://arxiv.org/abs/2007.08938} {\bibfield  {journal}
  {\bibinfo  {journal} {arXiv:2007.08938 [cond-mat]}\ } (\bibinfo {year}
  {2020})},\ \bibinfo {note} {arXiv: 2007.08938}\BibitemShut {NoStop}%
\bibitem [{\citenamefont {Zajac}\ \emph {et~al.}(2016)\citenamefont {Zajac},
  \citenamefont {Hazard}, \citenamefont {Mi}, \citenamefont {Nielsen},\ and\
  \citenamefont {Petta}}]{zajac_scalable_2016}%
  \BibitemOpen
  \bibfield  {author} {\bibinfo {author} {\bibfnamefont {D.~M.}\ \bibnamefont
  {Zajac}}, \bibinfo {author} {\bibfnamefont {T.~M.}\ \bibnamefont {Hazard}},
  \bibinfo {author} {\bibfnamefont {X.}~\bibnamefont {Mi}}, \bibinfo {author}
  {\bibfnamefont {E.}~\bibnamefont {Nielsen}},\ and\ \bibinfo {author}
  {\bibfnamefont {J.~R.}\ \bibnamefont {Petta}},\ }\href
  {https://doi.org/10.1103/PhysRevApplied.6.054013} {\bibfield  {journal}
  {\bibinfo  {journal} {Physical Review Applied}\ }\textbf {\bibinfo {volume}
  {6}},\ \bibinfo {pages} {054013} (\bibinfo {year} {2016})},\ \bibinfo {note}
  {publisher: American Physical Society}\BibitemShut {NoStop}%
\bibitem [{\citenamefont {Hensgens}\ \emph {et~al.}(2017)\citenamefont
  {Hensgens}, \citenamefont {Fujita}, \citenamefont {Janssen}, \citenamefont
  {Li}, \citenamefont {Van~Diepen}, \citenamefont {Reichl}, \citenamefont
  {Wegscheider}, \citenamefont {Das~Sarma},\ and\ \citenamefont
  {Vandersypen}}]{hensgens_quantum_2017}%
  \BibitemOpen
  \bibfield  {author} {\bibinfo {author} {\bibfnamefont {T.}~\bibnamefont
  {Hensgens}}, \bibinfo {author} {\bibfnamefont {T.}~\bibnamefont {Fujita}},
  \bibinfo {author} {\bibfnamefont {L.}~\bibnamefont {Janssen}}, \bibinfo
  {author} {\bibfnamefont {X.}~\bibnamefont {Li}}, \bibinfo {author}
  {\bibfnamefont {C.~J.}\ \bibnamefont {Van~Diepen}}, \bibinfo {author}
  {\bibfnamefont {C.}~\bibnamefont {Reichl}}, \bibinfo {author} {\bibfnamefont
  {W.}~\bibnamefont {Wegscheider}}, \bibinfo {author} {\bibfnamefont
  {S.}~\bibnamefont {Das~Sarma}},\ and\ \bibinfo {author} {\bibfnamefont
  {L.~M.~K.}\ \bibnamefont {Vandersypen}},\ }\href
  {https://doi.org/10.1038/nature23022} {\bibfield  {journal} {\bibinfo
  {journal} {Nature}\ }\textbf {\bibinfo {volume} {548}},\ \bibinfo {pages}
  {70} (\bibinfo {year} {2017})},\ \bibinfo {note} {number: 7665 Publisher:
  Nature Publishing Group}\BibitemShut {NoStop}%
\bibitem [{\citenamefont {Mills}\ \emph {et~al.}(2019)\citenamefont {Mills},
  \citenamefont {Zajac}, \citenamefont {Gullans}, \citenamefont {Schupp},
  \citenamefont {Hazard},\ and\ \citenamefont {Petta}}]{mills_shuttling_2019}%
  \BibitemOpen
  \bibfield  {author} {\bibinfo {author} {\bibfnamefont {A.~R.}\ \bibnamefont
  {Mills}}, \bibinfo {author} {\bibfnamefont {D.~M.}\ \bibnamefont {Zajac}},
  \bibinfo {author} {\bibfnamefont {M.~J.}\ \bibnamefont {Gullans}}, \bibinfo
  {author} {\bibfnamefont {F.~J.}\ \bibnamefont {Schupp}}, \bibinfo {author}
  {\bibfnamefont {T.~M.}\ \bibnamefont {Hazard}},\ and\ \bibinfo {author}
  {\bibfnamefont {J.~R.}\ \bibnamefont {Petta}},\ }\href
  {https://doi.org/10.1038/s41467-019-08970-z} {\bibfield  {journal} {\bibinfo
  {journal} {Nature Communications}\ }\textbf {\bibinfo {volume} {10}},\
  \bibinfo {pages} {1063} (\bibinfo {year} {2019})},\ \bibinfo {note} {number:
  1 Publisher: Nature Publishing Group}\BibitemShut {NoStop}%
\bibitem [{\citenamefont {Amico}\ \emph {et~al.}(2020)\citenamefont {Amico},
  \citenamefont {Boshier}, \citenamefont {Birkl}, \citenamefont {Minguzzi},
  \citenamefont {Miniatura}, \citenamefont {Kwek}, \citenamefont {Aghamalyan},
  \citenamefont {Ahufinger}, \citenamefont {Andrei}, \citenamefont {Arnold},
  \citenamefont {Baker}, \citenamefont {Bell}, \citenamefont {Bland},
  \citenamefont {Brantut}, \citenamefont {Cassettari}, \citenamefont {Chevy},
  \citenamefont {Citro}, \citenamefont {De~Palo}, \citenamefont {Dumke},
  \citenamefont {Edwards}, \citenamefont {Folman}, \citenamefont {Fortagh},
  \citenamefont {Gardiner}, \citenamefont {Garraway}, \citenamefont {Gauthier},
  \citenamefont {G{\"u}nther}, \citenamefont {Haug}, \citenamefont {Hufnagel},
  \citenamefont {Keil}, \citenamefont {von Klitzing}, \citenamefont {Ireland},
  \citenamefont {Lebrat}, \citenamefont {Li}, \citenamefont {Longchambon},
  \citenamefont {Mompart}, \citenamefont {Morsch}, \citenamefont {Naldesi},
  \citenamefont {Neely}, \citenamefont {Olshanii}, \citenamefont {Orignac},
  \citenamefont {Pandey}, \citenamefont {P{\'e}rez-Obiol}, \citenamefont
  {Perrin}, \citenamefont {Piroli}, \citenamefont {Polo}, \citenamefont
  {Pritchard}, \citenamefont {Proukakis}, \citenamefont {Rylands},
  \citenamefont {Rubinsztein-Dunlop}, \citenamefont {Scazza}, \citenamefont
  {Stringari}, \citenamefont {Tosto}, \citenamefont {Trombettoni},
  \citenamefont {Victorin}, \citenamefont {Xhani},\ and\ \citenamefont
  {Yakimenko}}]{amico_roadmap_2020}%
  \BibitemOpen
  \bibfield  {author} {\bibinfo {author} {\bibfnamefont {L.}~\bibnamefont
  {Amico}}, \bibinfo {author} {\bibfnamefont {M.}~\bibnamefont {Boshier}},
  \bibinfo {author} {\bibfnamefont {G.}~\bibnamefont {Birkl}}, \bibinfo
  {author} {\bibfnamefont {A.}~\bibnamefont {Minguzzi}}, \bibinfo {author}
  {\bibfnamefont {C.}~\bibnamefont {Miniatura}}, \bibinfo {author}
  {\bibfnamefont {L.-C.}\ \bibnamefont {Kwek}}, \bibinfo {author}
  {\bibfnamefont {D.}~\bibnamefont {Aghamalyan}}, \bibinfo {author}
  {\bibfnamefont {V.}~\bibnamefont {Ahufinger}}, \bibinfo {author}
  {\bibfnamefont {N.}~\bibnamefont {Andrei}}, \bibinfo {author} {\bibfnamefont
  {A.~S.}\ \bibnamefont {Arnold}}, \bibinfo {author} {\bibfnamefont
  {M.}~\bibnamefont {Baker}}, \bibinfo {author} {\bibfnamefont {T.~A.}\
  \bibnamefont {Bell}}, \bibinfo {author} {\bibfnamefont {T.}~\bibnamefont
  {Bland}}, \bibinfo {author} {\bibfnamefont {J.~P.}\ \bibnamefont {Brantut}},
  \bibinfo {author} {\bibfnamefont {D.}~\bibnamefont {Cassettari}}, \bibinfo
  {author} {\bibfnamefont {F.}~\bibnamefont {Chevy}}, \bibinfo {author}
  {\bibfnamefont {R.}~\bibnamefont {Citro}}, \bibinfo {author} {\bibfnamefont
  {S.}~\bibnamefont {De~Palo}}, \bibinfo {author} {\bibfnamefont
  {R.}~\bibnamefont {Dumke}}, \bibinfo {author} {\bibfnamefont
  {M.}~\bibnamefont {Edwards}}, \bibinfo {author} {\bibfnamefont
  {R.}~\bibnamefont {Folman}}, \bibinfo {author} {\bibfnamefont
  {J.}~\bibnamefont {Fortagh}}, \bibinfo {author} {\bibfnamefont {S.~A.}\
  \bibnamefont {Gardiner}}, \bibinfo {author} {\bibfnamefont {B.~M.}\
  \bibnamefont {Garraway}}, \bibinfo {author} {\bibfnamefont {G.}~\bibnamefont
  {Gauthier}}, \bibinfo {author} {\bibfnamefont {A.}~\bibnamefont
  {G{\"u}nther}}, \bibinfo {author} {\bibfnamefont {T.}~\bibnamefont {Haug}},
  \bibinfo {author} {\bibfnamefont {C.}~\bibnamefont {Hufnagel}}, \bibinfo
  {author} {\bibfnamefont {M.}~\bibnamefont {Keil}}, \bibinfo {author}
  {\bibfnamefont {W.}~\bibnamefont {von Klitzing}}, \bibinfo {author}
  {\bibfnamefont {P.}~\bibnamefont {Ireland}}, \bibinfo {author} {\bibfnamefont
  {M.}~\bibnamefont {Lebrat}}, \bibinfo {author} {\bibfnamefont
  {W.}~\bibnamefont {Li}}, \bibinfo {author} {\bibfnamefont {L.}~\bibnamefont
  {Longchambon}}, \bibinfo {author} {\bibfnamefont {J.}~\bibnamefont
  {Mompart}}, \bibinfo {author} {\bibfnamefont {O.}~\bibnamefont {Morsch}},
  \bibinfo {author} {\bibfnamefont {P.}~\bibnamefont {Naldesi}}, \bibinfo
  {author} {\bibfnamefont {T.~W.}\ \bibnamefont {Neely}}, \bibinfo {author}
  {\bibfnamefont {M.}~\bibnamefont {Olshanii}}, \bibinfo {author}
  {\bibfnamefont {E.}~\bibnamefont {Orignac}}, \bibinfo {author} {\bibfnamefont
  {S.}~\bibnamefont {Pandey}}, \bibinfo {author} {\bibfnamefont
  {A.}~\bibnamefont {P{\'e}rez-Obiol}}, \bibinfo {author} {\bibfnamefont
  {H.}~\bibnamefont {Perrin}}, \bibinfo {author} {\bibfnamefont
  {L.}~\bibnamefont {Piroli}}, \bibinfo {author} {\bibfnamefont
  {J.}~\bibnamefont {Polo}}, \bibinfo {author} {\bibfnamefont {A.~L.}\
  \bibnamefont {Pritchard}}, \bibinfo {author} {\bibfnamefont {N.~P.}\
  \bibnamefont {Proukakis}}, \bibinfo {author} {\bibfnamefont {C.}~\bibnamefont
  {Rylands}}, \bibinfo {author} {\bibfnamefont {H.}~\bibnamefont
  {Rubinsztein-Dunlop}}, \bibinfo {author} {\bibfnamefont {F.}~\bibnamefont
  {Scazza}}, \bibinfo {author} {\bibfnamefont {S.}~\bibnamefont {Stringari}},
  \bibinfo {author} {\bibfnamefont {F.}~\bibnamefont {Tosto}}, \bibinfo
  {author} {\bibfnamefont {A.}~\bibnamefont {Trombettoni}}, \bibinfo {author}
  {\bibfnamefont {N.}~\bibnamefont {Victorin}}, \bibinfo {author}
  {\bibfnamefont {K.}~\bibnamefont {Xhani}},\ and\ \bibinfo {author}
  {\bibfnamefont {A.}~\bibnamefont {Yakimenko}},\ }\href
  {http://arxiv.org/abs/2008.04439} {\bibfield  {journal} {\bibinfo  {journal}
  {arXiv:2008.04439 [cond-mat, physics:quant-ph]}\ } (\bibinfo {year}
  {2020})},\ \bibinfo {note} {arXiv: 2008.04439}\BibitemShut {NoStop}%
\bibitem [{\citenamefont {Amico}\ \emph {et~al.}(2005)\citenamefont {Amico},
  \citenamefont {Osterloh},\ and\ \citenamefont
  {Cataliotti}}]{amico_quantum_2005}%
  \BibitemOpen
  \bibfield  {author} {\bibinfo {author} {\bibfnamefont {L.}~\bibnamefont
  {Amico}}, \bibinfo {author} {\bibfnamefont {A.}~\bibnamefont {Osterloh}},\
  and\ \bibinfo {author} {\bibfnamefont {F.}~\bibnamefont {Cataliotti}},\
  }\href {https://doi.org/10.1103/PhysRevLett.95.063201} {\bibfield  {journal}
  {\bibinfo  {journal} {Physical Review Letters}\ }\textbf {\bibinfo {volume}
  {95}},\ \bibinfo {pages} {063201} (\bibinfo {year} {2005})},\ \bibinfo {note}
  {publisher: American Physical Society}\BibitemShut {NoStop}%
\bibitem [{\citenamefont {Seaman}\ \emph {et~al.}(2007)\citenamefont {Seaman},
  \citenamefont {Kr{\"a}mer}, \citenamefont {Anderson},\ and\ \citenamefont
  {Holland}}]{seaman_atomtronics_2007}%
  \BibitemOpen
  \bibfield  {author} {\bibinfo {author} {\bibfnamefont {B.~T.}\ \bibnamefont
  {Seaman}}, \bibinfo {author} {\bibfnamefont {M.}~\bibnamefont {Kr{\"a}mer}},
  \bibinfo {author} {\bibfnamefont {D.~Z.}\ \bibnamefont {Anderson}},\ and\
  \bibinfo {author} {\bibfnamefont {M.~J.}\ \bibnamefont {Holland}},\ }\href
  {https://doi.org/10.1103/PhysRevA.75.023615} {\bibfield  {journal} {\bibinfo
  {journal} {Physical Review A}\ }\textbf {\bibinfo {volume} {75}},\ \bibinfo
  {pages} {023615} (\bibinfo {year} {2007})},\ \bibinfo {note} {publisher:
  American Physical Society}\BibitemShut {NoStop}%
\bibitem [{\citenamefont {Stadler}\ \emph {et~al.}(2012)\citenamefont
  {Stadler}, \citenamefont {Krinner}, \citenamefont {Meineke}, \citenamefont
  {Brantut},\ and\ \citenamefont {Esslinger}}]{stadler_observing_2012}%
  \BibitemOpen
  \bibfield  {author} {\bibinfo {author} {\bibfnamefont {D.}~\bibnamefont
  {Stadler}}, \bibinfo {author} {\bibfnamefont {S.}~\bibnamefont {Krinner}},
  \bibinfo {author} {\bibfnamefont {J.}~\bibnamefont {Meineke}}, \bibinfo
  {author} {\bibfnamefont {J.-P.}\ \bibnamefont {Brantut}},\ and\ \bibinfo
  {author} {\bibfnamefont {T.}~\bibnamefont {Esslinger}},\ }\href
  {https://doi.org/10.1038/nature11613} {\bibfield  {journal} {\bibinfo
  {journal} {Nature}\ }\textbf {\bibinfo {volume} {491}},\ \bibinfo {pages}
  {736} (\bibinfo {year} {2012})}\BibitemShut {NoStop}%
\bibitem [{\citenamefont {Brantut}\ \emph {et~al.}(2012)\citenamefont
  {Brantut}, \citenamefont {Meineke}, \citenamefont {Stadler}, \citenamefont
  {Krinner},\ and\ \citenamefont {Esslinger}}]{brantut_conduction_2012}%
  \BibitemOpen
  \bibfield  {author} {\bibinfo {author} {\bibfnamefont {J.-P.}\ \bibnamefont
  {Brantut}}, \bibinfo {author} {\bibfnamefont {J.}~\bibnamefont {Meineke}},
  \bibinfo {author} {\bibfnamefont {D.}~\bibnamefont {Stadler}}, \bibinfo
  {author} {\bibfnamefont {S.}~\bibnamefont {Krinner}},\ and\ \bibinfo {author}
  {\bibfnamefont {T.}~\bibnamefont {Esslinger}},\ }\href
  {https://doi.org/10.1126/science.1223175} {\bibfield  {journal} {\bibinfo
  {journal} {Science}\ }\textbf {\bibinfo {volume} {337}},\ \bibinfo {pages}
  {1069} (\bibinfo {year} {2012})}\BibitemShut {NoStop}%
\bibitem [{\citenamefont {Brantut}\ \emph {et~al.}(2013)\citenamefont
  {Brantut}, \citenamefont {Grenier}, \citenamefont {Meineke}, \citenamefont
  {Stadler}, \citenamefont {Krinner}, \citenamefont {Kollath}, \citenamefont
  {Esslinger},\ and\ \citenamefont {Georges}}]{brantut_thermoelectric_2013}%
  \BibitemOpen
  \bibfield  {author} {\bibinfo {author} {\bibfnamefont {J.-P.}\ \bibnamefont
  {Brantut}}, \bibinfo {author} {\bibfnamefont {C.}~\bibnamefont {Grenier}},
  \bibinfo {author} {\bibfnamefont {J.}~\bibnamefont {Meineke}}, \bibinfo
  {author} {\bibfnamefont {D.}~\bibnamefont {Stadler}}, \bibinfo {author}
  {\bibfnamefont {S.}~\bibnamefont {Krinner}}, \bibinfo {author} {\bibfnamefont
  {C.}~\bibnamefont {Kollath}}, \bibinfo {author} {\bibfnamefont
  {T.}~\bibnamefont {Esslinger}},\ and\ \bibinfo {author} {\bibfnamefont
  {A.}~\bibnamefont {Georges}},\ }\href
  {https://doi.org/10.1126/science.1242308} {\bibfield  {journal} {\bibinfo
  {journal} {Science}\ }\textbf {\bibinfo {volume} {342}},\ \bibinfo {pages}
  {713} (\bibinfo {year} {2013})}\BibitemShut {NoStop}%
\bibitem [{\citenamefont {Krinner}\ \emph {et~al.}(2015)\citenamefont
  {Krinner}, \citenamefont {Stadler}, \citenamefont {Husmann}, \citenamefont
  {Brantut},\ and\ \citenamefont {Esslinger}}]{krinner_observation_2015}%
  \BibitemOpen
  \bibfield  {author} {\bibinfo {author} {\bibfnamefont {S.}~\bibnamefont
  {Krinner}}, \bibinfo {author} {\bibfnamefont {D.}~\bibnamefont {Stadler}},
  \bibinfo {author} {\bibfnamefont {D.}~\bibnamefont {Husmann}}, \bibinfo
  {author} {\bibfnamefont {J.-P.}\ \bibnamefont {Brantut}},\ and\ \bibinfo
  {author} {\bibfnamefont {T.}~\bibnamefont {Esslinger}},\ }\href
  {https://doi.org/10.1038/nature14049} {\bibfield  {journal} {\bibinfo
  {journal} {Nature}\ }\textbf {\bibinfo {volume} {517}},\ \bibinfo {pages}
  {64} (\bibinfo {year} {2015})}\BibitemShut {NoStop}%
\bibitem [{\citenamefont {Lebrat}\ \emph {et~al.}(2018)\citenamefont {Lebrat},
  \citenamefont {Gri{\v s}ins}, \citenamefont {Husmann}, \citenamefont
  {H{\"a}usler}, \citenamefont {Corman}, \citenamefont {Giamarchi},
  \citenamefont {Brantut},\ and\ \citenamefont {Esslinger}}]{lebrat_band_2018}%
  \BibitemOpen
  \bibfield  {author} {\bibinfo {author} {\bibfnamefont {M.}~\bibnamefont
  {Lebrat}}, \bibinfo {author} {\bibfnamefont {P.}~\bibnamefont {Gri{\v
  s}ins}}, \bibinfo {author} {\bibfnamefont {D.}~\bibnamefont {Husmann}},
  \bibinfo {author} {\bibfnamefont {S.}~\bibnamefont {H{\"a}usler}}, \bibinfo
  {author} {\bibfnamefont {L.}~\bibnamefont {Corman}}, \bibinfo {author}
  {\bibfnamefont {T.}~\bibnamefont {Giamarchi}}, \bibinfo {author}
  {\bibfnamefont {J.-P.}\ \bibnamefont {Brantut}},\ and\ \bibinfo {author}
  {\bibfnamefont {T.}~\bibnamefont {Esslinger}},\ }\href
  {https://doi.org/10.1103/PhysRevX.8.011053} {\bibfield  {journal} {\bibinfo
  {journal} {Physical Review X}\ }\textbf {\bibinfo {volume} {8}},\ \bibinfo
  {pages} {011053} (\bibinfo {year} {2018})}\BibitemShut {NoStop}%
\bibitem [{\citenamefont {Jepsen}\ \emph {et~al.}(2020)\citenamefont {Jepsen},
  \citenamefont {Amato-Grill}, \citenamefont {Dimitrova}, \citenamefont {Ho},
  \citenamefont {Demler},\ and\ \citenamefont {Ketterle}}]{jepsen_spin_2020}%
  \BibitemOpen
  \bibfield  {author} {\bibinfo {author} {\bibfnamefont {N.}~\bibnamefont
  {Jepsen}}, \bibinfo {author} {\bibfnamefont {J.}~\bibnamefont {Amato-Grill}},
  \bibinfo {author} {\bibfnamefont {I.}~\bibnamefont {Dimitrova}}, \bibinfo
  {author} {\bibfnamefont {W.~W.}\ \bibnamefont {Ho}}, \bibinfo {author}
  {\bibfnamefont {E.}~\bibnamefont {Demler}},\ and\ \bibinfo {author}
  {\bibfnamefont {W.}~\bibnamefont {Ketterle}},\ }\href
  {http://arxiv.org/abs/2005.09549} {\bibfield  {journal} {\bibinfo  {journal}
  {arXiv:2005.09549 [cond-mat, physics:physics, physics:quant-ph]}\ } (\bibinfo
  {year} {2020})},\ \bibinfo {note} {arXiv: 2005.09549}\BibitemShut {NoStop}%
\bibitem [{\citenamefont {Husmann}\ \emph {et~al.}(2015)\citenamefont
  {Husmann}, \citenamefont {Uchino}, \citenamefont {Krinner}, \citenamefont
  {Lebrat}, \citenamefont {Giamarchi}, \citenamefont {Esslinger},\ and\
  \citenamefont {Brantut}}]{ShunEsslingerGiamarchi_superfluidpointcontact}%
  \BibitemOpen
  \bibfield  {author} {\bibinfo {author} {\bibfnamefont {D.}~\bibnamefont
  {Husmann}}, \bibinfo {author} {\bibfnamefont {S.}~\bibnamefont {Uchino}},
  \bibinfo {author} {\bibfnamefont {S.}~\bibnamefont {Krinner}}, \bibinfo
  {author} {\bibfnamefont {M.}~\bibnamefont {Lebrat}}, \bibinfo {author}
  {\bibfnamefont {T.}~\bibnamefont {Giamarchi}}, \bibinfo {author}
  {\bibfnamefont {T.}~\bibnamefont {Esslinger}},\ and\ \bibinfo {author}
  {\bibfnamefont {J.-P.}\ \bibnamefont {Brantut}},\ }\href
  {https://doi.org/10.1126/science.aac9584} {\bibfield  {journal} {\bibinfo
  {journal} {Science}\ }\textbf {\bibinfo {volume} {350}},\ \bibinfo {pages}
  {1498} (\bibinfo {year} {2015})}\BibitemShut {NoStop}%
\bibitem [{\citenamefont {Eckel}\ \emph {et~al.}(2014)\citenamefont {Eckel},
  \citenamefont {Jendrzejewski}, \citenamefont {Kumar}, \citenamefont {Lobb},\
  and\ \citenamefont {Campbell}}]{eckel_interferometric_2014}%
  \BibitemOpen
  \bibfield  {author} {\bibinfo {author} {\bibfnamefont {S.}~\bibnamefont
  {Eckel}}, \bibinfo {author} {\bibfnamefont {F.}~\bibnamefont
  {Jendrzejewski}}, \bibinfo {author} {\bibfnamefont {A.}~\bibnamefont
  {Kumar}}, \bibinfo {author} {\bibfnamefont {C.~J.}\ \bibnamefont {Lobb}},\
  and\ \bibinfo {author} {\bibfnamefont {G.~K.}\ \bibnamefont {Campbell}},\
  }\href {https://doi.org/10.1103/PhysRevX.4.031052} {\bibfield  {journal}
  {\bibinfo  {journal} {Physical Review X}\ }\textbf {\bibinfo {volume} {4}},\
  \bibinfo {pages} {031052} (\bibinfo {year} {2014})}\BibitemShut {NoStop}%
\bibitem [{\citenamefont {Eckel}\ \emph {et~al.}(2016)\citenamefont {Eckel},
  \citenamefont {Lee}, \citenamefont {Jendrzejewski}, \citenamefont {Lobb},
  \citenamefont {Campbell},\ and\ \citenamefont {Hill}}]{eckel_contact_2016}%
  \BibitemOpen
  \bibfield  {author} {\bibinfo {author} {\bibfnamefont {S.}~\bibnamefont
  {Eckel}}, \bibinfo {author} {\bibfnamefont {J.~G.}\ \bibnamefont {Lee}},
  \bibinfo {author} {\bibfnamefont {F.}~\bibnamefont {Jendrzejewski}}, \bibinfo
  {author} {\bibfnamefont {C.~J.}\ \bibnamefont {Lobb}}, \bibinfo {author}
  {\bibfnamefont {G.~K.}\ \bibnamefont {Campbell}},\ and\ \bibinfo {author}
  {\bibfnamefont {W.~T.}\ \bibnamefont {Hill}},\ }\href
  {https://doi.org/10.1103/PhysRevA.93.063619} {\bibfield  {journal} {\bibinfo
  {journal} {Physical Review A}\ }\textbf {\bibinfo {volume} {93}},\ \bibinfo
  {pages} {063619} (\bibinfo {year} {2016})},\ \bibinfo {note} {publisher:
  American Physical Society}\BibitemShut {NoStop}%
\bibitem [{\citenamefont {Cominotti}\ \emph {et~al.}(2014)\citenamefont
  {Cominotti}, \citenamefont {Rossini}, \citenamefont {Rizzi}, \citenamefont
  {Hekking},\ and\ \citenamefont {Minguzzi}}]{cominotti_optimal_2014}%
  \BibitemOpen
  \bibfield  {author} {\bibinfo {author} {\bibfnamefont {M.}~\bibnamefont
  {Cominotti}}, \bibinfo {author} {\bibfnamefont {D.}~\bibnamefont {Rossini}},
  \bibinfo {author} {\bibfnamefont {M.}~\bibnamefont {Rizzi}}, \bibinfo
  {author} {\bibfnamefont {F.}~\bibnamefont {Hekking}},\ and\ \bibinfo {author}
  {\bibfnamefont {A.}~\bibnamefont {Minguzzi}},\ }\href
  {https://doi.org/10.1103/PhysRevLett.113.025301} {\bibfield  {journal}
  {\bibinfo  {journal} {Physical Review Letters}\ }\textbf {\bibinfo {volume}
  {113}},\ \bibinfo {pages} {025301} (\bibinfo {year} {2014})}\BibitemShut
  {NoStop}%
\bibitem [{\citenamefont {Gutman}\ \emph {et~al.}(2012)\citenamefont {Gutman},
  \citenamefont {Gefen},\ and\ \citenamefont {Mirlin}}]{gutman_cold_2012}%
  \BibitemOpen
  \bibfield  {author} {\bibinfo {author} {\bibfnamefont {D.~B.}\ \bibnamefont
  {Gutman}}, \bibinfo {author} {\bibfnamefont {Y.}~\bibnamefont {Gefen}},\ and\
  \bibinfo {author} {\bibfnamefont {A.~D.}\ \bibnamefont {Mirlin}},\ }\href
  {https://doi.org/10.1103/PhysRevB.85.125102} {\bibfield  {journal} {\bibinfo
  {journal} {Physical Review B}\ }\textbf {\bibinfo {volume} {85}},\ \bibinfo
  {pages} {125102} (\bibinfo {year} {2012})}\BibitemShut {NoStop}%
\bibitem [{\citenamefont {Filippone}\ \emph {et~al.}(2016)\citenamefont
  {Filippone}, \citenamefont {Hekking},\ and\ \citenamefont
  {Minguzzi}}]{Filippone2016b}%
  \BibitemOpen
  \bibfield  {author} {\bibinfo {author} {\bibfnamefont {M.}~\bibnamefont
  {Filippone}}, \bibinfo {author} {\bibfnamefont {F.}~\bibnamefont {Hekking}},\
  and\ \bibinfo {author} {\bibfnamefont {A.}~\bibnamefont {Minguzzi}},\ }\href
  {https://doi.org/10.1103/PhysRevA.93.011602} {\bibfield  {journal} {\bibinfo
  {journal} {Physical Review A}\ }\textbf {\bibinfo {volume} {93}},\ \bibinfo
  {pages} {011602} (\bibinfo {year} {2016})}\BibitemShut {NoStop}%
\bibitem [{\citenamefont {Papoular}\ \emph {et~al.}(2012)\citenamefont
  {Papoular}, \citenamefont {Ferrari}, \citenamefont {Pitaevskii},\ and\
  \citenamefont {Stringari}}]{papoular_increasing_2012}%
  \BibitemOpen
  \bibfield  {author} {\bibinfo {author} {\bibfnamefont {D.~J.}\ \bibnamefont
  {Papoular}}, \bibinfo {author} {\bibfnamefont {G.}~\bibnamefont {Ferrari}},
  \bibinfo {author} {\bibfnamefont {L.~P.}\ \bibnamefont {Pitaevskii}},\ and\
  \bibinfo {author} {\bibfnamefont {S.}~\bibnamefont {Stringari}},\ }\href
  {https://doi.org/10.1103/PhysRevLett.109.084501} {\bibfield  {journal}
  {\bibinfo  {journal} {Physical Review Letters}\ }\textbf {\bibinfo {volume}
  {109}},\ \bibinfo {pages} {084501} (\bibinfo {year} {2012})}\BibitemShut
  {NoStop}%
\bibitem [{\citenamefont {Simpson}\ \emph {et~al.}(2014)\citenamefont
  {Simpson}, \citenamefont {Gangardt}, \citenamefont {Lerner},\ and\
  \citenamefont {Kr{\"u}ger}}]{simpson_one-dimensional_2014}%
  \BibitemOpen
  \bibfield  {author} {\bibinfo {author} {\bibfnamefont {D.~P.}\ \bibnamefont
  {Simpson}}, \bibinfo {author} {\bibfnamefont {D.~M.}\ \bibnamefont
  {Gangardt}}, \bibinfo {author} {\bibfnamefont {I.~V.}\ \bibnamefont
  {Lerner}},\ and\ \bibinfo {author} {\bibfnamefont {P.}~\bibnamefont
  {Kr{\"u}ger}},\ }\href {https://doi.org/10.1103/PhysRevLett.112.100601}
  {\bibfield  {journal} {\bibinfo  {journal} {Physical Review Letters}\
  }\textbf {\bibinfo {volume} {112}},\ \bibinfo {pages} {100601} (\bibinfo
  {year} {2014})}\BibitemShut {NoStop}%
\bibitem [{\citenamefont {Salerno}\ \emph {et~al.}(2019)\citenamefont
  {Salerno}, \citenamefont {Price}, \citenamefont {Lebrat}, \citenamefont
  {H{\"a}usler}, \citenamefont {Esslinger}, \citenamefont {Corman},
  \citenamefont {Brantut},\ and\ \citenamefont {Goldman}}]{Salerno2019}%
  \BibitemOpen
  \bibfield  {author} {\bibinfo {author} {\bibfnamefont {G.}~\bibnamefont
  {Salerno}}, \bibinfo {author} {\bibfnamefont {H.~M.}\ \bibnamefont {Price}},
  \bibinfo {author} {\bibfnamefont {M.}~\bibnamefont {Lebrat}}, \bibinfo
  {author} {\bibfnamefont {S.}~\bibnamefont {H{\"a}usler}}, \bibinfo {author}
  {\bibfnamefont {T.}~\bibnamefont {Esslinger}}, \bibinfo {author}
  {\bibfnamefont {L.}~\bibnamefont {Corman}}, \bibinfo {author} {\bibfnamefont
  {J.-P.}\ \bibnamefont {Brantut}},\ and\ \bibinfo {author} {\bibfnamefont
  {N.}~\bibnamefont {Goldman}},\ }\href
  {https://doi.org/10.1103/PhysRevX.9.041001} {\bibfield  {journal} {\bibinfo
  {journal} {Physical Review X}\ }\textbf {\bibinfo {volume} {9}},\ \bibinfo
  {pages} {041001} (\bibinfo {year} {2019})}\BibitemShut {NoStop}%
\bibitem [{\citenamefont {Filippone}\ \emph {et~al.}(2019)\citenamefont
  {Filippone}, \citenamefont {Bardyn}, \citenamefont {Greschner},\ and\
  \citenamefont {Giamarchi}}]{Filippone2019}%
  \BibitemOpen
  \bibfield  {author} {\bibinfo {author} {\bibfnamefont {M.}~\bibnamefont
  {Filippone}}, \bibinfo {author} {\bibfnamefont {C.-E.}\ \bibnamefont
  {Bardyn}}, \bibinfo {author} {\bibfnamefont {S.}~\bibnamefont {Greschner}},\
  and\ \bibinfo {author} {\bibfnamefont {T.}~\bibnamefont {Giamarchi}},\ }\href
  {https://doi.org/10.1103/PhysRevLett.123.086803} {\bibfield  {journal}
  {\bibinfo  {journal} {Physical Review Letters}\ }\textbf {\bibinfo {volume}
  {123}},\ \bibinfo {pages} {086803} (\bibinfo {year} {2019})}\BibitemShut
  {NoStop}%
\bibitem [{\citenamefont {Greschner}\ \emph {et~al.}(2019)\citenamefont
  {Greschner}, \citenamefont {Filippone},\ and\ \citenamefont
  {Giamarchi}}]{Greschner2019}%
  \BibitemOpen
  \bibfield  {author} {\bibinfo {author} {\bibfnamefont {S.}~\bibnamefont
  {Greschner}}, \bibinfo {author} {\bibfnamefont {M.}~\bibnamefont
  {Filippone}},\ and\ \bibinfo {author} {\bibfnamefont {T.}~\bibnamefont
  {Giamarchi}},\ }\href {https://doi.org/10.1103/PhysRevLett.122.083402}
  {\bibfield  {journal} {\bibinfo  {journal} {Physical Review Letters}\
  }\textbf {\bibinfo {volume} {122}},\ \bibinfo {pages} {083402} (\bibinfo
  {year} {2019})},\ \Eprint {https://arxiv.org/abs/1809.10927}
  {arXiv:1809.10927} \BibitemShut {NoStop}%
\bibitem [{\citenamefont {Nietner}\ \emph {et~al.}(2014)\citenamefont
  {Nietner}, \citenamefont {Schaller},\ and\ \citenamefont
  {Brandes}}]{nietner_transport_2014}%
  \BibitemOpen
  \bibfield  {author} {\bibinfo {author} {\bibfnamefont {C.}~\bibnamefont
  {Nietner}}, \bibinfo {author} {\bibfnamefont {G.}~\bibnamefont {Schaller}},\
  and\ \bibinfo {author} {\bibfnamefont {T.}~\bibnamefont {Brandes}},\ }\href
  {https://doi.org/10.1103/PhysRevA.89.013605} {\bibfield  {journal} {\bibinfo
  {journal} {Physical Review A}\ }\textbf {\bibinfo {volume} {89}},\ \bibinfo
  {pages} {013605} (\bibinfo {year} {2014})}\BibitemShut {NoStop}%
\bibitem [{\citenamefont {Ran{\c c}on}\ \emph {et~al.}(2014)\citenamefont
  {Ran{\c c}on}, \citenamefont {Chin},\ and\ \citenamefont
  {Levin}}]{rancon_bosonic_2014}%
  \BibitemOpen
  \bibfield  {author} {\bibinfo {author} {\bibfnamefont {A.}~\bibnamefont
  {Ran{\c c}on}}, \bibinfo {author} {\bibfnamefont {C.}~\bibnamefont {Chin}},\
  and\ \bibinfo {author} {\bibfnamefont {K.}~\bibnamefont {Levin}},\ }\href
  {https://doi.org/10.1088/1367-2630/16/11/113072} {\bibfield  {journal}
  {\bibinfo  {journal} {New Journal of Physics}\ }\textbf {\bibinfo {volume}
  {16}},\ \bibinfo {pages} {113072} (\bibinfo {year} {2014})},\ \bibinfo {note}
  {publisher: IOP Publishing}\BibitemShut {NoStop}%
\bibitem [{\citenamefont {Meir}\ and\ \citenamefont
  {Wingreen}(1992)}]{MeirWingreenformula}%
  \BibitemOpen
  \bibfield  {author} {\bibinfo {author} {\bibfnamefont {Y.}~\bibnamefont
  {Meir}}\ and\ \bibinfo {author} {\bibfnamefont {N.~S.}\ \bibnamefont
  {Wingreen}},\ }\href {https://doi.org/10.1103/PhysRevLett.68.2512} {\bibfield
   {journal} {\bibinfo  {journal} {Phys. Rev. Lett.}\ }\textbf {\bibinfo
  {volume} {68}},\ \bibinfo {pages} {2512} (\bibinfo {year}
  {1992})}\BibitemShut {NoStop}%
\bibitem [{\citenamefont {Breuer}\ and\ \citenamefont
  {Petruccione}(2002)}]{BreuerPetruccione_book}%
  \BibitemOpen
  \bibfield  {author} {\bibinfo {author} {\bibfnamefont {H.~P.}\ \bibnamefont
  {Breuer}}\ and\ \bibinfo {author} {\bibfnamefont {F.}~\bibnamefont
  {Petruccione}},\ }\href@noop {} {\emph {\bibinfo {title} {The theory of open
  quantum systems}}}\ (\bibinfo  {publisher} {Oxford University Press},\
  \bibinfo {address} {Great Clarendon Street},\ \bibinfo {year}
  {2002})\BibitemShut {NoStop}%
\bibitem [{\citenamefont {Gardiner}\ and\ \citenamefont
  {Zoller}(2000)}]{GardinerZoller_quantumnoise}%
  \BibitemOpen
  \bibfield  {author} {\bibinfo {author} {\bibfnamefont {C.}~\bibnamefont
  {Gardiner}}\ and\ \bibinfo {author} {\bibfnamefont {P.}~\bibnamefont
  {Zoller}},\ }\href {https://books.google.fr/books?id=4bJ6MgEACAAJ} {\emph
  {\bibinfo {title} {Quantum Noise: A Handbook of Markovian and Non-Markovian
  Quantum Stochastic Methods with Applications to Quantum Optics}}},\ Springer
  series in synergetics\ (\bibinfo  {publisher} {Springer},\ \bibinfo {year}
  {2000})\BibitemShut {NoStop}%
\bibitem [{\citenamefont {Lindblad}(1976)}]{Lindblad_seminalpaper}%
  \BibitemOpen
  \bibfield  {author} {\bibinfo {author} {\bibfnamefont {G.}~\bibnamefont
  {Lindblad}},\ }\href {https://doi.org/10.1007/BF01608499} {\bibfield
  {journal} {\bibinfo  {journal} {Communications in Mathematical Physics}\
  }\textbf {\bibinfo {volume} {48}},\ \bibinfo {pages} {119} (\bibinfo {year}
  {1976})}\BibitemShut {NoStop}%
\bibitem [{\citenamefont {Gorini}\ \emph {et~al.}(1976)\citenamefont {Gorini},
  \citenamefont {Kossakowski},\ and\ \citenamefont {Sudarshan}}]{Gorini1976}%
  \BibitemOpen
  \bibfield  {author} {\bibinfo {author} {\bibfnamefont {V.}~\bibnamefont
  {Gorini}}, \bibinfo {author} {\bibfnamefont {A.}~\bibnamefont
  {Kossakowski}},\ and\ \bibinfo {author} {\bibfnamefont {E.~C.~G.}\
  \bibnamefont {Sudarshan}},\ }\href {https://doi.org/10.1063/1.522979}
  {\bibfield  {journal} {\bibinfo  {journal} {Journal of Mathematical Physics}\
  }\textbf {\bibinfo {volume} {17}},\ \bibinfo {pages} {821} (\bibinfo {year}
  {1976})}\BibitemShut {NoStop}%
\bibitem [{\citenamefont {Wichterich}\ \emph {et~al.}(2007)\citenamefont
  {Wichterich}, \citenamefont {Henrich}, \citenamefont {Breuer}, \citenamefont
  {Gemmer},\ and\ \citenamefont {Michel}}]{Breuer_HeattransportCPmaps}%
  \BibitemOpen
  \bibfield  {author} {\bibinfo {author} {\bibfnamefont {H.}~\bibnamefont
  {Wichterich}}, \bibinfo {author} {\bibfnamefont {M.~J.}\ \bibnamefont
  {Henrich}}, \bibinfo {author} {\bibfnamefont {H.-P.}\ \bibnamefont {Breuer}},
  \bibinfo {author} {\bibfnamefont {J.}~\bibnamefont {Gemmer}},\ and\ \bibinfo
  {author} {\bibfnamefont {M.}~\bibnamefont {Michel}},\ }\href
  {https://doi.org/10.1103/PhysRevE.76.031115} {\bibfield  {journal} {\bibinfo
  {journal} {Phys. Rev. E}\ }\textbf {\bibinfo {volume} {76}},\ \bibinfo
  {pages} {031115} (\bibinfo {year} {2007})}\BibitemShut {NoStop}%
\bibitem [{\citenamefont {Bertini}\ \emph {et~al.}(2020)\citenamefont
  {Bertini}, \citenamefont {Heidrich-Meisner}, \citenamefont {Karrasch},
  \citenamefont {Prosen}, \citenamefont {Steinigeweg},\ and\ \citenamefont
  {Znidaric}}]{Bertini2020}%
  \BibitemOpen
  \bibfield  {author} {\bibinfo {author} {\bibfnamefont {B.}~\bibnamefont
  {Bertini}}, \bibinfo {author} {\bibfnamefont {F.}~\bibnamefont
  {Heidrich-Meisner}}, \bibinfo {author} {\bibfnamefont {C.}~\bibnamefont
  {Karrasch}}, \bibinfo {author} {\bibfnamefont {T.}~\bibnamefont {Prosen}},
  \bibinfo {author} {\bibfnamefont {R.}~\bibnamefont {Steinigeweg}},\ and\
  \bibinfo {author} {\bibfnamefont {M.}~\bibnamefont {Znidaric}},\ }\href
  {http://arxiv.org/abs/2003.03334} {\bibfield  {journal} {\bibinfo  {journal}
  {arXiv:2003.03334 [cond-mat, physics:quant-ph]}\ } (\bibinfo {year}
  {2020})}\BibitemShut {NoStop}%
\bibitem [{\citenamefont
  {{\v{Z}}nidari{\v{c}}}(2010{\natexlab{a}})}]{Znidaric_MPS_XX_open}%
  \BibitemOpen
  \bibfield  {author} {\bibinfo {author} {\bibfnamefont {M.}~\bibnamefont
  {{\v{Z}}nidari{\v{c}}}},\ }\href
  {https://doi.org/10.1088/1751-8113/43/41/415004} {\bibfield  {journal}
  {\bibinfo  {journal} {Journal of Physics A: Mathematical and Theoretical}\
  }\textbf {\bibinfo {volume} {43}},\ \bibinfo {pages} {415004} (\bibinfo
  {year} {2010}{\natexlab{a}})}\BibitemShut {NoStop}%
\bibitem [{\citenamefont
  {{\v{Z}}nidari{\v{c}}}(2010{\natexlab{b}})}]{Znidaric__XXdeph}%
  \BibitemOpen
  \bibfield  {author} {\bibinfo {author} {\bibfnamefont {M.}~\bibnamefont
  {{\v{Z}}nidari{\v{c}}}},\ }\href
  {https://doi.org/10.1088/1742-5468/2010/05/l05002} {\bibfield  {journal}
  {\bibinfo  {journal} {Journal of Statistical Mechanics: Theory and
  Experiment}\ }\textbf {\bibinfo {volume} {2010}},\ \bibinfo {pages} {L05002}
  (\bibinfo {year} {2010}{\natexlab{b}})}\BibitemShut {NoStop}%
\bibitem [{\citenamefont {Prosen}(2011{\natexlab{a}})}]{Prosen_OpenXXZ}%
  \BibitemOpen
  \bibfield  {author} {\bibinfo {author} {\bibfnamefont {T.}~\bibnamefont
  {Prosen}},\ }\href {https://doi.org/10.1103/PhysRevLett.106.217206}
  {\bibfield  {journal} {\bibinfo  {journal} {Phys. Rev. Lett.}\ }\textbf
  {\bibinfo {volume} {106}},\ \bibinfo {pages} {217206} (\bibinfo {year}
  {2011}{\natexlab{a}})}\BibitemShut {NoStop}%
\bibitem [{\citenamefont {{Medvedyeva}}\ and\ \citenamefont
  {{Kehrein}}(2013)}]{Kehrein_Lindblad_fullcounting}%
  \BibitemOpen
  \bibfield  {author} {\bibinfo {author} {\bibfnamefont {M.~V.}\ \bibnamefont
  {{Medvedyeva}}}\ and\ \bibinfo {author} {\bibfnamefont {S.}~\bibnamefont
  {{Kehrein}}},\ }\href@noop {} {\bibfield  {journal} {\bibinfo  {journal}
  {arXiv e-prints}\ ,\ \bibinfo {eid} {arXiv:1310.4997}} (\bibinfo {year}
  {2013})},\ \Eprint {https://arxiv.org/abs/1310.4997} {arXiv:1310.4997
  [cond-mat.mes-hall]} \BibitemShut {NoStop}%
\bibitem [{\citenamefont {Karevski}\ \emph {et~al.}(2013)\citenamefont
  {Karevski}, \citenamefont {Popkov},\ and\ \citenamefont
  {Sch\"utz}}]{Schutz_MPSXXZLindblad}%
  \BibitemOpen
  \bibfield  {author} {\bibinfo {author} {\bibfnamefont {D.}~\bibnamefont
  {Karevski}}, \bibinfo {author} {\bibfnamefont {V.}~\bibnamefont {Popkov}},\
  and\ \bibinfo {author} {\bibfnamefont {G.~M.}\ \bibnamefont {Sch\"utz}},\
  }\href {https://doi.org/10.1103/PhysRevLett.110.047201} {\bibfield  {journal}
  {\bibinfo  {journal} {Phys. Rev. Lett.}\ }\textbf {\bibinfo {volume} {110}},\
  \bibinfo {pages} {047201} (\bibinfo {year} {2013})}\BibitemShut {NoStop}%
\bibitem [{\citenamefont {Bu\ifmmode~\check{c}\else \v{c}\fi{}a}\ and\
  \citenamefont {Prosen}(2014)}]{ProsenBuca_Countingstatistics}%
  \BibitemOpen
  \bibfield  {author} {\bibinfo {author} {\bibfnamefont {B.}~\bibnamefont
  {Bu\ifmmode~\check{c}\else \v{c}\fi{}a}}\ and\ \bibinfo {author}
  {\bibfnamefont {T.}~\bibnamefont {Prosen}},\ }\href
  {https://doi.org/10.1103/PhysRevLett.112.067201} {\bibfield  {journal}
  {\bibinfo  {journal} {Phys. Rev. Lett.}\ }\textbf {\bibinfo {volume} {112}},\
  \bibinfo {pages} {067201} (\bibinfo {year} {2014})}\BibitemShut {NoStop}%
\bibitem [{\citenamefont {Guimar\~aes}\ \emph {et~al.}(2016)\citenamefont
  {Guimar\~aes}, \citenamefont {Landi},\ and\ \citenamefont
  {de~Oliveira}}]{Guimaraes_NonequilibriummultisiteLindblad}%
  \BibitemOpen
  \bibfield  {author} {\bibinfo {author} {\bibfnamefont {P.~H.}\ \bibnamefont
  {Guimar\~aes}}, \bibinfo {author} {\bibfnamefont {G.~T.}\ \bibnamefont
  {Landi}},\ and\ \bibinfo {author} {\bibfnamefont {M.~J.}\ \bibnamefont
  {de~Oliveira}},\ }\href {https://doi.org/10.1103/PhysRevE.94.032139}
  {\bibfield  {journal} {\bibinfo  {journal} {Phys. Rev. E}\ }\textbf {\bibinfo
  {volume} {94}},\ \bibinfo {pages} {032139} (\bibinfo {year}
  {2016})}\BibitemShut {NoStop}%
\bibitem [{\citenamefont {Guo}\ and\ \citenamefont
  {Poletti}(2017)}]{Poletti_Quadraticopensystems}%
  \BibitemOpen
  \bibfield  {author} {\bibinfo {author} {\bibfnamefont {C.}~\bibnamefont
  {Guo}}\ and\ \bibinfo {author} {\bibfnamefont {D.}~\bibnamefont {Poletti}},\
  }\href {https://doi.org/10.1103/PhysRevA.95.052107} {\bibfield  {journal}
  {\bibinfo  {journal} {Phys. Rev. A}\ }\textbf {\bibinfo {volume} {95}},\
  \bibinfo {pages} {052107} (\bibinfo {year} {2017})}\BibitemShut {NoStop}%
\bibitem [{\citenamefont {{\v Z}nidari{\v
  c}}(2019)}]{znidaric_nonequilibrium_2019}%
  \BibitemOpen
  \bibfield  {author} {\bibinfo {author} {\bibfnamefont {M.}~\bibnamefont {{\v
  Z}nidari{\v c}}},\ }\href {https://doi.org/10.1103/PhysRevB.99.035143}
  {\bibfield  {journal} {\bibinfo  {journal} {Physical Review B}\ }\textbf
  {\bibinfo {volume} {99}},\ \bibinfo {pages} {035143} (\bibinfo {year}
  {2019})},\ \bibinfo {note} {publisher: American Physical Society}\BibitemShut
  {NoStop}%
\bibitem [{\citenamefont {Bernard}\ and\ \citenamefont
  {Jin}(2019)}]{BernardJin_QSSEP}%
  \BibitemOpen
  \bibfield  {author} {\bibinfo {author} {\bibfnamefont {D.}~\bibnamefont
  {Bernard}}\ and\ \bibinfo {author} {\bibfnamefont {T.}~\bibnamefont {Jin}},\
  }\href {https://doi.org/10.1103/PhysRevLett.123.080601} {\bibfield  {journal}
  {\bibinfo  {journal} {Phys. Rev. Lett.}\ }\textbf {\bibinfo {volume} {123}},\
  \bibinfo {pages} {080601} (\bibinfo {year} {2019})}\BibitemShut {NoStop}%
\bibitem [{\citenamefont {Debnath}\ \emph {et~al.}(2017)\citenamefont
  {Debnath}, \citenamefont {Mascarenhas},\ and\ \citenamefont
  {Savona}}]{debnath_nonequilibrium_2017}%
  \BibitemOpen
  \bibfield  {author} {\bibinfo {author} {\bibfnamefont {K.}~\bibnamefont
  {Debnath}}, \bibinfo {author} {\bibfnamefont {E.}~\bibnamefont
  {Mascarenhas}},\ and\ \bibinfo {author} {\bibfnamefont {V.}~\bibnamefont
  {Savona}},\ }\href {https://doi.org/10.1088/1367-2630/aa969e} {\bibfield
  {journal} {\bibinfo  {journal} {New Journal of Physics}\ }\textbf {\bibinfo
  {volume} {19}},\ \bibinfo {pages} {115006} (\bibinfo {year} {2017})},\
  \bibinfo {note} {publisher: IOP Publishing}\BibitemShut {NoStop}%
\bibitem [{\citenamefont {{Frassek}}\ \emph {et~al.}(2020)\citenamefont
  {{Frassek}}, \citenamefont {{Giardina}},\ and\ \citenamefont
  {{Kurchan}}}]{Kurchan_duality}%
  \BibitemOpen
  \bibfield  {author} {\bibinfo {author} {\bibfnamefont {R.}~\bibnamefont
  {{Frassek}}}, \bibinfo {author} {\bibfnamefont {C.}~\bibnamefont
  {{Giardina}}},\ and\ \bibinfo {author} {\bibfnamefont {J.}~\bibnamefont
  {{Kurchan}}},\ }\href@noop {} {\bibfield  {journal} {\bibinfo  {journal}
  {arXiv e-prints}\ ,\ \bibinfo {eid} {arXiv:2008.03476}} (\bibinfo {year}
  {2020})},\ \Eprint {https://arxiv.org/abs/2008.03476} {arXiv:2008.03476
  [cond-mat.stat-mech]} \BibitemShut {NoStop}%
\bibitem [{\citenamefont {Damanet}\ \emph
  {et~al.}(2019{\natexlab{a}})\citenamefont {Damanet}, \citenamefont
  {Mascarenhas}, \citenamefont {Pekker},\ and\ \citenamefont
  {Daley}}]{damanet_reservoir_2019}%
  \BibitemOpen
  \bibfield  {author} {\bibinfo {author} {\bibfnamefont {F.}~\bibnamefont
  {Damanet}}, \bibinfo {author} {\bibfnamefont {E.}~\bibnamefont
  {Mascarenhas}}, \bibinfo {author} {\bibfnamefont {D.}~\bibnamefont
  {Pekker}},\ and\ \bibinfo {author} {\bibfnamefont {A.~J.}\ \bibnamefont
  {Daley}},\ }\href {https://doi.org/10.1088/1367-2630/ab4f5d} {\bibfield
  {journal} {\bibinfo  {journal} {New Journal of Physics}\ }\textbf {\bibinfo
  {volume} {21}},\ \bibinfo {pages} {115001} (\bibinfo {year}
  {2019}{\natexlab{a}})},\ \bibinfo {note} {publisher: IOP
  Publishing}\BibitemShut {NoStop}%
\bibitem [{\citenamefont {Damanet}\ \emph
  {et~al.}(2019{\natexlab{b}})\citenamefont {Damanet}, \citenamefont
  {Mascarenhas}, \citenamefont {Pekker},\ and\ \citenamefont
  {Daley}}]{damanet_controlling_2019}%
  \BibitemOpen
  \bibfield  {author} {\bibinfo {author} {\bibfnamefont {F.}~\bibnamefont
  {Damanet}}, \bibinfo {author} {\bibfnamefont {E.}~\bibnamefont
  {Mascarenhas}}, \bibinfo {author} {\bibfnamefont {D.}~\bibnamefont
  {Pekker}},\ and\ \bibinfo {author} {\bibfnamefont {A.~J.}\ \bibnamefont
  {Daley}},\ }\href {https://doi.org/10.1103/PhysRevLett.123.180402} {\bibfield
   {journal} {\bibinfo  {journal} {Physical Review Letters}\ }\textbf {\bibinfo
  {volume} {123}},\ \bibinfo {pages} {180402} (\bibinfo {year}
  {2019}{\natexlab{b}})},\ \bibinfo {note} {publisher: American Physical
  Society}\BibitemShut {NoStop}%
\bibitem [{\citenamefont {Zerah-Harush}\ and\ \citenamefont
  {Dubi}(2020)}]{Dubi_DisorderInteractionQtransport}%
  \BibitemOpen
  \bibfield  {author} {\bibinfo {author} {\bibfnamefont {E.}~\bibnamefont
  {Zerah-Harush}}\ and\ \bibinfo {author} {\bibfnamefont {Y.}~\bibnamefont
  {Dubi}},\ }\href {https://doi.org/10.1103/PhysRevResearch.2.023294}
  {\bibfield  {journal} {\bibinfo  {journal} {Phys. Rev. Research}\ }\textbf
  {\bibinfo {volume} {2}},\ \bibinfo {pages} {023294} (\bibinfo {year}
  {2020})}\BibitemShut {NoStop}%
\bibitem [{\citenamefont {Prosen}(2008)}]{Prosen_thirdquantization}%
  \BibitemOpen
  \bibfield  {author} {\bibinfo {author} {\bibfnamefont {T.}~\bibnamefont
  {Prosen}},\ }\href {https://doi.org/10.1088/1367-2630/10/4/043026} {\bibfield
   {journal} {\bibinfo  {journal} {New Journal of Physics}\ }\textbf {\bibinfo
  {volume} {10}},\ \bibinfo {pages} {043026} (\bibinfo {year}
  {2008})}\BibitemShut {NoStop}%
\bibitem [{\citenamefont {Medvedyeva}\ \emph {et~al.}(2016)\citenamefont
  {Medvedyeva}, \citenamefont {Essler},\ and\ \citenamefont
  {Prosen}}]{ProsenEssler_Mapping}%
  \BibitemOpen
  \bibfield  {author} {\bibinfo {author} {\bibfnamefont {M.~V.}\ \bibnamefont
  {Medvedyeva}}, \bibinfo {author} {\bibfnamefont {F.~H.~L.}\ \bibnamefont
  {Essler}},\ and\ \bibinfo {author} {\bibfnamefont {T.~c.~v.}\ \bibnamefont
  {Prosen}},\ }\href {https://doi.org/10.1103/PhysRevLett.117.137202}
  {\bibfield  {journal} {\bibinfo  {journal} {Phys. Rev. Lett.}\ }\textbf
  {\bibinfo {volume} {117}},\ \bibinfo {pages} {137202} (\bibinfo {year}
  {2016})}\BibitemShut {NoStop}%
\bibitem [{\citenamefont {Ziolkowska}\ and\ \citenamefont
  {Essler}(2020)}]{Essler_YBLindblad}%
  \BibitemOpen
  \bibfield  {author} {\bibinfo {author} {\bibfnamefont {A.~A.}\ \bibnamefont
  {Ziolkowska}}\ and\ \bibinfo {author} {\bibfnamefont {F.~H.}\ \bibnamefont
  {Essler}},\ }\href {https://doi.org/10.21468/SciPostPhys.8.3.044} {\bibfield
  {journal} {\bibinfo  {journal} {SciPost Phys.}\ }\textbf {\bibinfo {volume}
  {8}},\ \bibinfo {pages} {44} (\bibinfo {year} {2020})}\BibitemShut {NoStop}%
\bibitem [{\citenamefont {{Bernard}}\ and\ \citenamefont
  {{Jin}}(2020)}]{BernardJin_SolutionQSSEPcontinue}%
  \BibitemOpen
  \bibfield  {author} {\bibinfo {author} {\bibfnamefont {D.}~\bibnamefont
  {{Bernard}}}\ and\ \bibinfo {author} {\bibfnamefont {T.}~\bibnamefont
  {{Jin}}},\ }\href@noop {} {\bibfield  {journal} {\bibinfo  {journal} {arXiv
  e-prints}\ ,\ \bibinfo {eid} {arXiv:2006.12222}} (\bibinfo {year} {2020})},\
  \Eprint {https://arxiv.org/abs/2006.12222} {arXiv:2006.12222 [math-ph]}
  \BibitemShut {NoStop}%
\bibitem [{\citenamefont {Zotos}\ \emph {et~al.}(1997)\citenamefont {Zotos},
  \citenamefont {Naef},\ and\ \citenamefont {Prelovsek}}]{Zotos1997}%
  \BibitemOpen
  \bibfield  {author} {\bibinfo {author} {\bibfnamefont {X.}~\bibnamefont
  {Zotos}}, \bibinfo {author} {\bibfnamefont {F.}~\bibnamefont {Naef}},\ and\
  \bibinfo {author} {\bibfnamefont {P.}~\bibnamefont {Prelovsek}},\ }\href
  {https://doi.org/10.1103/PhysRevB.55.11029} {\bibfield  {journal} {\bibinfo
  {journal} {Physical Review B}\ }\textbf {\bibinfo {volume} {55}},\ \bibinfo
  {pages} {11029} (\bibinfo {year} {1997})}\BibitemShut {NoStop}%
\bibitem [{\citenamefont {Zotos}(1999)}]{Zotos1999}%
  \BibitemOpen
  \bibfield  {author} {\bibinfo {author} {\bibfnamefont {X.}~\bibnamefont
  {Zotos}},\ }\href {https://doi.org/10.1103/PhysRevLett.82.1764} {\bibfield
  {journal} {\bibinfo  {journal} {Physical Review Letters}\ }\textbf {\bibinfo
  {volume} {82}},\ \bibinfo {pages} {1764} (\bibinfo {year}
  {1999})}\BibitemShut {NoStop}%
\bibitem [{\citenamefont {Prosen}(2011{\natexlab{b}})}]{Prosen2011}%
  \BibitemOpen
  \bibfield  {author} {\bibinfo {author} {\bibfnamefont {T.}~\bibnamefont
  {Prosen}},\ }\href {https://doi.org/10.1103/PhysRevLett.106.217206}
  {\bibfield  {journal} {\bibinfo  {journal} {Physical Review Letters}\
  }\textbf {\bibinfo {volume} {106}},\ \bibinfo {pages} {2} (\bibinfo {year}
  {2011}{\natexlab{b}})},\ \Eprint {https://arxiv.org/abs/1103.1350}
  {arXiv:1103.1350} \BibitemShut {NoStop}%
\bibitem [{\citenamefont {Gopalakrishnan}\ and\ \citenamefont
  {Vasseur}(2019)}]{Gopalakrishnan2019}%
  \BibitemOpen
  \bibfield  {author} {\bibinfo {author} {\bibfnamefont {S.}~\bibnamefont
  {Gopalakrishnan}}\ and\ \bibinfo {author} {\bibfnamefont {R.}~\bibnamefont
  {Vasseur}},\ }\href {https://doi.org/10.1103/PhysRevLett.122.127202}
  {\bibfield  {journal} {\bibinfo  {journal} {Physical Review Letters}\
  }\textbf {\bibinfo {volume} {122}},\ \bibinfo {pages} {127202} (\bibinfo
  {year} {2019})}\BibitemShut {NoStop}%
\bibitem [{\citenamefont {{\v{Z}}nidari{\v{c}}}(2011)}]{Znidaric2011}%
  \BibitemOpen
  \bibfield  {author} {\bibinfo {author} {\bibfnamefont {M.}~\bibnamefont
  {{\v{Z}}nidari{\v{c}}}},\ }\bibfield  {journal} {\bibinfo  {journal}
  {Physical Review Letters}\ }\textbf {\bibinfo {volume} {106}},\ \href
  {https://doi.org/10.1103/PhysRevLett.106.220601}
  {10.1103/PhysRevLett.106.220601} (\bibinfo {year} {2011}),\ \Eprint
  {https://arxiv.org/abs/1103.4094} {arXiv:1103.4094} \BibitemShut {NoStop}%
\bibitem [{\citenamefont {Ljubotina}\ \emph {et~al.}(2017)\citenamefont
  {Ljubotina}, \citenamefont {{\v{Z}}nidaric},\ and\ \citenamefont
  {Prosen}}]{Ljubotina2017}%
  \BibitemOpen
  \bibfield  {author} {\bibinfo {author} {\bibfnamefont {M.}~\bibnamefont
  {Ljubotina}}, \bibinfo {author} {\bibfnamefont {M.}~\bibnamefont
  {{\v{Z}}nidaric}},\ and\ \bibinfo {author} {\bibfnamefont {T.}~\bibnamefont
  {Prosen}},\ }\href {https://doi.org/10.1038/ncomms16117} {\bibfield
  {journal} {\bibinfo  {journal} {Nature Communications}\ }\textbf {\bibinfo
  {volume} {8}},\ \bibinfo {pages} {1} (\bibinfo {year} {2017})},\ \Eprint
  {https://arxiv.org/abs/1702.04210} {arXiv:1702.04210} \BibitemShut {NoStop}%
\bibitem [{\citenamefont {Kardar}\ \emph {et~al.}(1986)\citenamefont {Kardar},
  \citenamefont {Parisi},\ and\ \citenamefont {Zhang}}]{kardar_dynamic_1986}%
  \BibitemOpen
  \bibfield  {author} {\bibinfo {author} {\bibfnamefont {M.}~\bibnamefont
  {Kardar}}, \bibinfo {author} {\bibfnamefont {G.}~\bibnamefont {Parisi}},\
  and\ \bibinfo {author} {\bibfnamefont {Y.-C.}\ \bibnamefont {Zhang}},\ }\href
  {https://doi.org/10.1103/PhysRevLett.56.889} {\bibfield  {journal} {\bibinfo
  {journal} {Physical Review Letters}\ }\textbf {\bibinfo {volume} {56}},\
  \bibinfo {pages} {889} (\bibinfo {year} {1986})},\ \bibinfo {note}
  {publisher: American Physical Society}\BibitemShut {NoStop}%
\bibitem [{\citenamefont {Kriecherbauer}\ and\ \citenamefont
  {Krug}(2010)}]{kriecherbauer_pedestriantextquotesingles_2010}%
  \BibitemOpen
  \bibfield  {author} {\bibinfo {author} {\bibfnamefont {T.}~\bibnamefont
  {Kriecherbauer}}\ and\ \bibinfo {author} {\bibfnamefont {J.}~\bibnamefont
  {Krug}},\ }\href {https://doi.org/10.1088/1751-8113/43/40/403001} {\bibfield
  {journal} {\bibinfo  {journal} {Journal of Physics A: Mathematical and
  Theoretical}\ }\textbf {\bibinfo {volume} {43}},\ \bibinfo {pages} {403001}
  (\bibinfo {year} {2010})},\ \bibinfo {note} {publisher: IOP
  Publishing}\BibitemShut {NoStop}%
\bibitem [{\citenamefont {Ljubotina}\ \emph {et~al.}(2019)\citenamefont
  {Ljubotina}, \citenamefont {{\v Z}nidari{\v c}},\ and\ \citenamefont
  {Prosen}}]{ljubotina_kardar-parisi-zhang_2019}%
  \BibitemOpen
  \bibfield  {author} {\bibinfo {author} {\bibfnamefont {M.}~\bibnamefont
  {Ljubotina}}, \bibinfo {author} {\bibfnamefont {M.}~\bibnamefont {{\v
  Z}nidari{\v c}}},\ and\ \bibinfo {author} {\bibfnamefont {T.}~\bibnamefont
  {Prosen}},\ }\href {https://doi.org/10.1103/PhysRevLett.122.210602}
  {\bibfield  {journal} {\bibinfo  {journal} {Physical Review Letters}\
  }\textbf {\bibinfo {volume} {122}},\ \bibinfo {pages} {210602} (\bibinfo
  {year} {2019})},\ \bibinfo {note} {publisher: American Physical
  Society}\BibitemShut {NoStop}%
\bibitem [{\citenamefont {De~Nardis}\ \emph {et~al.}(2020)\citenamefont
  {De~Nardis}, \citenamefont {Medenjak}, \citenamefont {Karrasch},\ and\
  \citenamefont {Ilievski}}]{de_nardis_universality_2020}%
  \BibitemOpen
  \bibfield  {author} {\bibinfo {author} {\bibfnamefont {J.}~\bibnamefont
  {De~Nardis}}, \bibinfo {author} {\bibfnamefont {M.}~\bibnamefont {Medenjak}},
  \bibinfo {author} {\bibfnamefont {C.}~\bibnamefont {Karrasch}},\ and\
  \bibinfo {author} {\bibfnamefont {E.}~\bibnamefont {Ilievski}},\ }\href
  {https://doi.org/10.1103/PhysRevLett.124.210605} {\bibfield  {journal}
  {\bibinfo  {journal} {Physical Review Letters}\ }\textbf {\bibinfo {volume}
  {124}},\ \bibinfo {pages} {210605} (\bibinfo {year} {2020})},\ \bibinfo
  {note} {publisher: American Physical Society}\BibitemShut {NoStop}%
\bibitem [{\citenamefont {Jin}\ \emph {et~al.}(2020)\citenamefont {Jin},
  \citenamefont {Krajenbrink},\ and\ \citenamefont
  {Bernard}}]{JinKrajenbrinkBernard_QKPZ}%
  \BibitemOpen
  \bibfield  {author} {\bibinfo {author} {\bibfnamefont {T.}~\bibnamefont
  {Jin}}, \bibinfo {author} {\bibfnamefont {A.}~\bibnamefont {Krajenbrink}},\
  and\ \bibinfo {author} {\bibfnamefont {D.}~\bibnamefont {Bernard}},\ }\href
  {https://doi.org/10.1103/PhysRevLett.125.040603} {\bibfield  {journal}
  {\bibinfo  {journal} {Phys. Rev. Lett.}\ }\textbf {\bibinfo {volume} {125}},\
  \bibinfo {pages} {040603} (\bibinfo {year} {2020})}\BibitemShut {NoStop}%
\bibitem [{\citenamefont {Bernard}\ and\ \citenamefont
  {Doussal}(2020)}]{PierreDenis_QKPZ}%
  \BibitemOpen
  \bibfield  {author} {\bibinfo {author} {\bibfnamefont {D.}~\bibnamefont
  {Bernard}}\ and\ \bibinfo {author} {\bibfnamefont {P.~L.}\ \bibnamefont
  {Doussal}},\ }\href {https://doi.org/10.1209/0295-5075/131/10007} {\bibfield
  {journal} {\bibinfo  {journal} {{EPL} (Europhysics Letters)}\ }\textbf
  {\bibinfo {volume} {131}},\ \bibinfo {pages} {10007} (\bibinfo {year}
  {2020})}\BibitemShut {NoStop}%
\bibitem [{\citenamefont {{\v Z}nidari{\v c}}\ \emph
  {et~al.}(2016)\citenamefont {{\v Z}nidari{\v c}}, \citenamefont
  {Scardicchio},\ and\ \citenamefont {Varma}}]{znidaric_diffusive_2016}%
  \BibitemOpen
  \bibfield  {author} {\bibinfo {author} {\bibfnamefont {M.}~\bibnamefont {{\v
  Z}nidari{\v c}}}, \bibinfo {author} {\bibfnamefont {A.}~\bibnamefont
  {Scardicchio}},\ and\ \bibinfo {author} {\bibfnamefont {V.~K.}\ \bibnamefont
  {Varma}},\ }\href {https://doi.org/10.1103/PhysRevLett.117.040601} {\bibfield
   {journal} {\bibinfo  {journal} {Physical Review Letters}\ }\textbf {\bibinfo
  {volume} {117}},\ \bibinfo {pages} {040601} (\bibinfo {year}
  {2016})}\BibitemShut {NoStop}%
\bibitem [{\citenamefont {Mendoza-Arenas}\ \emph
  {et~al.}(2019{\natexlab{a}})\citenamefont {Mendoza-Arenas}, \citenamefont
  {{\v Z}nidari{\v c}}, \citenamefont {Varma}, \citenamefont {Goold},
  \citenamefont {Clark},\ and\ \citenamefont
  {Scardicchio}}]{mendoza-arenas_asymmetry_2019}%
  \BibitemOpen
  \bibfield  {author} {\bibinfo {author} {\bibfnamefont {J.~J.}\ \bibnamefont
  {Mendoza-Arenas}}, \bibinfo {author} {\bibfnamefont {M.}~\bibnamefont {{\v
  Z}nidari{\v c}}}, \bibinfo {author} {\bibfnamefont {V.~K.}\ \bibnamefont
  {Varma}}, \bibinfo {author} {\bibfnamefont {J.}~\bibnamefont {Goold}},
  \bibinfo {author} {\bibfnamefont {S.~R.}\ \bibnamefont {Clark}},\ and\
  \bibinfo {author} {\bibfnamefont {A.}~\bibnamefont {Scardicchio}},\ }\href
  {https://doi.org/10.1103/PhysRevB.99.094435} {\bibfield  {journal} {\bibinfo
  {journal} {Physical Review B}\ }\textbf {\bibinfo {volume} {99}},\ \bibinfo
  {pages} {094435} (\bibinfo {year} {2019}{\natexlab{a}})},\ \bibinfo {note}
  {publisher: American Physical Society}\BibitemShut {NoStop}%
\bibitem [{\citenamefont {{\v Z}nidari{\v c}}\ and\ \citenamefont
  {Ljubotina}(2018)}]{znidaric_interaction_2018}%
  \BibitemOpen
  \bibfield  {author} {\bibinfo {author} {\bibfnamefont {M.}~\bibnamefont {{\v
  Z}nidari{\v c}}}\ and\ \bibinfo {author} {\bibfnamefont {M.}~\bibnamefont
  {Ljubotina}},\ }\href@noop {} {\bibfield  {journal} {\bibinfo  {journal}
  {Proceedings of the National Academy of Sciences}\ }\textbf {\bibinfo
  {volume} {115}},\ \bibinfo {pages} {4595} (\bibinfo {year}
  {2018})}\BibitemShut {NoStop}%
\bibitem [{\citenamefont {Brenes}\ \emph {et~al.}(2018)\citenamefont {Brenes},
  \citenamefont {Mascarenhas}, \citenamefont {Rigol},\ and\ \citenamefont
  {Goold}}]{Brenes2018}%
  \BibitemOpen
  \bibfield  {author} {\bibinfo {author} {\bibfnamefont {M.}~\bibnamefont
  {Brenes}}, \bibinfo {author} {\bibfnamefont {E.}~\bibnamefont {Mascarenhas}},
  \bibinfo {author} {\bibfnamefont {M.}~\bibnamefont {Rigol}},\ and\ \bibinfo
  {author} {\bibfnamefont {J.}~\bibnamefont {Goold}},\ }\href
  {https://doi.org/10.1103/PhysRevB.98.235128} {\bibfield  {journal} {\bibinfo
  {journal} {Physical Review B}\ }\textbf {\bibinfo {volume} {98}},\ \bibinfo
  {pages} {235128} (\bibinfo {year} {2018})}\BibitemShut {NoStop}%
\bibitem [{\citenamefont {Brenes}\ \emph
  {et~al.}(2020{\natexlab{a}})\citenamefont {Brenes}, \citenamefont {LeBlond},
  \citenamefont {Goold},\ and\ \citenamefont {Rigol}}]{Brenes2020}%
  \BibitemOpen
  \bibfield  {author} {\bibinfo {author} {\bibfnamefont {M.}~\bibnamefont
  {Brenes}}, \bibinfo {author} {\bibfnamefont {T.}~\bibnamefont {LeBlond}},
  \bibinfo {author} {\bibfnamefont {J.}~\bibnamefont {Goold}},\ and\ \bibinfo
  {author} {\bibfnamefont {M.}~\bibnamefont {Rigol}},\ }\href
  {http://arxiv.org/abs/2004.04755} {\  (\bibinfo {year}
  {2020}{\natexlab{a}})},\ \Eprint {https://arxiv.org/abs/2004.04755}
  {arXiv:2004.04755} \BibitemShut {NoStop}%
\bibitem [{\citenamefont {Brenes}\ \emph
  {et~al.}(2020{\natexlab{b}})\citenamefont {Brenes}, \citenamefont {Goold},\
  and\ \citenamefont {Rigol}}]{Brenes2020b}%
  \BibitemOpen
  \bibfield  {author} {\bibinfo {author} {\bibfnamefont {M.}~\bibnamefont
  {Brenes}}, \bibinfo {author} {\bibfnamefont {J.}~\bibnamefont {Goold}},\ and\
  \bibinfo {author} {\bibfnamefont {M.}~\bibnamefont {Rigol}},\ }\href
  {http://arxiv.org/abs/2005.12309} {\bibfield  {journal} {\bibinfo  {journal}
  {arXiv:2005.12309 [cond-mat]}\ } (\bibinfo {year}
  {2020}{\natexlab{b}})}\BibitemShut {NoStop}%
\bibitem [{\citenamefont {Znidaric}(2020)}]{znidaric_weak_2020}%
  \BibitemOpen
  \bibfield  {author} {\bibinfo {author} {\bibfnamefont {M.}~\bibnamefont
  {Znidaric}},\ }\href {http://arxiv.org/abs/2006.09793} {\bibfield  {journal}
  {\bibinfo  {journal} {arXiv:2006.09793 [cond-mat, physics:nlin,
  physics:quant-ph]}\ } (\bibinfo {year} {2020})},\ \bibinfo {note} {arXiv:
  2006.09793}\BibitemShut {NoStop}%
\bibitem [{\citenamefont {Ferreira}\ and\ \citenamefont
  {Filippone}(2020)}]{ferreira_ballistic--diffusive_2020}%
  \BibitemOpen
  \bibfield  {author} {\bibinfo {author} {\bibfnamefont {J.~S.}\ \bibnamefont
  {Ferreira}}\ and\ \bibinfo {author} {\bibfnamefont {M.}~\bibnamefont
  {Filippone}},\ }\href {http://arxiv.org/abs/2006.13891} {\bibfield  {journal}
  {\bibinfo  {journal} {arXiv:2006.13891 [cond-mat, physics:quant-ph]}\ }
  (\bibinfo {year} {2020})},\ \bibinfo {note} {arXiv: 2006.13891}\BibitemShut
  {NoStop}%
\bibitem [{\citenamefont {Dorda}\ \emph {et~al.}(2017)\citenamefont {Dorda},
  \citenamefont {Sorantin}, \citenamefont {Linden},\ and\ \citenamefont
  {Arrigoni}}]{dorda_optimized_2017}%
  \BibitemOpen
  \bibfield  {author} {\bibinfo {author} {\bibfnamefont {A.}~\bibnamefont
  {Dorda}}, \bibinfo {author} {\bibfnamefont {M.}~\bibnamefont {Sorantin}},
  \bibinfo {author} {\bibfnamefont {W.~v.~d.}\ \bibnamefont {Linden}},\ and\
  \bibinfo {author} {\bibfnamefont {E.}~\bibnamefont {Arrigoni}},\ }\href
  {https://doi.org/10.1088/1367-2630/aa6ccc} {\bibfield  {journal} {\bibinfo
  {journal} {New Journal of Physics}\ }\textbf {\bibinfo {volume} {19}},\
  \bibinfo {pages} {063005} (\bibinfo {year} {2017})},\ \bibinfo {note}
  {publisher: IOP Publishing}\BibitemShut {NoStop}%
\bibitem [{\citenamefont {Arrigoni}\ and\ \citenamefont
  {Dorda}(2018)}]{Arrigoni_Keldysh}%
  \BibitemOpen
  \bibfield  {author} {\bibinfo {author} {\bibfnamefont {E.}~\bibnamefont
  {Arrigoni}}\ and\ \bibinfo {author} {\bibfnamefont {A.}~\bibnamefont
  {Dorda}},\ }in\ \href {https://doi.org/10.1007/978-3-319-94956-7_4} {\emph
  {\bibinfo {booktitle} {Out-of-Equilibrium Physics of Correlated Electron
  Systems}}}\ (\bibinfo  {publisher} {Springer International Publishing},\
  \bibinfo {year} {2018})\ pp.\ \bibinfo {pages} {121--188}\BibitemShut
  {NoStop}%
\bibitem [{\citenamefont {Kamenev}(2011)}]{kamenev_fieldtheorybook}%
  \BibitemOpen
  \bibfield  {author} {\bibinfo {author} {\bibfnamefont {A.}~\bibnamefont
  {Kamenev}},\ }\href {https://doi.org/10.1017/CBO9781139003667} {\emph
  {\bibinfo {title} {Field Theory of Non-Equilibrium Systems}}}\ (\bibinfo
  {publisher} {Cambridge University Press},\ \bibinfo {year}
  {2011})\BibitemShut {NoStop}%
\bibitem [{\citenamefont {Sieberer}\ \emph {et~al.}(2016)\citenamefont
  {Sieberer}, \citenamefont {Buchhold},\ and\ \citenamefont
  {Diehl}}]{Diehl_KeldyshLindblad}%
  \BibitemOpen
  \bibfield  {author} {\bibinfo {author} {\bibfnamefont {L.~M.}\ \bibnamefont
  {Sieberer}}, \bibinfo {author} {\bibfnamefont {M.}~\bibnamefont {Buchhold}},\
  and\ \bibinfo {author} {\bibfnamefont {S.}~\bibnamefont {Diehl}},\ }\href
  {https://doi.org/10.1088/0034-4885/79/9/096001} {\bibfield  {journal}
  {\bibinfo  {journal} {Reports on Progress in Physics}\ }\textbf {\bibinfo
  {volume} {79}},\ \bibinfo {pages} {096001} (\bibinfo {year}
  {2016})}\BibitemShut {NoStop}%
\bibitem [{\citenamefont {Keldysh}(1965)}]{keldysh_1965}%
  \BibitemOpen
  \bibfield  {author} {\bibinfo {author} {\bibfnamefont {L.~V.}\ \bibnamefont
  {Keldysh}},\ }\href
  {http://www.jetp.ac.ru/cgi-bin/e/index/r/47/4/p1515?a=list} {\bibfield
  {journal} {\bibinfo  {journal} {JETP}\ }\textbf {\bibinfo {volume} {20}},\
  \bibinfo {pages} {1018} (\bibinfo {year} {1965})}\BibitemShut {NoStop}%
\bibitem [{\citenamefont {Altland}\ and\ \citenamefont
  {Simons}(2006)}]{altland_condensed_2006}%
  \BibitemOpen
  \bibfield  {author} {\bibinfo {author} {\bibfnamefont {A.}~\bibnamefont
  {Altland}}\ and\ \bibinfo {author} {\bibfnamefont {B.}~\bibnamefont
  {Simons}},\ }\href@noop {} {\emph {\bibinfo {title} {Condensed matter field
  theory}}}\ (\bibinfo  {publisher} {Cambridge University Press},\ \bibinfo
  {year} {2006})\BibitemShut {NoStop}%
\bibitem [{\citenamefont {Larkin}\ and\ \citenamefont
  {Ovchinnikov}(1977)}]{Larkin_vortices_supra}%
  \BibitemOpen
  \bibfield  {author} {\bibinfo {author} {\bibfnamefont {A.}~\bibnamefont
  {Larkin}}\ and\ \bibinfo {author} {\bibfnamefont {Y.~N.}\ \bibnamefont
  {Ovchinnikov}},\ }\href {http://www.jetp.ac.ru/cgi-bin/dn/e_046_01_0155.pdf}
  {\bibfield  {journal} {\bibinfo  {journal} {Journal of Experimental and
  Theoretical Physics}\ }\textbf {\bibinfo {volume} {46}},\ \bibinfo {pages}
  {155} (\bibinfo {year} {1977})}\BibitemShut {NoStop}%
\bibitem [{\citenamefont {Caroli}\ \emph {et~al.}(1971)\citenamefont {Caroli},
  \citenamefont {Combescot}, \citenamefont {Nozieres},\ and\ \citenamefont
  {Saint-James}}]{CaroliNozieres_tunneling_current}%
  \BibitemOpen
  \bibfield  {author} {\bibinfo {author} {\bibfnamefont {C.}~\bibnamefont
  {Caroli}}, \bibinfo {author} {\bibfnamefont {R.}~\bibnamefont {Combescot}},
  \bibinfo {author} {\bibfnamefont {P.}~\bibnamefont {Nozieres}},\ and\
  \bibinfo {author} {\bibfnamefont {D.}~\bibnamefont {Saint-James}},\ }\href
  {https://doi.org/10.1088/0022-3719/4/8/018} {\bibfield  {journal} {\bibinfo
  {journal} {Journal of Physics C: Solid State Physics}\ }\textbf {\bibinfo
  {volume} {4}},\ \bibinfo {pages} {916} (\bibinfo {year} {1971})}\BibitemShut
  {NoStop}%
\bibitem [{\citenamefont {Breit}\ and\ \citenamefont
  {Wigner}(1936)}]{breit_capture_1936}%
  \BibitemOpen
  \bibfield  {author} {\bibinfo {author} {\bibfnamefont {G.}~\bibnamefont
  {Breit}}\ and\ \bibinfo {author} {\bibfnamefont {E.}~\bibnamefont {Wigner}},\
  }\href {https://doi.org/10.1103/PhysRev.49.519} {\bibfield  {journal}
  {\bibinfo  {journal} {Physical Review}\ }\textbf {\bibinfo {volume} {49}},\
  \bibinfo {pages} {519} (\bibinfo {year} {1936})},\ \bibinfo {note}
  {publisher: American Physical Society}\BibitemShut {NoStop}%
\bibitem [{\citenamefont {Tan}(2019)}]{Tan_InverseTridiagonal}%
  \BibitemOpen
  \bibfield  {author} {\bibinfo {author} {\bibfnamefont {L.~S.~L.}\
  \bibnamefont {Tan}},\ }\href {https://doi.org/10.1093/imamat/hxz010}
  {\bibfield  {journal} {\bibinfo  {journal} {IMA Journal of Applied
  Mathematics}\ }\textbf {\bibinfo {volume} {84}},\ \bibinfo {pages} {679}
  (\bibinfo {year} {2019})},\ \Eprint
  {https://arxiv.org/abs/https://academic.oup.com/imamat/article-pdf/84/4/679/29027895/hxz010.pdf}
  {https://academic.oup.com/imamat/article-pdf/84/4/679/29027895/hxz010.pdf}
  \BibitemShut {NoStop}%
\bibitem [{\citenamefont {Dorda}\ \emph {et~al.}(2014)\citenamefont {Dorda},
  \citenamefont {Nuss}, \citenamefont {von~der Linden},\ and\ \citenamefont
  {Arrigoni}}]{DordaArrigoni_Impurity2014}%
  \BibitemOpen
  \bibfield  {author} {\bibinfo {author} {\bibfnamefont {A.}~\bibnamefont
  {Dorda}}, \bibinfo {author} {\bibfnamefont {M.}~\bibnamefont {Nuss}},
  \bibinfo {author} {\bibfnamefont {W.}~\bibnamefont {von~der Linden}},\ and\
  \bibinfo {author} {\bibfnamefont {E.}~\bibnamefont {Arrigoni}},\ }\href
  {https://doi.org/10.1103/PhysRevB.89.165105} {\bibfield  {journal} {\bibinfo
  {journal} {Phys. Rev. B}\ }\textbf {\bibinfo {volume} {89}},\ \bibinfo
  {pages} {165105} (\bibinfo {year} {2014})}\BibitemShut {NoStop}%
\bibitem [{\citenamefont {{\v{Z}}nidari{\v{c}}}\ and\ \citenamefont
  {Horvat}(2013)}]{Znidaric_dephasing}%
  \BibitemOpen
  \bibfield  {author} {\bibinfo {author} {\bibfnamefont {M.}~\bibnamefont
  {{\v{Z}}nidari{\v{c}}}}\ and\ \bibinfo {author} {\bibfnamefont
  {M.}~\bibnamefont {Horvat}},\ }\href
  {https://doi.org/10.1140/epjb/e2012-30730-9} {\bibfield  {journal} {\bibinfo
  {journal} {The European Physical Journal B}\ }\textbf {\bibinfo {volume}
  {86}},\ \bibinfo {pages} {67} (\bibinfo {year} {2013})}\BibitemShut {NoStop}%
\bibitem [{\citenamefont
  {{\v{Z}}nidari{\v{c}}}(2010{\natexlab{c}})}]{Znidaric__dephasingXXZ}%
  \BibitemOpen
  \bibfield  {author} {\bibinfo {author} {\bibfnamefont {M.}~\bibnamefont
  {{\v{Z}}nidari{\v{c}}}},\ }\href
  {https://doi.org/10.1088/1367-2630/12/4/043001} {\bibfield  {journal}
  {\bibinfo  {journal} {New Journal of Physics}\ }\textbf {\bibinfo {volume}
  {12}},\ \bibinfo {pages} {043001} (\bibinfo {year}
  {2010}{\natexlab{c}})}\BibitemShut {NoStop}%
\bibitem [{\citenamefont {Bauer}\ \emph {et~al.}(2017)\citenamefont {Bauer},
  \citenamefont {Bernard},\ and\ \citenamefont
  {Jin}}]{BauerBernardJin_Stoqdissipative}%
  \BibitemOpen
  \bibfield  {author} {\bibinfo {author} {\bibfnamefont {M.}~\bibnamefont
  {Bauer}}, \bibinfo {author} {\bibfnamefont {D.}~\bibnamefont {Bernard}},\
  and\ \bibinfo {author} {\bibfnamefont {T.}~\bibnamefont {Jin}},\ }\href
  {https://doi.org/10.21468/SciPostPhys.3.5.033} {\bibfield  {journal}
  {\bibinfo  {journal} {SciPost Phys.}\ }\textbf {\bibinfo {volume} {3}},\
  \bibinfo {pages} {033} (\bibinfo {year} {2017})}\BibitemShut {NoStop}%
\bibitem [{\citenamefont {Dolgirev}\ \emph {et~al.}(2020)\citenamefont
  {Dolgirev}, \citenamefont {Marino}, \citenamefont {Sels},\ and\ \citenamefont
  {Demler}}]{dolgirev_non-gaussian_2020}%
  \BibitemOpen
  \bibfield  {author} {\bibinfo {author} {\bibfnamefont {P.~E.}\ \bibnamefont
  {Dolgirev}}, \bibinfo {author} {\bibfnamefont {J.}~\bibnamefont {Marino}},
  \bibinfo {author} {\bibfnamefont {D.}~\bibnamefont {Sels}},\ and\ \bibinfo
  {author} {\bibfnamefont {E.}~\bibnamefont {Demler}},\ }\href
  {https://arxiv.org/abs/2004.07797v3} {\  (\bibinfo {year}
  {2020})}\BibitemShut {NoStop}%
\bibitem [{\citenamefont {Hewson}(1993)}]{hewson_kondo_1993}%
  \BibitemOpen
  \bibfield  {author} {\bibinfo {author} {\bibfnamefont {A.~C.}\ \bibnamefont
  {Hewson}},\ }\href@noop {} {\emph {\bibinfo {title} {The {Kondo} {Problem} to
  {Heavy} {Fermions}}}}\ (\bibinfo  {publisher} {Cambridge University Press,
  Cambridge},\ \bibinfo {year} {1993})\BibitemShut {NoStop}%
\bibitem [{\citenamefont {Bulla}\ \emph {et~al.}(2008)\citenamefont {Bulla},
  \citenamefont {Costi},\ and\ \citenamefont
  {Pruschke}}]{bulla_numerical_2008}%
  \BibitemOpen
  \bibfield  {author} {\bibinfo {author} {\bibfnamefont {R.}~\bibnamefont
  {Bulla}}, \bibinfo {author} {\bibfnamefont {T.~A.}\ \bibnamefont {Costi}},\
  and\ \bibinfo {author} {\bibfnamefont {T.}~\bibnamefont {Pruschke}},\ }\href
  {https://doi.org/10.1103/RevModPhys.80.395} {\bibfield  {journal} {\bibinfo
  {journal} {Reviews of Modern Physics}\ }\textbf {\bibinfo {volume} {80}},\
  \bibinfo {pages} {395} (\bibinfo {year} {2008})}\BibitemShut {NoStop}%
\bibitem [{\citenamefont {Gull}\ \emph {et~al.}(2011)\citenamefont {Gull},
  \citenamefont {Millis}, \citenamefont {Lichtenstein}, \citenamefont
  {Rubtsov}, \citenamefont {Troyer},\ and\ \citenamefont
  {Werner}}]{Gull_qimpurity}%
  \BibitemOpen
  \bibfield  {author} {\bibinfo {author} {\bibfnamefont {E.}~\bibnamefont
  {Gull}}, \bibinfo {author} {\bibfnamefont {A.~J.}\ \bibnamefont {Millis}},
  \bibinfo {author} {\bibfnamefont {A.~I.}\ \bibnamefont {Lichtenstein}},
  \bibinfo {author} {\bibfnamefont {A.~N.}\ \bibnamefont {Rubtsov}}, \bibinfo
  {author} {\bibfnamefont {M.}~\bibnamefont {Troyer}},\ and\ \bibinfo {author}
  {\bibfnamefont {P.}~\bibnamefont {Werner}},\ }\href
  {https://doi.org/10.1103/RevModPhys.83.349} {\bibfield  {journal} {\bibinfo
  {journal} {Rev. Mod. Phys.}\ }\textbf {\bibinfo {volume} {83}},\ \bibinfo
  {pages} {349} (\bibinfo {year} {2011})}\BibitemShut {NoStop}%
\bibitem [{\citenamefont {Mendoza-Arenas}\ \emph
  {et~al.}(2019{\natexlab{b}})\citenamefont {Mendoza-Arenas}, \citenamefont
  {{\v{Z}}nidari{\v{c}}}, \citenamefont {Varma}, \citenamefont {Goold},
  \citenamefont {Clark},\ and\ \citenamefont
  {Scardicchio}}]{MendozaArenas2019}%
  \BibitemOpen
  \bibfield  {author} {\bibinfo {author} {\bibfnamefont {J.~J.}\ \bibnamefont
  {Mendoza-Arenas}}, \bibinfo {author} {\bibfnamefont {M.}~\bibnamefont
  {{\v{Z}}nidari{\v{c}}}}, \bibinfo {author} {\bibfnamefont {V.~K.}\
  \bibnamefont {Varma}}, \bibinfo {author} {\bibfnamefont {J.}~\bibnamefont
  {Goold}}, \bibinfo {author} {\bibfnamefont {S.~R.}\ \bibnamefont {Clark}},\
  and\ \bibinfo {author} {\bibfnamefont {A.}~\bibnamefont {Scardicchio}},\
  }\href {https://doi.org/10.1103/PhysRevB.99.094435} {\bibfield  {journal}
  {\bibinfo  {journal} {Physical Review B}\ }\textbf {\bibinfo {volume} {99}},\
  \bibinfo {pages} {094435} (\bibinfo {year} {2019}{\natexlab{b}})},\ \Eprint
  {https://arxiv.org/abs/1803.11555} {arXiv:1803.11555} \BibitemShut {NoStop}%
\end{thebibliography}%

\end{document}